\documentclass[twocolumn]{emulateapj}

\usepackage{lscape}
\def\ie{i.e.}

\def\deg{\ifmmode^\circ\else$^\circ$\fi}

\def\ah{\ifmmode{^\textrm{\scriptsize h}}\else{$^\textrm{\scriptsize h}$}\fi}
\def\am{\ifmmode{^\textrm{\scriptsize m}}\else{$^\textrm{\scriptsize m}$}\fi}
\def\as{\ifmmode{^\textrm{\scriptsize s}}\else{$^\textrm{\scriptsize s}$}\fi}

\def\zf{\ifmmode{z_{\rm f}}\else{$z_{\rm f}$}\fi}
\def\Ks{\ifmmode{K_{\rm s}}\else{$K_{\rm s}$}\fi}
\def\Mcut{\ifmmode{M_{\rm cut}}\else{$M_{\rm cut}$}\fi}
\def\ncmod{{\tt ncmod}}
\def\min{\ifmmode{m_\mathrm{in}}\else{$m_\mathrm{in}$}\fi}
\def\mout{\ifmmode{m_\mathrm{out}}\else{$m_\mathrm{out}$}\fi}

\def\min{\ifmmode{m_\mathrm{in}}\else{$m_\mathrm{in}$}\fi}
\def\mout{\ifmmode{m_\mathrm{out}}\else{$m_\mathrm{out}$}\fi}

\def\Pinout{\ifmmode{P_\mathrm{in,out}}\else{$P_\mathrm{in,out}$}\fi}
\def\Pminmout{\ifmmode{P(\min;\mout)}\else{$P(\min;\mout)$}\fi}
\def\Pmoutmin{\ifmmode{P(\mout;\min)}\else{$P(\mout;\min)$}\fi}
\def\Psminmout{\ifmmode{P_s(\min;\mout)}\else{$P_s(\min;\mout)$}\fi}
\def\Psmoutmin{\ifmmode{P_s(\mout;\min)}\else{$P_s(\mout;\min)$}\fi}
\def\Esmout{\ifmmode{E_s(\mout)}\else{$E_s(\mout)$}\fi}
\def\Emout{\ifmmode{E(\mout)}\else{$E(\mout)$}\fi}

\slugcomment{04 Nov 2005 version, accepted for publication in ApJ.}

\shortauthors{Eliche-Moral et al.}
\shorttitle{U and B  Number Counts in the Groth Strip}
\begin{document}

\title{GOYA Survey:\\
U and B Number Counts in the Groth-Westphal Strip\altaffilmark{1}}
\author{M.~Carmen Eliche-Moral\altaffilmark{2}, Marc Balcells\altaffilmark{2}, 
Mercedes Prieto\altaffilmark{2,3}, C\'esar E.~Garc\'{\i}a-Dab\'o\altaffilmark{2}, 
Peter Erwin\altaffilmark{2}, and David Crist\'obal-Hornillos\altaffilmark{2}}

\altaffiltext{1}{Based on observations made with the Isaac Newton Telescope
operated on the island of La Palma by the Isaac Newton Group of Telescopes
in the Spanish Observatorio del Roque de los Muchachos of the Instituto de
Astrof\'{\i}sica de Canarias.}

\altaffiltext{2}{Instituto de Astrof\'{\i}sica
de Canarias, C/ V\'{\i}a L\'actea, E-38200 La Laguna, Canary Islands, 
Spain} 
\altaffiltext{3}{Departamento de Astrof\'{\i}sica, Universidad de La Laguna, 
Avda.\ Astrof\'{\i}sico Fco.\ S\'anchez, E-38200 La Laguna, Canary Islands, 
Spain\\
{\tt e-mail\/}: mcem@iac.es; balcells@iac.es; mpm@iac.es; enrique.garcia@iac.es; erwin@iac.es; dch@iac.es}

\begin{abstract}
We present $U$ and $B$ galaxy differential number counts from a field of
$\sim$900 arcmin$^2$, based on GOYA Survey imaging of the HST Groth-Westphal
strip. Source detection efficiency corrections as a function of the object size
have been applied.  A variation of the half-exposure image method has been devised to identify and remove spurious detections.  Achieved 50\% detection efficiencies are 24.8 mag in $U$ and 25.5 mag in $B$ in the Vega system. Number count slopes are $d\log (N)/dm =  0.50 \pm 0.02$ for $B$=21.0-24.5, and 
$d\log (N)/dm =  0.48 \pm 0.03$ for $U$=21.0-24.0. Simple number count models are presented that  simultaneously reproduce the counts over 15 mag in $U$ and $B$,  and over 10 mag in \Ks, using a $\Lambda$-dominated cosmology and SDSS local luminosity functions. Only by setting a recent $\zf\sim 1.5$ formation redshift for early-type, red galaxies do the models reproduce the change of slope observed at $\Ks = 17.5$ in NIR counts.  A moderate optical depth ($\tau_B = 0.6$) for all galaxy types ensures that the recent  formation for ellipticals does not leave a signature in the $U$ or $B$ number counts, which are featureless at intermediate magnitudes. No ad-hoc disappearing populations are needed to explain the counts if number evolution is introduced using an observationally-based $z$-evolution of the merger fraction.  
 
\end{abstract}

\keywords{ catalogs --- cosmology: observations ---  
galaxies: evolution --- galaxies: photometry --- surveys }

\section{Introduction}
\label{Sec:introduction}

\begin{deluxetable*}{llcrl}
\tabletypesize{\scriptsize}
\tablewidth{0pt}
\tableheadfrac{0.01}
\tablecaption{Depths and Areas Reached by Recent Number Counts 
Studies in $U$ and/or $B$\label{Tab:otrascuentas}}
\tablehead{\colhead{Ref.} & Filter & \colhead{Depth} 
& \colhead{Area} & \colhead{Comments}\\
&&\colhead{(mag)}&(arcmin$^2$)& \\
\colhead{(1)} &(2)&\colhead{(3)}&\colhead{(4)}&\colhead{(5)} }  
\startdata
Present work & $U$ & 24.8 & 846 & 2.5m INT/WFC \\
 &$B$ &25.5 &888 & 2.5m INT/WFC\\
 \citet{Capak04} & $U$& 25.4 &720   &Hawaii HDF-N\\
    	         & $B$& 26.1& 720  &Hawaii HDF-N \\
\citet{Radovich04} & $U$&  24.4& 2520& VIRMOS Deep Imaging Survey\\
\citet{Huang01b} & $B$ & 22.5 & 1080 & CADIS Survey \\  
\citet{Kummel01} & $B_{j}$ & 24.5 & 3857 & Northern Ecliptic Pole Field  \\  
\citet{Metcalfe01} & $B_{Harris}$ &  27.5 &  49 & William Herschel Deep Field\\
		&   $U_{RGO}$  &26.5 & 49& William Herschel Deep Field\\
		&   $F300W$ & 27.6 & 5.7&  WFPC2/HDF-N\\
		&   $F450W$ & 28.6& 5.7 & WFPC2/HDF-N\\
	        &   $F300W$ & 26.9 &5.7 &    WFPC2/HDF-S\\
		&   $F450W$ & 28.1  & 5.7 &  WFPC2/HDF-S\\
\citet{Yasuda01} & $u'$ \& $g$ & 21.0 & $1.584\cdot 10^{6}$ & Sloan Digital Sky 
Survey \\  
\citet{Gardner00} & $UV$ & 29.0 & $\sim $20 & HDFs (STIS/HST) \\  
& $FUV$ & 30.0 &  $\sim $20  & HDFs (STIS/HST) \\  
\citet{Crawford00} & $B$ & 26.0 & 165.6 & 2.5m Du Pond Telescope \\  
\citet{Driver98} & $B_{450}$ & 29.0 & 5.7 & HST (WFPC2) \\  
\citet{Hogg97} & $U$ & 25.5 & 81 & 5m Hale Telescope \\  
\citet{Williams96} & $U_{300}$ & 28.0 & 5.7 & HDF (HST/WFPC2) \\  
& $B_{450}$ & 29.0 & 5.7 & HDF (HST/WFPC2)\\  
\citet{Metcalfe95} & $B$ & 27.5 & 19.7 & 2.5m INT \\  
&  & 28.0 & 3.5 & 4.5m WHT \\[-0.2cm]
\enddata
\tablecomments{Depths have been converted to magnitudes in the Vega system.}
\end{deluxetable*}

The first deep CCD measurements and automatic detection algorithms revealed 
an excess of faint galaxies over the simple 
extrapolation of the tendency of local galaxies \citep{Tyson88}. 
Differences between the measured surface density of galaxies and the predicted 
extrapolation of the local luminosity function (LF) can be related to 
changes in the volume element, to evolution of 
the spectral energy distribution of galaxies, or to the effects of merging. 
Some authors have used LF evolution 
to match number count models to optical data, either in density 
($\phi^\ast$) or in luminosity ($M^\ast$)  
\citep[see][among others]{Lilly91,Metcalfe95}, or number 
evolution by collapse \citep{Glazebrook94,Fried01}; while other works 
insert a population of blue dwarfs that vanishes at z$\sim$0.4
\citep{Babulrees92}. 




The excess in number counts over non-evolution models is more pronounced 
as bluer filters are used \citep[see, e.g.,][]{Odewahn96} . Thus, modeling optical and NIR number counts simultaneously provides additional constraints on 
galaxy evolution.  \citet{Broadhurst92} resolved the optical/NIR difference 
in number counts by invoking merging and an enhancement of the star formation rate in galaxies at moderate redshifts.  \citet{Gardner96} reproduce $B$,
$V$, $I$, and $K$ number counts  until $B\sim 20$ and $K\sim 16$ mag,  
using passive-evolutionary models with a high $B$ normalization. These models, as well as others \citep[e.g.,][]{Kauffmann94,Nakata99,McCracken00,Huang01a}, had at their disposal shallow or noisy NIR count data coming from different sources which often disagree with each other.  With deeper observations, the discrepancy between NIR and optical counts became more pronounced. A degeneracy between the effects of galaxy evolution and cosmology gives rise to different interpretations even of the same observations. Using data from several authors, \citet{Pozzetti96} found that a pure luminosity evolution (PLE) model in an open Universe ($\Omega\sim 0$) fits number counts, colours and redshift distributions reasonably well in $U$, $b_j$, $r_f$, $I$, and $K$. \citet{Huang01b} showed that $B$ and $K$ number counts from Calar Alto Deep Imaging Survey (CADIS) are better reproduced by passive evolution models than by no-evolution ones, and that an open Universe is preferred to an Einstein-de Sitter (EdS), $\Omega = 1$, Universe. Totani and collaborators fitted very deep optical and NIR data from the Hubble Deep Field (HDF) and the Subaru Deep Field (SDF) respectively \citep{Totani00,Totani01}, including selection effects in the model. Although both data sets were well reproduced with a PLE model in a flat $\Lambda$-dominated Universe, optical number counts needed a mild merger rate ($\mu\sim 1$), while NIR ones were incompatible with merging. \citet{Nagashima02} have fitted the same data as Totani et al.\ using a semianalytical model (SAM) that includes selection effects. Their results rule out the standard CDM, low-density model, and favour a flat $\Lambda$-dominated Universe or a low-density, open Universe. 

To some degree, the various interpretations are probably affected by
field-to-field variations, as well as by
differing data reduction and analysis techniques. It has been a general trait that models that fit optical data 
need to be modified to fit the faint end of the NIR counts. 
One of the key problems revealed by recent, high-quality NIR count data is
the slope change in NIR number counts  at $K = 17.5$ (hereafter, the "knee").  This feature has been reported by several authors 
\citep{Gardner93,Bershady98,McCracken00,Cristobal03} and is now well established.   While CH03 provide a model that reproduces the $\Ks = 17.5$ knee, no model has been yet presented that simultaneously 
reproduces this NIR feature and the counts in blue bands, which do not show a knee at intermediate magnitudes.  

We are carrying out a deep optical-NIR survey as part of a wide project 
for studying galaxy evolution and formation, the GOYA 
Survey\footnote{Known as the 'COSMOS Project' up to 2004 February.  GOYA Project home page:\\
http://www.iac.es/proyect/GOYAiac/GOYAiac.html}. In this paper we present 
$U$ and $B$ number counts over a 
$\sim$900 arcmin$^2$ area of sky, covering one of the GOYA survey fields, 
the Groth-Westphal Strip (GWS). Our $U$ number counts present one of the highest product depth$\times$area reached at the moment (see Table \ref{Tab:otrascuentas}). 
The present $U$ and $B$ galaxy number counts are complementary to the K$_S$ number 
counts published by our team (CH03), so we have fitted a
number count model to our optical ($U$ and $B$) and NIR ($K$) data over
the GWS to put new constraints to the different ingredients of the 
galaxy number count models.

The paper is organized as follows. The GOYA Survey and the 
GWS are described in \S\ref{Sec:goya}. Comments on the observations are 
in \S\ref{Sec:observations}, while reduction is described at 
\S\ref{Sec:reduction}. Source extraction and estimation of detection 
efficiency, reliability, and Galactic extinction are presented in 
\S\ref{Sec:sourceextraction}, which summarizes the catalog
generation process. In \S\ref{Sec:stargalaxy}, we explain the 
procedure to subtract stars counts. Final $U$ and 
$B$ galaxy number counts over the GWS field and modeling 
are presented and discussed in \S\ref{Sec:results} and \S\ref{Sec:modeling}. 
A brief summary is given in \S\ref{Sec:summary}.
We use a $\Omega_M = 0.3$, 
$\Omega_\Lambda = 0.7$, $H_0 = 70$ km s$^{-1}$ Mpc$^{-1}$ cosmology.

\begin{figure}[!ht]
\begin{center}
\plotone{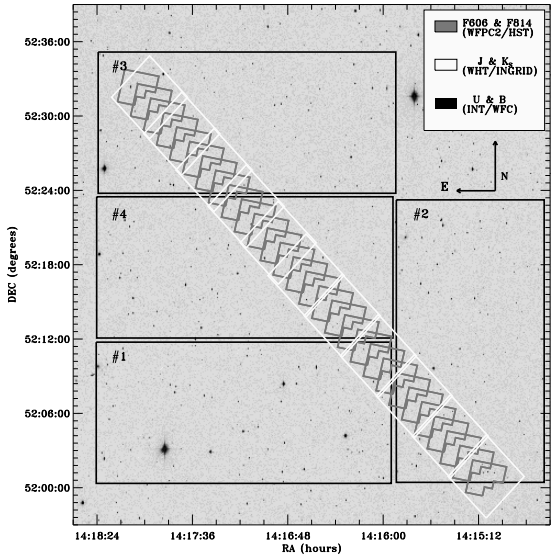}
\caption{Observations of GOYA Survey 
at the GWS region over a DSS image. 
The 28 HST/WFPC2 pointings defining the GWS in $F606W$ and $F814W$ are shown in grey \citep[see][]{Ratnatunga95}, 
while $J$ and $K_S$ WHT/INGRID fields are shown in white (see CH03 for more information about $K_S$ data). The GWS runs diagonally across 
the wide 45\arcmin$\times$45\arcmin\  field marked in black, which corresponds 
to the INT/WFC field ($U$ and $B$ data, presented here). Numbers in black indicate positions and orientations of the 4 chips of WFC.}\label{Fig:fields}
\end{center}
\end{figure}

\section{GOYA Survey}
\label{Sec:goya}

The GOYA Survey is described in detail elsewhere 
\citep[see][CH03, and references therein]{Balcells02}, 
so we proceed to make a brief introduction of the survey. 
GOYA (Galaxy Origins and
Young Assembly) is a wide project for studying 
galaxy formation and evolution with EMIR, the NIR multiobject spectrograph that will be operated on the 10 m GTC 
\citep[see][]{Balcells98,Balcells00,Balcells02}.

The GOYA photometric Survey is a multi-color survey in six broad band filters 
($U$, $B$, $V$, $I$, $J$, $K_{s}$), covering $\sim$0.5 deg$^2$ of sky 
in several fields, with target depths of $U$=$B$=$V$=$I$=26, and $J$=$K$=22 (AB mags). Its principal aim is to generate a galaxy database for sample selection and characterization for subsequent NIR spectroscopy with EMIR.


The $U$ and $B$ imaging presented here from INT/WFC cover the GWS field \citep{Groth94}. GOYA Survey has also reduced and analysed data over this field in NIR filters from WHT/INGRID ($J$ and $K_S$, see CH03), and in visible filters from HST/WFPC2 \citep[$F606W$ and $F814W$, see][]{Ratnatunga95}. Originally, GWS field was defined as 28 HST/WFPC2 pointings extended along a 45 arcmin strip, centered at $\alpha = 14\ifmmode{^\textrm{\scriptsize h}}\else{$^\textrm{\scriptsize h}$}\fi16\ifmmode{^\textrm{\scriptsize m}}\else{$^\textrm{\scriptsize m}$}\fi38\fs8$ and $\delta$ = 52
\arcdeg 16\arcmin 52\arcsec (J2000.0) and inclined 
40\arcdeg 3\arcmin 48\arcsec to the North. 
It has an area of $\sim$150 arcmin$^{2}$ of sky. $F606W$ and $F814W$ data in the field are provided by the DEEP database\footnote{
DEEP Project Home Page: \\http://deep.ucolick.org/} 
\citep[see][]{Phillips97,Simard02,Weiner05}, as well as morphology and photometry. Exposure times were 4,400 s in $F814W$ and 2,800 s in $F606W$ for 27 pointings, and 25.2 ks in both
WFPC2 filters for a single pointing. In Figure \ref{Fig:fields}, covered areas in the 
available six filters of the GOYA survey are plotted over
a DSS\footnote{Digitized Sky Survey (DSS):\\ 
http://archive.stsci.edu/dss/index.html} image of the GWS sky
region. Compared to other existing optical-NIR surveys, GOYA offers a 
notable increase in the depth$\times$area product in several filters, 
compiling complementary photometry in six optical-NIR 
bands, and morphological and 
surface brightness information from high-resolution HST/WFPC2 images. 

\begin{deluxetable}{lllll}
\tabletypesize{\scriptsize}
\tablewidth{0pt}
\tableheadfrac{0.01}
\tablecaption{INT/WFC Main Parameters\tablenotemark{*}\label{Tab:camera}}
\tablehead{\colhead{Parameter} & \colhead{Value}}
\startdata
Collecting area&\multicolumn{4}{l}{5.07 m$^2$}\\
Focal ratio at WFC & \multicolumn{4}{l}{3.29}\\
Field of view	& \multicolumn{4}{l}{Irregular, 
$\sim$34\arcmin$\times$34\arcmin}\\
Detector type 		& \multicolumn{4}{l}{4 EEV CCDs}\\
Detector format& \multicolumn{4}{l}{2048$\times$4096 pixels}\\
Pixel size  	&\multicolumn{4}{l}{13.5$\times$13.5 $\mu$m}\\
Pixel scale	 &\multicolumn{4}{l}{$0.3334$\,arcsec\,pixel$^{-1}$}\\
Field co-planity  & \multicolumn{4}{l}{$\pm $20 $\mu$}   \\
Readout time & \multicolumn{4}{l}{56 s (full camera) slow mode} \\

Cosmic ray counts & \multicolumn{4}{l}{$\sim$ 2\,000 per hour per chip}\\
\ (at sea level)& \multicolumn{4}{l}{ }\\
Chip planity      & \multicolumn{4}{l}{6-10 $\mu m$}\\
Bad pixels    &  \multicolumn{4}{l}{$\le 2\%$}\\\\[-0.4cm]\\\hline\hline\\[-0.1cm]
\multicolumn{1}{c}{Parameter }                         & CCD\#1  & CCD\#2 & CCD\#3  & CCD\#4 \\\hline\\[-0.1cm]
Gain (e$^-$\,ADU$^{-1}$)  & 2.8 & 3.0  &2.5  &2.9 \\
Bias (ADU)			& 1527  & 1590  & 1623 & 1644  \\
Readout noise (e$^-$)  & 6.4  &6.9  & 5.5  &  5.8 \\
Quantum efficiency  & 67\% & 72\% & 62\% &61\% \\
\ ($U$, @ -120\deg C) &  &   &  & \\
Quantum efficiency  & 80\% & 87\% & 80\% & 78\%\\
\ ($B$, @ -120\deg C) &  &   &  & \\
Dark current  & 3.8 & 3.3 & 2.9 &2.0\\
\ (ADU/hr, @ -120\deg C)&  &   &  & \\ \\[-0.4cm]\\\hline\hline\\[-0.1cm]

 \multicolumn{1}{l}{Filters} & \multicolumn{2}{l}{Peak (\AA)}&\multicolumn{2}{l}{Width (\AA)}\\\hline\\[-0.1cm]
\multicolumn{1}{l}{$U_{RGO}$} & \multicolumn{2}{l}{3518}&\multicolumn{2}{l}{638} \\
\multicolumn{1}{l}{$B_{KPNO}$} & \multicolumn{2}{l}{4407}&\multicolumn{2}{l}{1022} \\[-0.2cm]
\enddata
\tablenotetext{*}{INT/WFC information: \\
http://www.ing.iac.es/Astronomy/instruments/wfc/index.html}
\end{deluxetable}

\section{Observations}
\label{Sec:observations}

$U$ and $B$ observations were obtained during one run in May 2002, using the
Wide Field Camera (WFC) mounted on the prime focus of the 2.5m
Isaac Newton Telescope (INT) at Roque de Los Muchachos Observatory, in La Palma.
The camera consists of a mosaic of four CCDs (see Figure 
\ref{Fig:fields}), each of them
with 2,048$\times$4,096 pixels, giving an irregular field of view of
approximately 34\arcmin$\times$34\arcmin, and a pixel scale of 
0.3334\arcsec /pixel. The main camera and filter characteristics are listed in Table \ref{Tab:camera}. The average interchip spacing is $\sim 1$\arcsec. In order to cover these gaps and to facilitate bad pixel correction in the final stacked images, we dithered the exposures by $\sim$10\arcsec\ in each band, but only in the N-S direction. Therefore, gaps between CCDs \#1, \#3 and \#4 were removed, but not the spacing between CCD\#2 and the others (see Figure \ref{Fig:fields}). The dithering was set to a low value to maximize the area of  maximum exposure time and to asses this last to be uniform. Each image was exposed for 1,800 s, except one $B$ exposure of 1,300 s, for a total integration time of 14,400 s in $U$ and 10,300 in $B$. Effective exposure times must be lower due to the loss of light because of high cirrus at the beginning of the night. Dome and twilight flat-field images were obtained for mapping different pixel responses, and zero exposure frames were taken to
estimate the bias structure in each CCD.

The North-East corner of the field of view suffers from serious vignetting. 
We therefore offset our pointing, and, as a result, miss a fraction of the 
WFPC2 frame corresponding to the South-West end of the GWS.

Atmospheric turbulence produced an average PSF of FWHM$\sim$1.3\arcsec\ in $U$ and $\sim$1.2\arcsec\ in $B$ in the final stacked images, stable over the whole run, with small fluctuations across the field. The maximum ratio between the PSF FWHMs of the inner and outer parts of the mosaic was $\sim 1.2$ in both bands. Standard photometric star fields \citep{Landolt92} were observed during the night at different airmasses in order to correct final images for atmospheric extinction and to determine the photometric zero point for each filter.  Attained limiting magnitudes at 50\% detection efficiency are 24.8 mag in $U$ and 25.5 mag in $B$ in the Vega system (see \S\ref{Sec:sourceextraction} for a description of the efficiency and reliability analysis).

\section{Reduction}
\label{Sec:reduction}

\begin{figure*}
\begin{center}
\plottwo{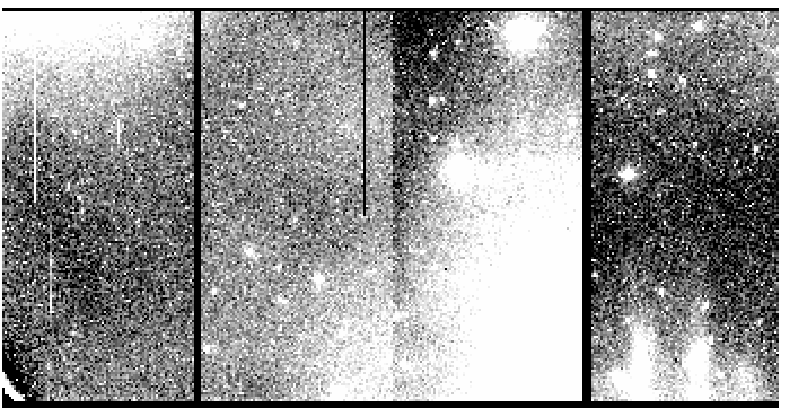}{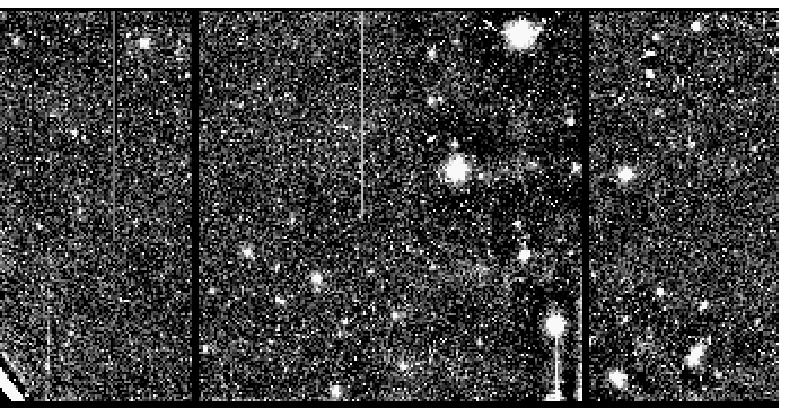}
\caption{
\emph{Left}: Typical diffuse structure in one of the $B$ exposures, due to 
saturated stars' light and reflections in the internal optics of the 
telescope. Beginning from the left, CCDs ID numbers are: 
CCD\#3, CCD\#4, CCD\#1, and 
CCD\#2. Vigneting at left corner of CCD\#3 could not be completely removed 
by flat-fielding. \emph{Right}: The same $B$ pointing once the diffuse pattern 
was subtracted. Grey scaling in both figures is not the same, in order to 
emphasize small residual structures near very bright stars' haloes that 
resulted after diffuse pattern removal, of $\sim$1\% sky level.
\label{Fig:diffuse}}
\end{center}\end{figure*}

\subsection{Pre-reduction}
\label{Sec:prereduction} 
Basic reduction was carried out using a package
specially designed by us for reducing INT/WFC images, and based 
on the {\tt MSCRED\/}\footnote{{\tt MSCRED\/} Home Page: \\
http://iraf.noao.edu/scripts/irafref?mscred} package in IRAF 
\citep{Valdes02,Valdes01,Valdes99,Valdes98a,Valdes98b}. 
CCD mosaic exposures were bias- and dark-corrected
using overscan columns, because the bias structures were lower than  
0.1\%. The ING group\footnote{INT Home Page: \\http://www.ing.iac.es} 
have reported departures from linearity of $\sim$2\% starting at 
$\sim$50,000 ADUs. CASU 
INT linearity coefficients\footnote{
CASU INT Wide Field Survey Home Page:\\
http://www.ast.cam.ac.uk/$\sim$wfcsur/index.php} were used to correct 
for this, which are known to be
very stable over periods of years, and are precise to 0.2\%
in the 0-50 K count range. Images were corrected for 
pixel-to pixel-response using flat-fields from the combination of twilight and dome exposures. Vigneting was completely corrected in CCD\#2, but 
residuals at the North-East corner of CCD\#3 remained after flatfielding 
($\sim$5\% the sky level in $B$ band and $\sim$20\% in $U$). 

After flat-fielding, a \emph{super-flat} image could not be constructed from our data because of  scattered light from
internal optics and saturated stars, which introduced diffuse variable 
patterns of more than 50\% of the sky level over scales of several arcmins.
The 10\arcsec\ of our dither pattern was insufficient for filtering out 
such diffuse patterns when constructing the superflat. 
Diffuse light patterns in wide field cameras may affect the photometry. 
\citet{Lauer97} found that diffuse light affects on small 
scales when combining images, and Capaccioli et al.\ (2001)
\footnote{Capaccioli, M., et al\@.2001, The Capodimonte Deep Field: 
Data reduction and first results on galaxy clusters identification, 
Osservatorio astronomico di Capodimonte (Napoli: Istituto Nazionale di 
Astrofisica), 
\\http://www.na.astro.it/oacdf/OACDFPAP/OACDFPAP.html} found that 
it inserted a $\sim$3\% error in the wide bands of the Capodimonte Deep 
Field. \citet{Manfroid01} showed 
that the zero point changes across the WFI mosaic were a consequence 
of scattered light, inserting an additional error of 0.1 
mag in the $U$ zero point of the VIRMOS photometry 
\citep{Radovich04}. 
In addition to the above problems, we argue that variable sky levels 
insert errors in the structural parameters estimated from isophotal 
analysis. Thus, removing diffuse light in mosaics 
is needed in order to get a reliable photometry. 
\citet{Valdes00} argued that a 
theoretical model of the response of the 
detector-telescope system was necessary to remove diffuse light in the 
NOAO mosaic. 
We built a model surface of the sky by fitting 1-D splines to image in both directions consecutively, using rejection algorithms to suppress objects from the fitting. 
A first aproximation to the diffuse light pattern was obtained by fitting 1-D splines to all the rows of each image, independently from row to row. The posible row-to-row residuals of this first model were smoothed out when we performed a second fit to all the columns of this first model. 
The result was the desired model surface of the sky, with low RMS in small areas of $5\times 5$ pixels (RMS$\lesssim 0.01$). After substracting each sky model to its corresponding image, diffuse light residuals were reduced to less than 1\% of the sky in the whole field of all the exposures (see Figure \ref{Fig:diffuse}). We verified that aperture photometry of stars was unaffected to better than 0.007 mag typically, obtaining higher differences in the regions with the most complex scattered light structures ($\lesssim 0.015$ mag).

From regions of images which were not affected by vigneting or diffuse 
patterns, it could be deduced that the illumination pattern was less than 
0.5\% the sky level. Diffuse pattern residuals are of the same order, so we did not correct for it.

\subsection{Photometry}
\label{Sec:photometry}

Several Landolt (1992) fields were taken for photometric calibration. 
Colors of 
the Landolt stars covered a wide range ($-1.2 <U-B< 2.0$). 
Fitted calibration equations included zero point, 
atmospheric extinction, and color terms, as follows:
\begin{equation}
m_{\mathrm{U}}=U+u_{0}+u_{1}\cdot X+u_{2}\cdot (U-B)
\end{equation} 
\begin{equation}
m_{\mathrm{B}}=B+b_{0}+b_{1}\cdot X+b_{2}\cdot (U-B),
\end{equation} 
\noindent where $m_U$ and $m_B$ are the instrumental magnitudes in the $U$- and $B$-bands 
respectively; $U$ and $B$ are 
the Johnson magnitudes; $u_0$ and $b_0$ represent the zero points; 
$u_1$ and $b_1$ the extinction coefficients; $u_2$ and $b_2$ are the 
color-term coefficients; and $X$ represents the airmass.

Standard stars were positioned 
in the center of the WFC field, on CCD\#4, which is free of vigneting. 
This calibration applies directly 
to the rest of CCDs because all the 
CCDs have been converted to "mean count" units, multiplying the 
flat-fields by the constant factor $N/\langle G\rangle$ ; 
where $N$ is the mean level of the flat-field and 
$\langle G\rangle$ is the mean gain of the 4 CCDs 
(see their values in Table \ref{Tab:camera}). 
Our photometric solution was derived from star fields exposed during the 
second half of the night, given that high cirrus was present during the 
first half. It was applied to the entire dataset by previously scaling 
science frames from the first half of the night to a reference exposure 
from the second half. Photometric calibration results are shown at 
Table \ref{Tab:photometry}.   
Final RMS residuals are 0.09 mag for the $U$-band and 
0.06 mag for the $B$-band. Color terms and zero points were similar to those from the web page of the INT Wide Field Survey. 
Estimates for the $U$ and $B$ extinction terms derived from the 
theoretical extinction curve for La Palma \citep{King85}, as well as 
the extinction measurements of the 
Carlsberg\footnote{Carlsberg Telescope Web: 
\\http://www.ast.cam.ac.uk/$\sim$dwe/SFR/camc\_extinction.html} and 
Mercator\footnote{Mercator telescope Web: 
\\http://www.mercator.iac.es/extinction/extin\_previous.html} 
telescopes for the night of the run, were similar to those obtained 
from our fits.

We applied zero-point and extinction correction to all of our sources, but color terms were only applied to those sources which had counterparts in both filters. Photometric errors were obtained by quadrature-sum of photometric calibration error, Poisson noise and Galactic extinction errors. 
Error values for sources brighter than the 50\% 
limiting magnitude are typically $\sim$0.10 mag in $U$ and 
$\sim$0.05 mag in $B$.

\begin{deluxetable}{cccccc}
\tablewidth{0pt}
\tabletypesize{\small}
\tableheadfrac{0.01}
\tablecaption{Photometric Calibration Coefficients for $U$ and $B$\label{Tab:photometry}}
\tablehead{
Filter  &  Zero Point\tablenotemark{a}& Extinction\tablenotemark{b}& 
Color Term\tablenotemark{c} & RMS\tablenotemark{d}  & X\tablenotemark{e} }
\startdata
$U$ & 23.64$\pm$0.16   & 0.52$\pm$0.11   & -0.13$\pm$0.04 & 0.093  & 1.24 \\
$B$ & 25.34$\pm$0.12  & 0.25$\pm$0.09   & -0.09$\pm$0.03 & 0.062  & 1.62 \\[-0.2cm]
\enddata
\tablenotetext{a}{Magnitudes per ADU s$^{-1}$, in the Vega System}
\tablenotetext{b}{Extinction coefficient (mag airmass$^{-1}$)}
\tablenotetext{c}{Color coefficient (adimensional)}
\tablenotetext{d}{The rms of photometric calibration fit}
\tablenotetext{e}{Airmass of the combined images taken as reference in each filter}

\end{deluxetable}

\subsection{Astrometry}
\label{Sec:astrometry}

Astrometry was performed using IRAF astrometric tasks. The Guide Star Catalog 
II\footnote{Guide Star Catalog II: 
\\http://www-gsss.stsci.edu/gsc/gsc2/GSC2home.htm} (GSC-II) 
provided the coordinates of stars in the field. As with most wide-field devices, the WFC is known to have an important "pincushion" distortion that scales 
as $r^{3}$, where $r$ is the distance to the optical axis \citep{Taylor00}. 
The World Coordinate System (WCS) inserted by the telescope in the 
image headers included linear terms only, with a code which was unrecognizable by IRAF tasks. Therefore, after creating a new primary 
linear WCS in each CCD, we proceeded to describe WFC distortions 
using {\tt msctpeak\/} \citep{Valdes00, Valdes97}, with the TNX 
projection, which includes high-order polynomial terms to the  
tangential projection fit. The {\tt msctpeak\/} task has the disadvantage of 
fitting distortion terms in the transformed coordinate plane, $\eta$ and $\xi$ 
\citep[see][ for a detailed description of WCS coding]{Calabretta02,
Greisen02}, where the 
$r^{3}$ dependence is diluted in a complex combination of cross-terms. 
As the fitting algorithm is not able to give the right weight to 
each term, astrometry at the CCD edges is not improved by the inclusion 
of higher-order terms. 
This fact became evident when we combined the 4 separate WFC frames into 
single images: residuals of up to $\sim$1\arcsec\ 
changed sign abruptly at the chip-to-chip transitions. 
Our conclusion was that a single fit to the entire field was needed. 
Combined single images were created using the relative positions 
and rotations between the four WFC CCDs reported in 
\citet{Taylor00}. Discontinuities in the astrometric solutions 
indicated that errors of $\sim$1\arcsec\ are present in the chip 
separations presented by \citet{Taylor00}. We estimated corrections by  
procedure, offseting the CCD relative positions until we found a combination 
which minimized the astrometric RMS in the whole field. 
We list corrected values in Appendix \ref{Append:offsets}. The achieved RMS astrometric error is down to 0.3\arcsec\ over the entire $\sim$36\arcmin$\times$36\arcmin\ field. 
We have also detected a global rotation of our WCS with respect to 
that which is in the DEEP database for the 
F606W and F814W HST/WFPC2 images of the GWS, which is not related with the 
different celestial reference frames. Our coordinate system is rotated $\sim 0.075^o$ in the tangential plane to the sky respect to the DEEP coordinate system, to the North-East direction. This rotation is centered at $\alpha = 14.28333^\textrm{\scriptsize h}$, $\delta = 52.35$\arcdeg. 
The problem could be due to the small number of stars that 
were used for calibrating the HST images (just 4 for the entire 
GWS, Groth priv.~com.). Our catalogues include source coordinates in both the DEEP and GSC-II systems, in order to facilitate cross-correlation 
with the DEEP database information.




\begin{figure}
\begin{center}
\plotone{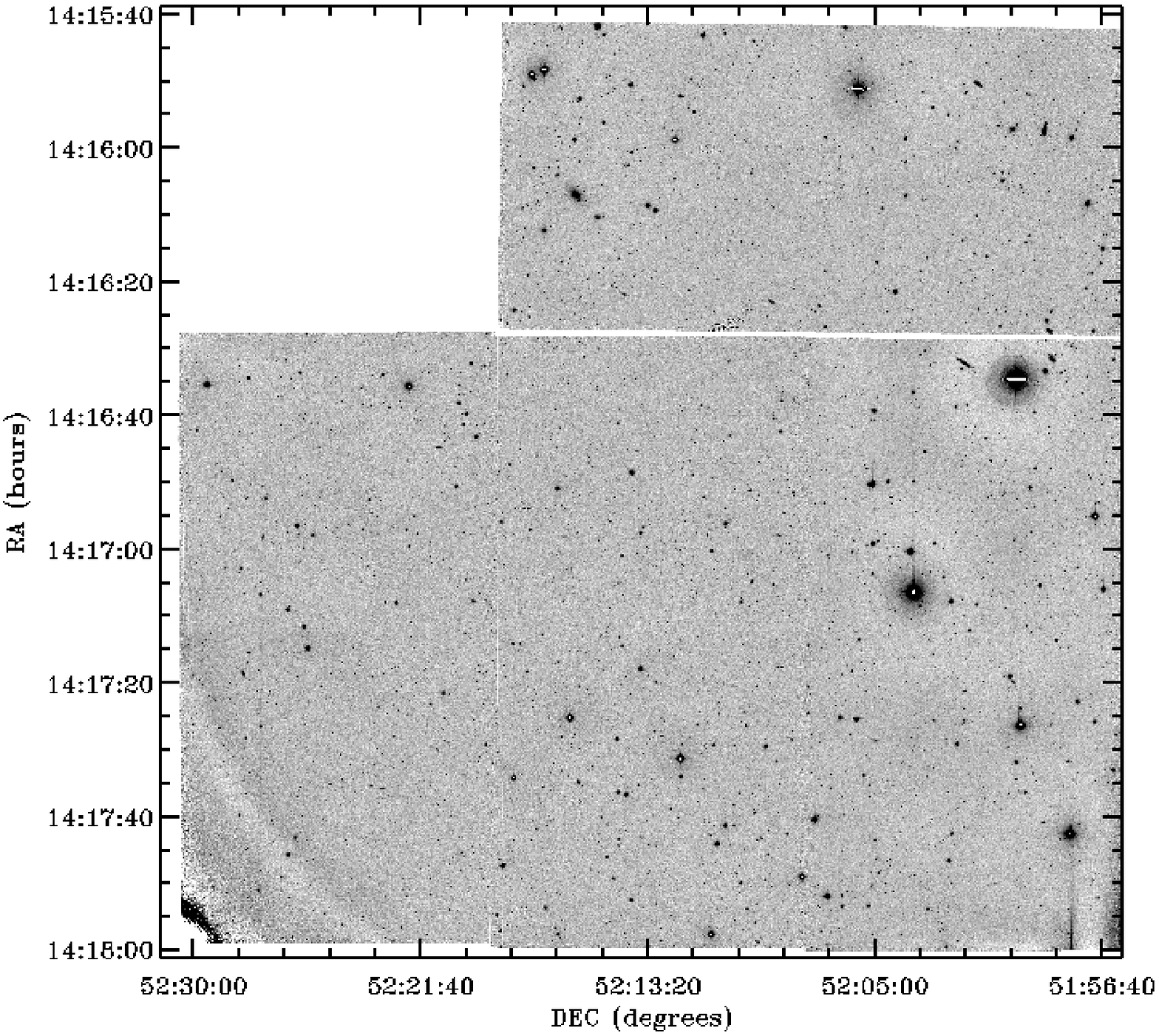}
\plotone{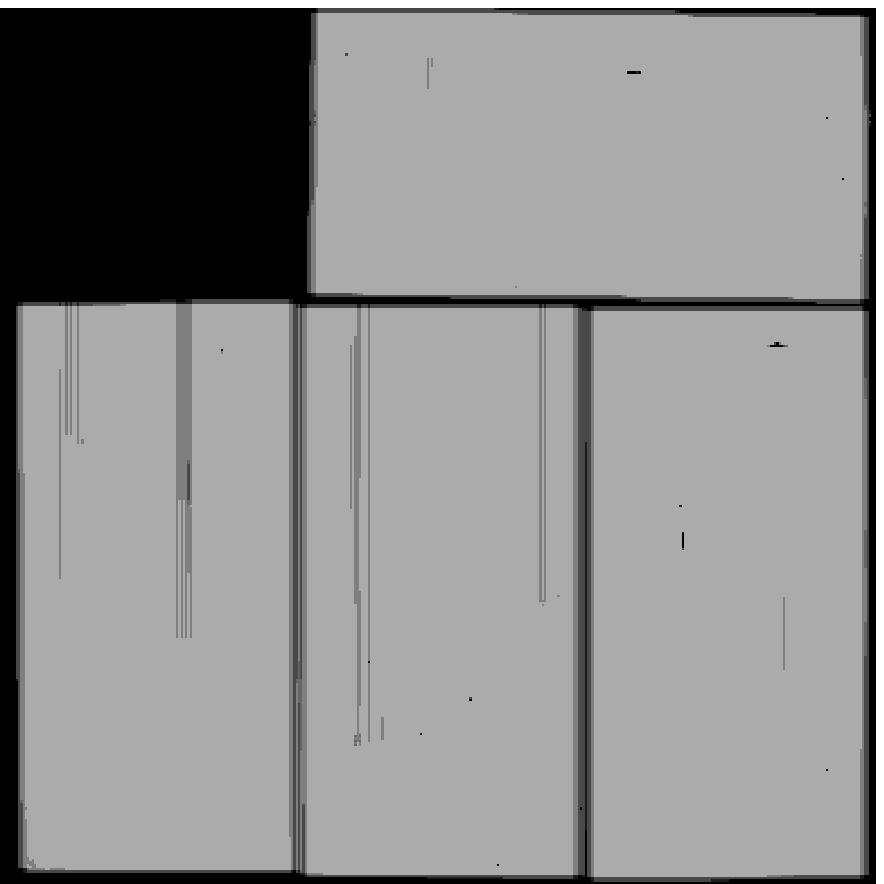}
\end{center}
\caption{\emph{Upper panel}: Final $U$ stacked image over the GWS field. $B$ final image 
is similar. \emph{Lower panel}: $U$ exposure-time map produced after stacking.}
\label{Fig:finalimage}
\end{figure}

\begin{deluxetable*}{cccccccc}
\tabletypesize{\scriptsize}
\tableheadfrac{0.01}
\tablewidth{0pt}
\tablecaption{Main Characteristics of the Final Stacked $U$ and $B$ 
Images\label{Tab:finalimages}}

\tablehead{\colhead{Filter} & \colhead{Dithering} & \colhead{Tot.\,exp.} & \colhead{FWHM} & 
\colhead{Limiting magnitude} & \colhead{r$_e$}   & \colhead{Det. Thresh.}  &
  \colhead{Area for counting} \\
  & & \colhead{(s)}    & \colhead{(arcsec)} & \colhead{(mag)}&\colhead{(arcsec)} &
 \colhead{(sky $\sigma$)} & \colhead{(arcmin$^2$)} \\
 \colhead{(1)}&\colhead{(2)}&\colhead{(3)}&\colhead{(4)}&\colhead{(5)}&
 \colhead{(6)}&\colhead{(7)}&\colhead{(8)}}
\startdata 

$U$ &  $8\times 1800$  &14\, 400 & 1.3  &  24.8 &  1.5, 3 & 0.6  &   846\\
$B$ &  $1300+4\times 1800$  &10\, 300 &  1.2 &  25.5 &  1.5, 3 & 0.6  &   888\\[-0.2cm]
\enddata
\tablecomments{Col.\,(6) shows the effective radius, $r_{\mathrm{e}}$, for group division 
to define "point", "intermediate-size", and "large" objects 
(a detailed description is at \S\ref{Sec:synthetic} and 
\S\ref{Sec:real}). 
Col.\,(7) is the detection threshold used in {\tt SExtractor\/} in sky $\sigma$ units. The final maximum-exposed area over which number 
counts have been computed in each band is listed in Col.\,(8) (see \S\ref{Sec:results}).}
\end{deluxetable*}

\subsection{Coaddition}
\label{Sec:coaddition}

Once the astrometric solution and the flux scaling factors were computed, 
{\tt MSCRED\/} tasks were used for stacking images. 
After subtracting the background, 
images were re-pixeled in order to create single images of the exposures 
on an uniform grid on the sky, free from distortions. Resampling was 
done using a "sinc" function multiplied by an interpolation 
kernel, which maintains the statistical characteristics of the sky noise 
\citep{Valdes98a}. After 
removing atmospheric refraction effects and applying the flux scaling 
factor (see \S\ref{Sec:photometry}), 
stacking of images in $U$ and $B$ was done by computing the 
exptime-weigthed average of 
the images, a process which preserves the Poissonian nature of the 
sky noise \citep{Valdes98a}. {\tt MSCRED\/} algorithms were used 
in order to reject cosmic rays and  
satellite tracks, masks of saturated defective pixels, and bleed 
trails. In Figure \ref{Fig:finalimage} we show final 
$U$-band combined image, and its exposure time map. The $B$ stacked 
image is similar, but $\sim$0.7 mag deeper. Both images cover an 
irregular-shaped field over the GWS 
of $\sim$40\arcmin$\times$40\arcmin , with an inter-ga,p 
non-covered zone between CCD\#2 and the other three. 
The main $U$ and $B$ image characteristics are listed in Table \ref{Tab:finalimages}. 

\section{Source extraction}
\label{Sec:sourceextraction}

A first estimate of the detection limiting magnitude at 3$\sigma$ level 
for point-like sources can be done considering that:
 
\begin{equation}   
\mathrm{mag}(3\sigma)=a_0 -2.5\cdot\log(3\sigma\sqrt[2]{A}),
\end{equation}   
\noindent where a$_0$ is the zero point in each band, and $A$ is the 
area of the selected aperture. We obtained $U_{\mathrm{lim}}\sim$24.4 and 
$B_{\mathrm{lim}}\sim$25.9 at 3$\sigma$ for an aperture of 1.0\arcsec. But 
a more accurate determination of the photometric limits of the survey 
must be done. 

Catalogues were obtained using {\tt SExtractor\/} \citep{Bertin96}, 
which basically considers as a detection every group of 
connected pixels above a fixed detection threshold, 
{\tt DETECT\_THRESH\/}, after filtering the 
image with a detection kernel. CH03 indicate that the detection 
kernel and the minimum area are well constrained as a function of seeing.  
The minimum area was fixed to the area of a symmetrical 
source with a diameter equal to one fourth the FWHM of the stars in the image.
On the other hand, fixing the detection threshold is more difficult because 
number of detected sources will increase as 
we lower the threshold, but so will the 
spurious detections (noise peaks and source substructures identified as sources). A compromise betwen maximization of 
detections and minimization of the spurious fraction must be found. 

Incompleteness effects are quantified through Monte Carlo methods: the behaviour of the detection algorithm is characterized studying how it detects an \emph{a priori} known source distribution, inserted in known positions in the science frames, or in synthetic images that simulate the noise characteristics of the science ones. Previous studies of this nature find that efficiency and reliability estimates based on synthetic images tend to overestimate the efficiency and underestimate the spurious fraction, probably due to overly regular shapes of artificial sources, and to the fact that the real sky noise is not extrictly Poissonian (Bershady et al.\ 1998; CH03).  
Nevertheless, the study with synthetic images is useful to narrow down 
the range of {\tt DETECT\_THRESH\/} values for the more detailed study with 
the science images. Thus, we have measured efficiencies and spurious 
fractions first on synthetic images (see \S\ref{Sec:synthetic}), and after that, we have carried out a more detailed study of incompleteness effects using the science frames, which is going to be taken as the definitive one for completeness correction (\S\ref{Sec:real}). For spurious characterization when  using the science images, we have slightly modified the method of CH03 (\S\ref{Sec:reliability}). 

{\tt SExtractor\/} is more efficient at detecting high surface brightness sources, \ie, compact sources have a higher probability of being found than 
extended sources at the same magnitude. We determine this dependence in the definitive study (with science frames) by 
carrying out the efficiency analysis over three different 
groups of object sizes, as CH03 describe. Bins are defined using the half-light radius of stars in the images, $r_{\mathrm{e,stars}}$: objects 
with $r_{\mathrm{e}}\le 1.5\,r_{\mathrm{e,stars}}$ are "point-like" sources, those with $1.5\,r_{\mathrm{e,stars}}\le r_{\mathrm{e}}\le 3\,r_{\mathrm{e,stars}}$ are "intermediate-sized" objects, and 
those with $r_{\mathrm{e}}\ge 3\,r_{\mathrm{e,stars}}$ are "large" objects. Therefore, our efficiency study using science frames measures detection efficiency not only as a function of the source magnitude, but also of its size (\S\ref{Sec:efficiency}). 

Differential number counts are usually corrected 
for efficiency using the \emph{efficiency matrix}, \Pminmout, or the \emph{efficiency function}, $E(\mout)$. Following \citet{Yan98}, the element $\Pminmout \equiv \Pinout$ of the efficiency matrix is defined as the probability of objects with an original magnitude \min\ to be  detected with magnitude \mout: 
\begin{equation}
\label{eq:efficiencymatrix}
\Pminmout \equiv \frac{N_\mathrm{det}(\min,\mout)}{N_\mathrm{orig}(\min)},
\end{equation}
\noindent where $N_\mathrm{det}(\min,\mout)$ is the number of objects with an original magnitude \min\ which are recovered with magnitude \mout, and  $N_\mathrm{orig}(\min)$ is the total number of objects that originally had magnitude \min. Notice that this definition is independent of the original source distribution. This is due to the fact that the probability for an object with an original magnitude \min\ to be recovered by {\tt SExtractor\/} 
in \mout\ only depends on its size, its original input magnitude \min, and the noise characteristics of the image, but not on the total number of objects with original magnitude \min. The efficiency matrix accounts for the incompleteness effects which are intrinsic to the detection algorithm, the flat-fielding and sky-subtraction errors. Depending on the way this matrix is computed, it also can include magnitude errors caused by crowding.

The \emph{functional efficiency}, \Emout, is usually defined as the fraction of sources detected with magnitude \mout\ irrespective of their input magnitude, from the total number of objets that originally had magnitude \mout:

\begin{equation}
\label{eq:efficiencyfunction}
\Emout \equiv \frac{N_\mathrm{det}(\mout)}{N_\mathrm{orig}(\mout)},
\end{equation}    
where $N_\mathrm{det}(\mout)$ is the number of sources which are detected with magnitud \mout, while $N_\mathrm{orig}(\mout)$ is going to represent the number of sources which originally had magnitude \mout. Therefore, the functional efficiency is defined as a "detection rate", and it is going to depend strongly on the initial number of input sources at each magnitude bin. 

From the previous definitions, the relation between \Emout\ and \Pminmout\ is given by

\begin{equation}
\label{eq:relation}
\Emout = \frac{\sum_{\forall \min}N_\mathrm{det}(\min,\mout)}{N_\mathrm{orig}(\mout)} 
= \frac{\sum_{\forall \min}\Pminmout\,N_\mathrm{orig}(\min)}{N_\mathrm{orig}(\mout)}.
\end{equation}    
\noindent Note that if and only if $N_\mathrm{orig}(\min)$ is constant for all \min, then 
$\Emout = \sum_{\forall \min}\Pminmout $.

The functional efficiency, \Emout, is used more often than the efficiency 
matrix \Pminmout\ in number-count studies \citep{Radovich04,Metcalfe01,Bershady98}, due to the instability which usually arises when inverting \Pminmout\ for applying the efficiency correction. Nevertheless, as pointed out by \citet{Hogg97}, the advantage of using \Pminmout\ is that it corrects not only for completeness errors due to the loss of sources, but also for photometric errors. 

We have corrected number counts for efficiency 
using the two methods: the efficiency matrix, \Pminmout, and the functional efficiency, \Emout, in the study with science frames; while we have only used the functional efficiency method in the study with synthetic images, because this is only used for a first estimation of the {\tt DETECT\_THRESH\/} range. As it will be commented later, the matricial method became very unstable at faint magnitudes when used to correct star number counts for efficiency (see \S\ref{Sec:stargalaxy}). Thus, we decided to 
apply the definitive efficiency corrections using the functional efficiency \Emout\ computed with the study with science frames. Nevertheless, the 
efficiency matrices were essential when we performed our variation of the 
reliability analysis by CH03, as described in \S\ref{Sec:reliability}.

Once we computed \Pminmout\ and \Emout\ using the science frames, the efficiency corrections were applied as follows.  We obtained the initial catalogues by running {\tt SExtractor\/} onto the final $U$ and $B$ images. Notice that the efficiency corrections must be applied to the number counts obtained from catalogues \emph{after removing spurious sources}. This is because the efficiency matrices and functions are computed considering only the number of sources which are recovered from an original input distribution, \emph{without counting the number of spurious detections}. Then, after rejecting the spurious sources in both catalogues as described in \S\ref{Sec:reliability}, we proceeded to count sources by magnitude intervals according to the 3 size groups defined above. Differential number counts can be corrected for completeness by the following two methods:
\begin{enumerate} 

\item  \emph{Using the efficiency function \Emout}. Efficiency-corrected number counts per magnitude bin and unit area are given by 

\begin{equation}  
N_{\mathrm{orig},s}(\mout)=\frac{N_{\mathrm{det},s}(\mout)}{0.5\cdot A\cdot E_{s}(\mout)},  
\label{eq:efficiency} \end{equation}  
\noindent where we have defined magnitude bins of 0.5 mag; $s=1,2,3$ refers to the size group; $N_{\mathrm{orig},s}(\mout)$ are the efficiency-corrected number counts for size group $s$ and magnitude bin \mout; 
$N_{\mathrm{det},s}(\mout)$ are the detected counts for that size group and magnitude bin \emph{corrected for spurious detections}; 
$A$ is the area over which we have made galaxy number counting; and 
$E_{s}(\mout)$  is the functional efficiency for the same size group and magnitude bin. Finally, contributions of each size group are
added up to give completeness-corrected total number counts:

\begin{equation}  \label{eq:sum}
N_{\mathrm{orig,total}}(\mout)=\sum_{s=1}^{3}N_{\mathrm{orig},s}(\mout) .
\end{equation}  
  
\item  \emph{Using the efficiency matrix \Pminmout}. 
From the definition of \Pminmout, the number of sources 
detected with magnitude \mout\ irrespective of their input magnitude without having spurious detections into account, is given by

\begin{equation}  
N_{\mathrm{det},s}(\mout)= \sum_{\forall \min} N_{\mathrm{orig},s}(\min)\cdot \Psminmout ,  
\end{equation}  
\noindent where $N_{\mathrm{orig},s}(\min)$ represents the source number 
originally injected at \min\ for the size group $s$; \Psminmout\ is the 
efficiency matrix element \Pminmout\ for size group $s$; and $N_{\mathrm{det},s}(\mout)$ are the detected counts for that size group and magnitude bin \emph{corrected for spurious detections}, as above. 
Thus, the efficiency-corrected number counts per magnitude and unit area are 

\begin{equation}  \label{eq:inversion}
N_{\mathrm{orig},s}(\min)^{\prime}=\\\frac{1}{0.5\cdot A}
\cdot \left\{  \sum_{\forall \mout} N_{\mathrm{det},s}(\mout)\cdot {P}^{-1}_{s} (\mout,\min) \right\}. 
\end{equation}  
\noindent We have defined ${P}^{-1}_{s} (\mout,\min)$ as the element $(\mout,\min)$ of the inverted matrix of \Pminmout, for objects from the size group $s$. Finally, contributions from the three size groups are added up to obtain total efficiency-corrected number counts at each band, using equation 
(\ref{eq:sum}).
\end{enumerate}

Error estimation for both methods is described in detail in Appendix \ref{Append:errors}. Briefly, when using the efficiency function, the \Emout\ errors are quadratically added to those from counting statistics, while errors 
using the efficiency matrices \Pminmout\ are quantified by estimating 
how number counts would change if maximum errors from counting statistics 
and from efficiency matrices are separately considered in equation (\ref{eq:inversion}).

\begin{figure*}\begin{center}
\epsscale{0.8}
\plotone{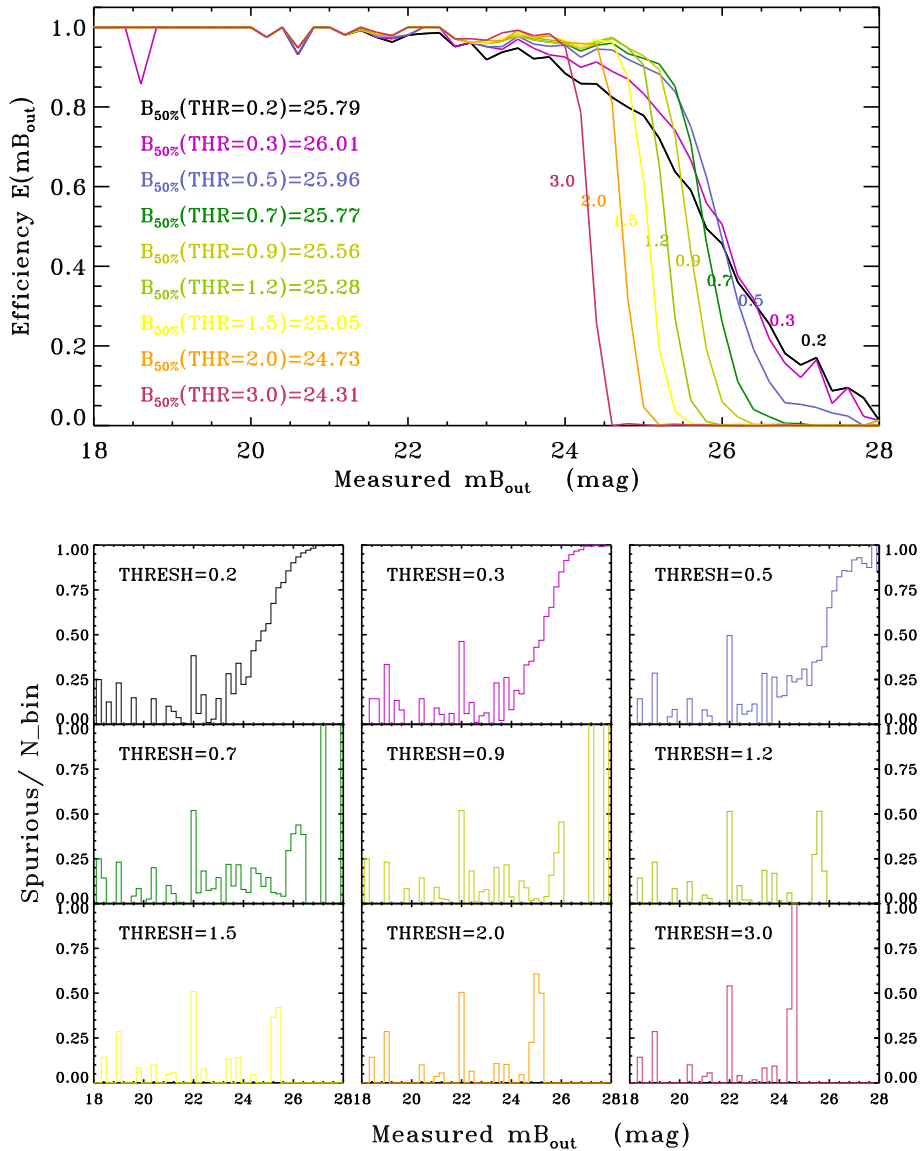}
\caption{Comparison of detection efficiencies \Emout\ and spurious fractions 
as a function of the detection magnitude \mout, obtained using the synthetic $B$ image for several {\tt DETECT\_THRESH\/} values. Drawn curves include all 
simulated objects, irrespective of their size. The situation at the $U$ 
synthetic image is completely analogous.\emph{Upper panel}: Efficiencies as a function of the magnitude of detection. \emph{Lower panel}:  Fractions of spurious sources detected at each magnitude interval relative to the total number of sources inserted at this magnitude.}\label{Fig:synthetic}
\end{center}
\end{figure*}

\subsection{Efficiency and Reliability Analysis for Synthetic Images}
\label{Sec:synthetic}

As commented in the previous section, the study of the efficiency and reliability using synthetic images must be interpreted with care, due to the difficulty in reproducing the sky conditions of the image and the profiles of simulated sources, which usually are regular in excess. So, we must remark that we have used this analysis using synthetic images only to narrow down 
the range of {\tt DETECT\_THRESH\/} values for the definitive study with 
the science images, which is more detailed and trustworthy. 

Using the {\tt artdata} package in IRAF, we created these artificial images 
with Poisson background noise of same RMS as our real images. Synthetic disk galaxies and stars of known magnitudes and sizes were added at given positions. 
Magnitudes span the range for stars and galaxies in our science frames. The number of input galaxies at each magnitude bin was chosen to reproduce an initial guess for the slope of the counts in each band, while the number of stars was the estimated using the \citet{Bahcall81a} model for the Galactic coordinates of the GWS. The sizes of all the stars and of half of the galaxies were set to the PSF FWHM in each band, while the other half of the synthetic galaxies was twice bigger. Detection efficiencies and spurious fractions as a function of source magnitude were determined by running {\tt SExtractor\/} with different {\tt DETECT\_THRESH\/} values \citep[see][for more details of this procedure]{Radovich04,Metcalfe01,Lin98}. In Figure \ref{Fig:synthetic} we show the efficiency functions, \Emout, and spurious fractions for several {\tt DETECT\_THRESH\/} values, computed using the artificial $B$ image of the GWS, as defined in \S\ref{Sec:sourceextraction}. The magnitude of 50\% efficiency ($m_{50\%}$) increases as we lower the detection threshold, while spurious detections quickly rise at magnitudes approaching $m_{50\%}$. From Figure \ref{Fig:synthetic} and the corresponding distributions for the $U$ band which behave similarly, we concluded that detection thresholds ranging from 0.4 to 0.7 provided a good compromise between high detection efficiency and low spurious fraction in both bands. Thus, we decided to use the following {\tt DETECT\_THRESH\/} values for the definitive study of incompleteness using the science frames: 0.4, 0.5, 0.6, and 0.7.

Notice that spurious fractions can be easily determined when using synthetic frames, because all the sources in the artificial images have been inserted by us. But this is not the same when using science frames, where we have a mixture of those sources we insert and of those which were there originally. In fact, the spurious analysis when using science frames is more complex (see \S\ref{Sec:reliability}).

\subsection{Efficiency and Reliability Analysis for Real Images}
\label{Sec:real}

\begin{figure*}\begin{center}
\epsscale{1}
\plotone{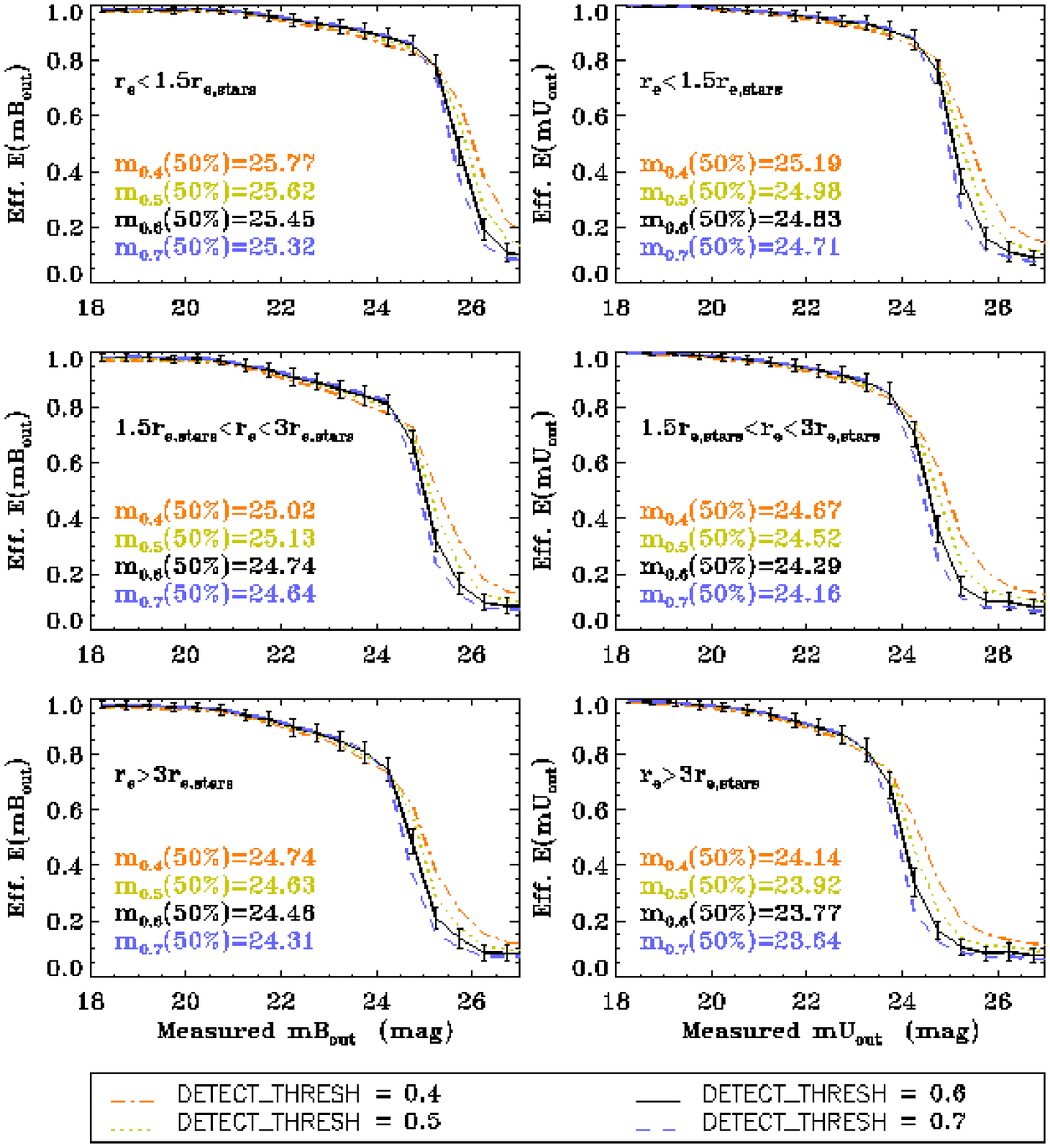}
\caption{ 
\emph{Left panels}: Detection efficiency functions, \Emout, given as 
the ratio of detected to input sources as a function of detection magnitude \mout, for the three apparent size groups in the $B$ image, using {\tt DETECT\_THRESH\/}=0.4, 0.5, 0.6, and 0.7. Errors are plotted for {\tt DETECT\_THRESH\/}=0.6 only, because they are similar for all the detection thresholds. Each line style corresponds to one of the {\tt DETECT\_THRESH\/} values according to the legend. The 50\% efficiency magnitudes for each size group are listed in each frame.
\emph{Right panels}: The same for the $U$ image.
}\label{Fig:efficiency}\end{center}
\end{figure*}

\begin{figure*}\begin{center}
\epsscale{0.8}
\plotone{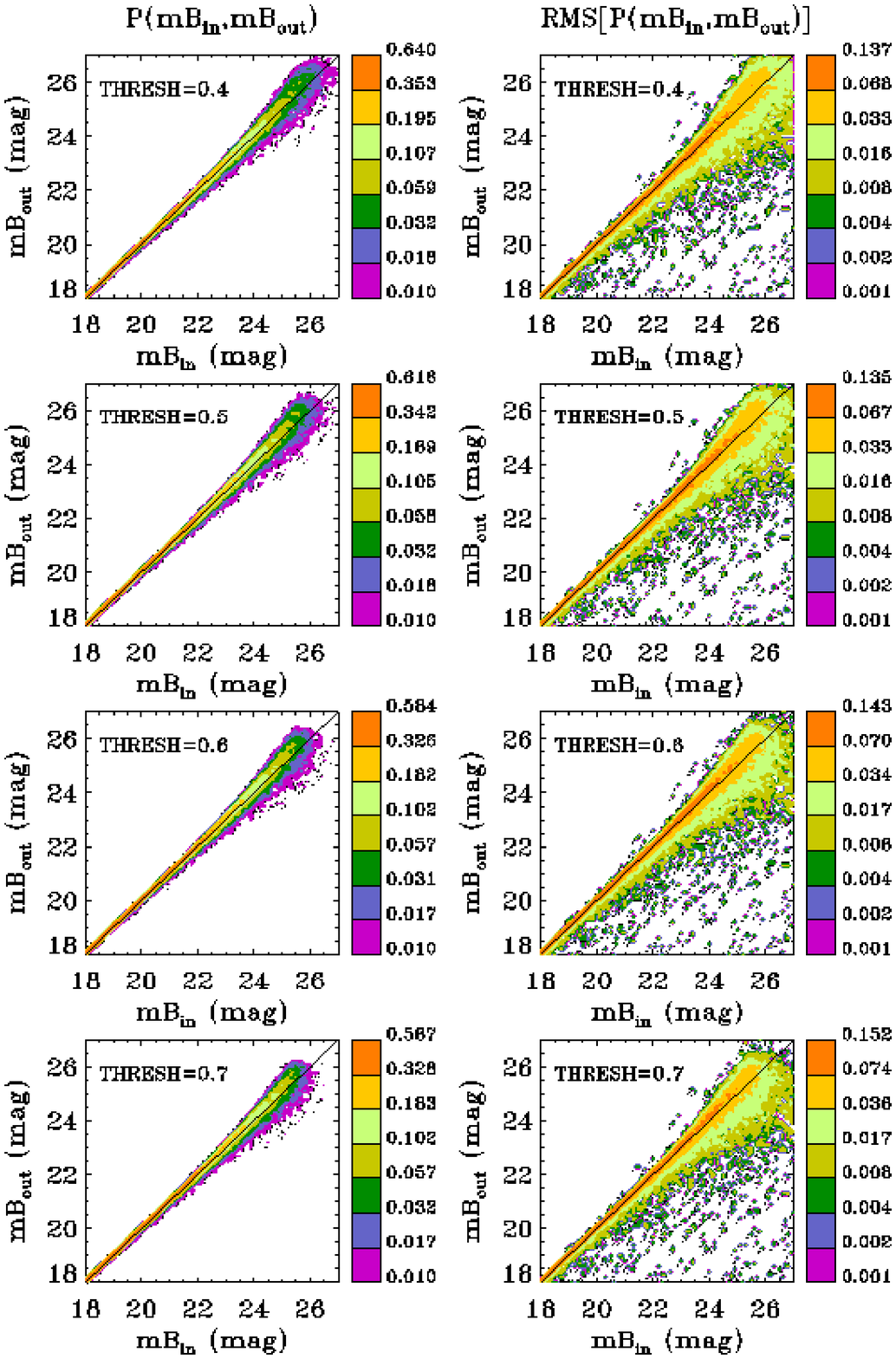}
\caption{Probability matrices \Pminmout\  and their corresponding standard
deviation matrices for point-like objects in the $B$-band, using the four {\tt DETECT\_THRESH\/} values
we are analysing. \emph{Left panels}: 
Probability matrices \Pminmout\ of finding an injected source with initial 
magnitude \min\ at output magnitude \mout, for point-like 
objects in the $B$ image, using {\tt DETECT\_THRESH\/}=0.4, 0.5, 0.6, and 0.7. 
Solid line indicates $mB_\mathrm{in}= mB_\mathrm{out}$. \emph{Right panels}: 
Standard deviation matrices associated to the efficiency matrices at the 
left. Similar patterns are found for all size groups and for the 
$U$ image.}\label{Fig:matrices}
\end{center}
\end{figure*}

\subsubsection{Efficiency}
\label{Sec:efficiency}
We have carried out an extensive series of simulations to quantify the detection efficiency using the real images \citep[see][CH03, among others, for a detailed description of the procedure]{Hogg97,Huang01b,Kummel01}. Firstly, we found the brightest source present in our images, in each one of the three size groups defined in \S\ref{Sec:sourceextraction}. Then, these three selected sources were inserted several times in the science frames following a flat magnitude distribution (18$\le B$ (or $U$) $\le$28) at  random locations. No constraints were imposed on the source positions, in order to include the effects of source confusion in the computation of \Pminmout\ and \Emout. This process was repeated 50 times for each size group, injecting 2,000 sources in each simulation, for a total of 300,000 objects. {\tt SExtractor\/} was run on each simulated 
image with {\tt DETECT\_THRESH\/}=0.4, 0.5, 0.6, and 0.7. 
For each simulation, we computed the efficiency matrix, \Pminmout, and the functional efficiency, \Emout\ (see their definitions in \S\ref{Sec:sourceextraction}).

In Figure \ref{Fig:efficiency}, we show the average \Emout\ 
from all the simulations, as a function of magnitude and size group, 
for the four selected values of {\tt DETECT\_THRESH\/}, in both $B$ and $U$ bands. Error bars are the quadratic addition of the RMS of all simulations 
\citep{Bevington69}. In all the cases, \Emout\ shows a gentle decline, 
followed by an abrupt drop near the detection limit. The 
50\% efficiency magnitudes 
for a given object size differ by as much as $\sim$0.6 mag when changing 
{\tt DETECT\_THRESH\/}; while they are $\sim$0.7 mag deeper for point sources 
than for large objects of the same magnitude. These trends are similar 
to those found by CH03 and \citet{Bershady98}. We have also found that our synthetic frames overestimate the efficiency of {\tt SExtractor\/} on to the science images, a fact that corroborates what \citet{Bershady98} and CH03 reported. It can be noticed just comparing the 50\% efficiency magnitudes obtained using synthetic frames for {\tt DETECT\_THRESH\/}=0.5,0.7 in $B$ (see the Figure \ref{Fig:synthetic}), with the achieved ones using the science frames (see the Figure \ref{Fig:efficiency}).

The sizes of the selected objects were approximately typical for their size groups, except for the object representing the "intermediate-sized" objects in $B$, which is large into the range of its size bin. Nevertheless, we have checked that this fact does not underestimate the typical detection efficiency of its size group for magnitudes less than the 70\% efficiency magnitude for a fixed {\tt DETECT\_THRESH\/}. For higher magnitudes, the curve of the efficiency function would be displaced $\lesssim 0.15$ mag to fainter magnitudes. But the changes in the results would be negligible in the last significant bin of magnitude (the 50\% efficiency magnitude bin), because we have defined bins of 0.5 mag and the effciency drops from 70\% to 50\% in less than 0.5 mags (see Figure \ref{Fig:efficiency}). Therefore, we can consider that the surface brightness of each selected brightest object represents the typical surface brightness in its correponding size group.

$B$-band average \Pminmout\ values for point-like objects are plotted in  
Figure \ref{Fig:matrices} for the four detection threshold 
values we studied. Errors associated to these matrices are also 
shown in the figure; they have been computed in the same way as the 
\Emout\ errors. Similar behaviours are seen for the three size groups, and for both filters. 

As we have commented above, we corrected number counts for incompleteness effects using the efficiency matrix, \Pminmout, and the functional 
efficiency, \Emout. The former became very unstable at faint magnitudes when used to correct star number counts for efficiency, as described in \S\ref{Sec:stargalaxy}. So, we decided to apply the definitive efficiency corrections using the functional efficiency \Emout\ only. The efficiency matrices were useful when we performed our variation of the 
reliability analysis by CH03 (see \S\ref{Sec:reliability}).

We must remark that the \Pminmout\ determination does not depend on the used input source distribution, and thus, it can be used for correcting for completeness whatever the real input distribution is. But this is not true for \Emout, which would have need an input source distribution as similar as possible to the real one. Nevertheless, this is not feasible when using science frames, because these images are already very populated by real objects. An aditional population of inserted objects as large as the original one would crowd the image excessively, and as a result, computed efficiency matrices and functions would overestimate the effects of source confusion. Thus, an additional error arises when number counts are corrected using an efficiency function which has been computed with an input source distribution flat in magnitudes (instead of an input one that reproduces the initial slope of the counts). The efficiency function error that arises at each magnitude bin can be estimated as 
\begin{equation}
\label{eq:errorefficiencyfunction}
\Delta \Emout=\frac{\Emout-{\widetilde{E}(\mout)}}{\Emout},
\end{equation}
\noindent where \Emout\ is our efficiency function computed using the flat input distribution of sources, and $\widetilde{E}(\mout)$ represents the efficiency function that would be obtained using an input distribution with the initial slope of the counts (i.e., the \emph{correct} one). We have estimated $\widetilde{E}(\mout)$ at each magnitude bin as follows: let us consider an input source distribution identical to the one we detect once the spurious sources have been subtracted for each size group $s$, $N_{\mathrm{slope},s}(\mout)$. As \Pminmout\ is independent of the used input source distribution, the \emph{correct} distribution of sources we are going to detect is $N_{\mathrm{det},s}(\mout)= \sum_{\forall \min} N_{\mathrm{slope},s}(\min)\Psminmout$. Moreover, if we had used the  \emph{correct} efficiency function $\widetilde{E}(\mout)$, the detected source distribution would have been the same: $N_{\mathrm{det},s}(\mout)=N_{\mathrm{slope},s}(\mout)\cdot\widetilde{E}(\mout)$. So, we can deduce  $\widetilde{E}(\mout)$ from the two previous expressions:
\begin{equation}
\label{eq:correctefficiencyfunction}
\widetilde{E}(\mout)=\frac{\sum_{\forall \min} N_{\mathrm{slope},s}(\min)\Psminmout}{N_{\mathrm{slope},s}(\mout)},
\end{equation}
\noindent We have estimated the error that arises in using \Emout\ instead of $\widetilde{E}(\mout)$ through equations (\ref{eq:correctefficiencyfunction}) and (\ref{eq:errorefficiencyfunction}), and considering the detected source distribution once the spurious have been removed as the input distribution $N_{\mathrm{slope},s}(\mout)$. The maximum error $\Delta \Emout$ reached in both bands (and thus in the number counts) is $\lesssim 10$\% for magnitudes  $m<m_\mathrm{50\%}$ in each size group, which is less than the total error from statistical counting and efficiency errors using the computed \Emout. Nevertheless, the $\widetilde{E}(\mout)$ determination has itself a high uncertainty which would overestimate efficiency errors, and that arises from the division by $N_{\mathrm{slope},s}(\mout)$ in those magnitude bins where we have low statistical significance, and from the accumulation of the errors from $\Psminmout$ and $N_{\mathrm{slope},s}(\mout)$  due to the error propagation of equation (\ref{eq:correctefficiencyfunction}). Therefore, we decided to ignore this error and use \Emout\ instead $\widetilde{E}(\mout)$ for correcting number counts.

\begin{figure*}\begin{center}
\plotone{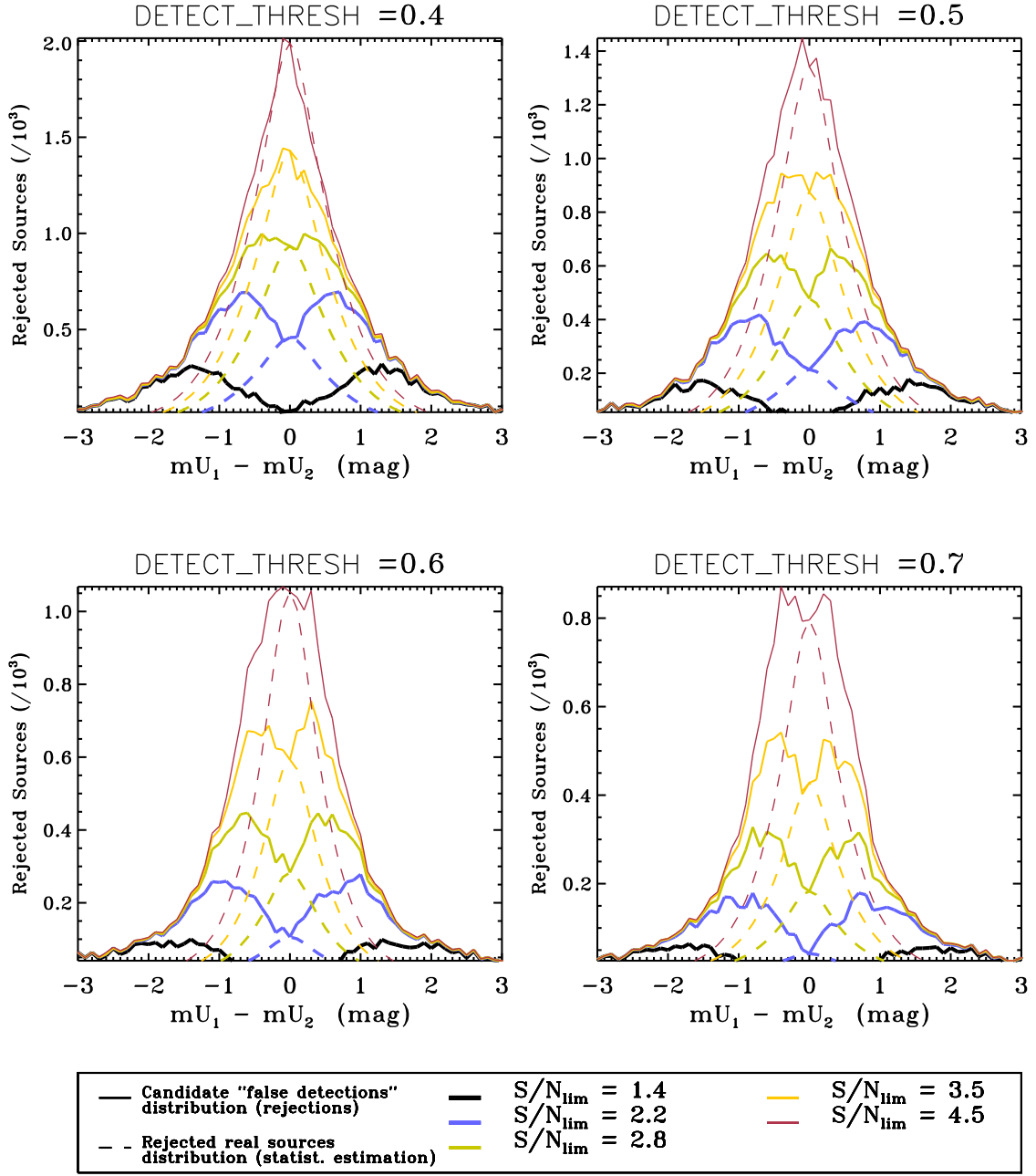}
\caption{Histograms of magnitude differences between the half-time images for
candidate "false detections", and their corresponding estimated real source
distributions. \emph{Solid lines}: Histograms of magnitude differences between 
the half-time images for candidate "false detections" with 
different $(S/N)_\mathrm{lim}$ and for the {\tt DETECT\_THRESH\/} values we are 
analizing. Corresponding $(S/N)_\mathrm{lim}$ are 1.4, 2.2, 2.8, 3.5, and 4.5, 
according to the legend in the figure. 
\emph{Dashed lines}: Estimated real source distributions compatible 
with each one of the the rejected source histograms.}\label{Fig:reliability}\end{center}
\end{figure*}

\subsubsection{Reliability}
\label{Sec:reliability}

Reliability has been characterized using the method described in CH03, 
based on that used by \citet{Bershady98}. It consists on creating two half-time exposured images from exclusive halves of the data. As spurious sources are basically noise correlated structures or peaks that appear in one of the subexposures, they must appear only in one of the two half-time images, the one that was built including the correponding subexposure. By running {\tt SExtractor\/} in double-image mode, photometry on both half-time images is measured at the positions of sources detected by {\tt SExtractor\/} in the total-time image. Those spurious sources detected in the total-time image will have very different magnitudes when measured in the half-time images, as they have a high probability of appearing only in one of the two half-time images; while the real sources will have very similar magnitudes when measured in the two half-time images, because they have a high probability of appearing in both half-time images with approximately the same flux. We must notice here that  this method could not be applied to $B$ data directly due to their dithering sequence (see Table \ref{Tab:finalimages}). However, the results obtained through this method in the $U$ band can be used for characterizing reliability in the $B$ image, as we will comment below. So the $(S/N)_\mathrm{lim}$ method for characterising {\tt SExtractor\/} reliability was carried out on the $U$ image only, and extrapolated to the $B$ image at the end.
 
In order to identify spurious detections, CH03 used a signal-to-noise ratio ($S/N$) criterion. Hereandafter, we will define the $S/N$ of a source as the ratio between its flux and its flux error measured by {\tt SExtractor\/}. The method consists on assuming that all the sources with $S/N$ below a given limit, $(S/N)_\mathrm{lim}$, in one or in both of the half-time images are considered as "false detections", and hence, rejected from the catalogue. Nevertheless, into this group of sources labelled as "false detections", we will be rejecting sources that actually are spurious detections (henceforth called "rejected truly spurious"), and sources that in fact are real objects whose $S/N$'s also obey the imposed criterion (called "rejected real sources", hereandafter). As we increase the value of the imposed $(S/N)_\mathrm{lim}$, the probability of being rejecting more truly spurious sources increases. But, at the same time, the probability of being rejecting a real source as a "false detection" also raises. There will be a limiting $(S/N)^\prime _{\mathrm{lim}}$ equal to the one of the highest $S/N$ that the truly spurious sources exhibit in the catalogue, so that it rejects most of them from it. Using $(S/N)_\mathrm{lim}> (S/N)^\prime_{\mathrm{lim}}$ will not reject more truly spurious, because the majority of them have already been rejected with the lower $(S/N)^\prime _{\mathrm{lim}}$, but it will reject more real sources as "false detections" (those with $(S/N)^\prime_{\mathrm{lim}}<S/N< (S/N)_\mathrm{lim}$). This is due to the fact that the population of real sources has higher $S/N$ on average than the truly spurious population, which is mostly related to the sky noise. Therefore, the fraction of rejected truly spurious sources will become constant for values $(S/N)_\mathrm{lim} > (S/N)^\prime_{\mathrm{lim}}$ in each {\tt DETECT\_THRESH\/}, and this constant fraction can be considered as the maximum fraction of truly spurious sources in the catalogue. 

For summarizing, for a fixed {\tt DETECT\_THRESH\/}, low values of the $(S/N)_\mathrm{lim}$ will reject a low number of sources as "false detections" from the catalogue, but with a high confidence level in that the majority of them will be truly spurious instead of real sources. On the other hand, high values of the $(S/N)_\mathrm{lim}$ will reject a high number of sources as "false detections", but we will have an unknown mixture of real and truly spurious sources into the rejected group. For $(S/N)_\mathrm{lim}$ values higher than an unknown limiting $(S/N)^\prime_{\mathrm{lim}}$, the majority of the truly spurious will be labelled as "false detections" (so the number of truly spurious which are rejected will be approximately constant), but the number of real sources labelled as "false detections" will keep increasing.

Moreover, when a good discrimination between real and truly spurious
sources is achieved for a given $(S/N)_\mathrm{lim}$, the majority of sources which are rejected as "false detections" will be truly spurious. Therefore, they will have very different magnitudes  when measured in the half-time images, because a false detection has a negligible probability of being detected in both half-time images with similar magnitudes. If we plot the histogram of $\Delta m\equiv m_\mathrm{half,1}-m_\mathrm{half,2}$ for the rejected sources with this fixed $(S/N)_\mathrm{lim}$, it will show a bimodal structure, as the majority of the rejected sources are truly spurious. On the other hand, if the fixed $(S/N)_\mathrm{lim}$ does not distinguish properly between real and truly spurious sources, there will be two populations mixed in the $\Delta m$ histogram: the rejected truly spurious sources, which will draw the bimodal structure, and the rejected real sources, which will contribute to a narrow single-peaked distribution centered in $\Delta m=0$, because the real sources have a high probability of having very similar magnitudes in the half-time images. 

In Figure \ref{Fig:reliability}, the histograms of magnitude differences between the half-time images for candidate "false detections" in the $U$ band are drawn with solid lines, for different $(S/N)_\mathrm{lim}$ and for the four {\tt DETECT\_THRESH\/} values we decided to study (see \S\ref{Sec:synthetic}). Seventeen values ranging from $(S/N)_\mathrm{lim}$=0.2 to $(S/N)_\mathrm{lim}$=10.0 were studied, but we have plotted results which correspond to $(S/N)_\mathrm{lim}$=1.4, 2.2, 2.8, 3.5, and 4.5 for clarity. When a good discrimination between real and spurious sources is achieved, the histograms show a bimodal structure, as for the lower values of $(S/N)_\mathrm{lim}$. Conversely, as we raise $(S/N)_\mathrm{lim}$, sources with higher $S/N$, and hence with more probability of being real, are classified as "false detections". This situation is responsible for the single-peaked, centered distributions of $\Delta m=0$ in the cases of high $(S/N)_\mathrm{lim}$. As can be seen from the figure, selecting the combination {\tt DETECT\_THRESH\/}-$(S/N)_\mathrm{lim}$ from these histograms is tricky, because histograms evolve from double to single-peaked gradually; \ie, there is always a population of rejected real sources mixed up with the distribution of rejected truly spurious detections, considered as "false detections" by this method. Until now, the selection {\tt DETECT\_THRESH\/}-$(S/N)_\mathrm{lim}$ was subjective, depending on which histogram seemed to be more double-peaked and populated at the same time. 

We have developed a self-consistent procedure for choosing the most 
adequate {\tt DETECT\_THRESH\/}-$(S/N)_\mathrm{lim}$ combination in an objective way, in order to minimize the number of real sources which are rejected as "false detections", and to maximize the number of rejected truly spurious sources and also the number of real detections that remain in the catalogues. The procedure is described in detail in \citet{Eliche05}, so we briefly summarize necessary information for this paper. The method consists on estimating statistically the population of real sources which is mixed up with the one of truly spurious sources in the histograms of "false detections" of Figure \ref{Fig:reliability}. For computing it, the following justified assumptions are used in the $U$ band: 
\begin{enumerate}
\item The detected source distribution directly obtained by {\tt SExtractor\/}, $N_\mathrm{det}(m)$, was taken as a first approximation of the distribution of detected real sources, $N_\mathrm{det,real}(m)$, being $m$ the detection magnitude. 

\item The distribution $N_\mathrm{det,real}(m)$ was truncated for magnitudes  $mU\ge 26$ mag, because 50\% efficiency magnitudes are less than $mU=25$ mag for all the {\tt DETECT\_THRESH\/} values we are analysing. Therefore, a source with $mU\gtrsim 26$ will have a $\sim 100$\% probability of being a truly spurious detection, and these magnitude bins will not contribute to the real population in the histograms of magnitude differences of Figure \ref{Fig:reliability}.

\item We have also considered that all the sources at the histograms of $\Delta mU$ that exhibit $\Delta mU=mU_\mathrm{out,1}-mU_\mathrm{out,2}=0$ are real detections. This is justified by the fact that the Poissonian probability of detecting two spurious sources at the same position with the same magnitude at both half-time images is negligible. 

\end{enumerate}

Once the distribution of real detections compatible with each histogram of magnitude differences is computed  for each $(S/N)_\mathrm{lim}$-{\tt DETECT\_THRESH\/} combination, we can easily estimate the fractions of real sources and truly spurious we are rejecting from the total number of detections, as well as the fraction of truly spurious sources that remain in the catalogue in each case. Notice that we can estimate also the limiting $(S/N)^\prime _{\mathrm{lim}}$ so that using $(S/N)_\mathrm{lim}> (S/N)^\prime_{\mathrm{lim}}$ will not reject more truly spurious. This will be the $(S/N)_\mathrm{lim}$ value from which the number of rejected truly spurious remains constant although we raise it.

In Figure \ref{Fig:reliability} we have overplotted with dashed lines the estimated distributions of real sources compatible with each histogram of magnitude differences in the half-time images. As $(S/N)_\mathrm{lim}$ increases, the estimated real distribution approaches the corresponding histogram, because most of the rejected "false detections" are real sources. It is remarkable that the widths of these real source distributions (FWHM$\sim$0.4-0.7 mag for all the {\tt DETECT\_THRESH\/}-$(S/N)_\mathrm{lim}$ combinations) are consistent with the photometric error for sources at those S/N on the half-time images, a fact that supports their robustness. 

In Table \ref{Tab:fractions}, we have compared the estimated fractions of rejected real sources, rejected truly spurious sources, and truly spurious that are non-rejected (\ie, that remain in the catalogue), for the four values of {\tt DETECT\_THRESH\/} and five of the seventeen values of $(S/N)_\mathrm{lim}$ we are analysing in the $U$ band: $S/N$=1.8, 2.2, 2.5, 7, and 10. The procedure for getting the values from Table \ref{Tab:fractions}, the arguments to set the $(S/N)_\mathrm{lim}$-{\tt DETECT\_THRESH\/} combination of values, and the results are given in detail in EB05. Thus, we will just explain them briefly here. 

Firstly, notice that the fraction of truly spurious sources becomes constant for $(S/N)_\mathrm{lim}=6,10$ in each {\tt DETECT\_THRESH\/} in the Table. In fact, it is constant for $(S/N)^\prime _\mathrm{lim}> 2.2$ in {\tt DETECT\_THRESH\/}=0.4, for $(S/N)^\prime _\mathrm{lim}> 3.0$ in {\tt DETECT\_THRESH\/}=0.5, for $(S/N)^\prime _\mathrm{lim}> 3.5$ in {\tt DETECT\_THRESH\/}=0.6, and for $(S/N)^\prime _\mathrm{lim}> 4.0$ in {\tt DETECT\_THRESH\/}=0.7. This means that the maximum number of truly spurious that are rejected in each {\tt DETECT\_THRESH\/} is reached with the corresponding $S/N^\prime _\mathrm{lim}$ value. Then, no advantage arises in using $(S/N)_\mathrm{lim}> (S/N)^\prime _\mathrm{lim}$. We have also found that the maximum fraction of truly spurious that are rejected for all the $(S/N)_\mathrm{lim}$ is the same for {\tt DETECT\_THRESH\/}$\leq$0.6 ($\sim 29$\%), but different for {\tt DETECT\_THRESH\/}=0.7 ($\sim 25$\%), as you can infer from Table \ref{Tab:fractions}. Therefore, {\tt DETECT\_THRESH\/}=0.6 seemed to be the limit between two different behaviours in detection. Although the total fraction of rejected truly spurious (respect to the total number of detections) is higher if we  use {\tt DETECT\_THRESH\/}$\leq$0.6 than using {\tt DETECT\_THRESH\/}=0.7, it can be controlled, and it is possible to reach deeper magnitudes than using {\tt DETECT\_THRESH\/}=0.7. With an adequate choice of the $(S/N)_\mathrm{lim}$ (see Table \ref{Tab:fractions}), we can get similar fractions of truly spurious rejections, of real source rejections, and of truly spurious remaining in the catalogues, and increase the number of detections. Thus, we discarded the value {\tt DETECT\_THRESH\/}=0.7. Moreover, $(S/N)_\mathrm{lim}$=2.2 gave a good compromise between rejected real and false detections for {\tt DETECT\_THRESH\/}$\leq$0.6. So we finally decided to use {\tt DETECT\_THRESH\/}=0.6 and $(S/N)_\mathrm{lim}$=2.2 for the 
$U$ filter, and hence, the 50\% efficiency magnitude for point-like 
sources is $U_{\mathrm{DET}=0.6}$(Eff=50\%)=24.83 mag (see Figure 
\ref{Fig:efficiency}). This selection allowed us to reject $\sim$22\% 
of truly spurious sources in the catalogue (see the values in Table 
\ref{Tab:fractions}). As the maximum fraction of truly spurious sources 
for {\tt DETECT\_THRESH\/}=0.6 was estimated to be $\sim$29\% of the 
total number of detections, then its is easily deduced that approximately 
$\sim$7\% of the final catalogue are truly spurious detections, the bulk of them at magnitudes fainter than the 50\% efficiency magnitude. 
With this choice, $\sim$3\% of the real sources are rejected from the catalogue. Using  $(S/N)_\mathrm{lim}$=2.2 in half-time images for rejection, 
sources considered as real detections have roughly $(S/N)_\mathrm{lim}\gtrsim$3.1 in the total-time image.
 
As commented previously, this method could not be applied to $B$ data 
directly because it was not possible to obtain two half-time images from the dithering sequence (see Table \ref{Tab:finalimages}). If we would have created two complementary images without using all the subexposures in the $B$ band, these two images would have been exposed less than $t_\mathrm{total}/2$. Therefore, we would have had a lower confidence in detecting spurious sources than in the $U$ band, and we also would have lost the spurious sources corresponding to those subexposures not used for creating the two complementary images. Nevertheless, the results for the $U$ filter can be used for estimating the best combination of {\tt DETECT\_THRESH\/}-$(S/N)_\mathrm{lim}$ for the $B$ image as follows. The $B$ and $U$ images had similar flux characteristics, source distributions and Poissonian sky noises, and hence, the detection threshold for $U$ image should be valid for the $B$ image too. This is because, although $\sigma _{sky}$ is higher in the $B$ image, $B$ is deeper than $U$, and their efficiency matrices and functions behave similarly 
(see Figures \,\ref{Fig:efficiency} and \ref{Fig:matrices}). Therefore, we use  
{\tt DETECT\_THRESH\/}=0.6 for the $B$-band also, and obtain a 50\% efficiency magnitude of $B_{\mathrm{DET}=0.6}$(Eff=50\%)=25.46 mag (see Figure \ref{Fig:efficiency}). In order to establish the $(S/N)_\mathrm{lim}$ for the $B$ image, we have considered that detections only depend on the $S/N$ of the sources once the minimum area has been fixed. When the $S/N$ of a source is below the fixed $(S/N)_\mathrm{lim}$, its flux per unit time and area 
must be lower than the flux that corresponds to that $(S/N)_\mathrm{lim}$. 
In EB05, the equations that relate the flux of a source and its  
$S/N$ are shown as a function of several image parameters. 
This flux can be expressed in units of the sky sigma of the image. Therefore, a source with $(S/N)_\mathrm{lim}$=2.2 in the $U$-band with the minimum area and for $t=t_{total}/2$ would have a flux per pixel and per unit time equal to 1.73 times the sky sigma of the total-time $U$ image. Extrapolation of this result to the $B$ image is justified due to the similarity of $U$ and $B$ images, and hence we have considered that a source in the $B$-band will only be considered a real detection if its flux per pixel is greater than 1.73 $\sigma _{sky}$ of the $B$ image, which implies a $(S/N)_\mathrm{lim}$=2.8 
for a source with the minimum area in the $B$ total-time image. 

Finally, we used the identified "false detections"in the $U$ band with the selected combination of {\tt DETECT\_THRESH\/}-$(S/N)_\mathrm{lim}$ for correcting the corresponding catalogue (extracted using the selected {\tt DETECT\_THRESH\/}=0.6) for spurious detections. Hereandafter, we are going to call "spurious sources or spurious detections" to all the "false detections" that have been rejected. As the same method can not be applied to the $B$ band, we corrected for spurious detections the $B$ catalogue obtained with {\tt DETECT\_THRESH\/}=0.6 rejecting all the sources with $(S/N)_\mathrm{lim}\leq 2.8$ in the $B$ total-time image. 

Our procedure for estimating the distribution of real sources  
in the $\Delta m$ histograms is able to fix objectively the 
{\tt DETECT\_THRESH\/}-$(S/N)_\mathrm{lim}$ combination 
which minimizes the number of rejected real sources, 
maximizing the number of true spurious rejections, and reaching  
the deepest limiting magnitude at the same time. In fact, the 
subjetive selection of $(S/N)_\mathrm{lim}$ by CH03 leads them to remove 
the 13.7\% of the sources with $S/N>5$ in their data, while we have rejected 
$\lesssim 1$\% of sources with $S/N>5$, $\lesssim 9$\% of sources with $3<S/N<5$, and $\sim 13$\% of sources with $S/N<3$. Our method allows us to estimate that only  $\sim$3\% of real sources with $S/N<3.1$ are rejected from our catalogues.

\subsection{Galactic Extinction}
\label{Sec:extinction}

Even at the high Galactic latitude of the GWS field ($b$=60\deg), 
Galactic extinction affects $U$- and $B$-band number counts.  
Lack of extinction corrections probably explain the differences 
between published $U$-band number count data \citep{Heidt03,Radovich04}.  
We have computed Galactic extinction corrections 
for each source, in both the $U$ and $B$ filters, using the 
\citet{Schlegel98} extinction maps\footnote{Dust maps and software for computing
the extinction corrections for each source were downloaded from: 
\\http://astron.berkeley.edu/davis/dust/local/local.html}. The average
value is $E(B-V)=0.011037\pm 0.000004$, ranging from 
$0.008\le E(B-V)\le 0.018$, 
which translates into a maximum Galactic extinction correction of 
$A_U\sim 0.11$ mag and $A_B\sim 0.08$ mag.

\begin{deluxetable*}{lrrrrrrr}
\tabletypesize{\scriptsize}
\tableheadfrac{0.01}
\tablewidth{0pt}
\tablecaption{$U$ Differential Star Counts in the GWS Field\label{Tab:starcountsu}}
\tablehead{\colhead{U}  & \colhead{$N_{\mathrm{star,raw}}$} & \colhead{$N_{\mathrm{star},1}$} & 
\colhead{$\sigma _{\mathrm{u},1}$} & 
\colhead{$\sigma _{\mathrm{l},1}$} & \colhead{$N_{\mathrm{star},2}$} & \colhead{$\sigma _{\mathrm{u},2}$} & 
\colhead{$\sigma _{\mathrm{l},2}$}\\
\colhead{(1)}&\colhead{(2)}&\colhead{(3)}&\colhead{(4)}&\colhead{(5)}
&\colhead{(6)}&\colhead{(7)}&\colhead{(8)}}
\startdata 
18.25 .............. &    144.68 &    144.68 &    171.87 &    108.97 &    226.34 &    167.88 &    103.84\\
18.75 .............. &    238.30 &    238.30 &    200.20 &    139.78 &    461.55 &    212.70 &    153.56\\
19.25 .............. &    246.81 &    246.81 &    134.46 &     65.15 &     25.01 &    119.91 &     48.68\\
19.75 .............. &    314.90 &    314.90 &    223.96 &    164.95 &    635.51 &    239.36 &    181.48\\
20.25 .............. &    151.74 &    153.85 &    150.60 &  83.49 &	79.77 &    136.88 &	68.50\\
20.75 .............. &    556.38 &    567.87 &    228.70 &    168.52 &    619.46 &    237.36 &    177.83\\
21.25 .............. &    404.64 &    415.41 &    205.63 & 143.64 &    365.02 &    198.07 &    135.82\\
21.75 .............. &    657.54 &    682.67 &    247.63 &    187.18 &    725.96 &    253.45 &    193.00\\
22.25 .............. &    708.12 &    741.54 &    256.88 & 196.22 &    595.24 &    239.57 &    179.28\\
22.75 .............. &   1315.08 &   1395.82 &    332.97 &    273.78 &   1588.32 &    362.57 &    300.18\\
23.25 .............. &   1719.72 &   1842.21 &    376.94 &    318.20 &   1257.36 &    282.47 &    235.81\\
23.75 .............. &   2174.93 &   2395.14 &    432.02 &    372.85 &   4076.67 &    694.77 &    600.13\\
24.25 .............. &   2377.25 &   2713.88 &    469.13 &    408.58 &  \multicolumn{1}{c}{...} & \multicolumn{1}{c}{...} &  \multicolumn{1}{c}{...}\\
24.75 .............. &   2225.51 &   2927.68 &    535.40 &    466.94 &  \multicolumn{1}{c}{...} & \multicolumn{1}{c}{...}&   \multicolumn{1}{c}{...}\\
25.25 .............. &   1466.82 &   3990.05 &   1011.32 &    881.00 &  \multicolumn{1}{c}{...} & \multicolumn{1}{c}{...} &  \multicolumn{1}{c}{...}\\[-0.2cm]
\enddata
\tablecomments{Col.\,(2) are $U$ raw star counts in units of 
N mag$^{-1}$ deg$^{-2}$ (once the spurious where removed from the catalogue); Col.\,(3) shows efficiency-corrected $U$ star counts by the functional method in units of N mag$^{-1}$ deg$^{-2}$ (denoted by subindex 1); Col. (4) and (5) are the 1-$\sigma$ confidence upper and lower errors associated to the functional-efficiency method; Col.\,(6) shows efficiency-corrected $U$ star counts using the matricial method in units of N mag$^{-1}$ deg$^{-2}$ (denoted by subindex 2);  Col. (7) and (8) are the 1-$\sigma$ confidence upper and lower errors associated to the matricial-efficiency method.}
\end{deluxetable*}

\begin{deluxetable*}{lrrrrrrr}
\tabletypesize{\scriptsize}
\tableheadfrac{0.01}
\tablewidth{0pt}
\tablecaption{$B$ Differential Star Counts in the GWS Field\label{Tab:starcountsb}}
\tablehead{\colhead{$B$}  & \colhead{$N_{\mathrm{star,raw}}$} & \colhead{$N_{\mathrm{star},1}$} & 
\colhead{$\sigma _{\mathrm{u},1}$} & 
\colhead{$\sigma _{\mathrm{l},1}$} & \colhead{$N_{\mathrm{star},2}$} & \colhead{$\sigma _{\mathrm{u},2}$} & 
\colhead{$\sigma _{\mathrm{l},2}$}\\
\colhead{(1)}&\colhead{(2)}&\colhead{(3)}&\colhead{(4)}&\colhead{(5)}
&\colhead{(6)}&\colhead{(7)}&\colhead{(8)}}
\startdata 
19.75 ..............  &    322.77 &    322.77 &    225.84 &    166.34 &    611.48 &    237.29 &    177.20\\
20.25 ..............  &    204.02 &    206.57 &    164.21 &     98.61 &    179.91 &    158.92 &     92.89\\
20.75 ..............  &    408.04 &    415.95 &    205.89 &    143.82 &    387.84 &    202.26 &    140.04\\
21.25 ..............  &    765.08 &    786.34 &    260.79 &    200.95 &    817.84 &    264.97 &    205.34\\
21.75 ..............  &    714.07 &    743.40 &    257.47 &    196.64 &    736.53 &    256.42 &    195.49\\
22.25 ..............  &    765.08 &    806.99 &    267.99 &    206.69 &    819.00 &    268.92 &    207.41\\
22.75 ..............  &    867.09 &    928.63 &    285.79 &    224.09 &    806.13 &    271.33 &    209.65\\
23.25 ..............  &   1224.13 &   1327.75 &    332.46 &    271.29 &   1732.40 &    379.34 &    315.42\\
23.75 ..............  &   1224.13 &   1351.64 &    339.39 &    277.33 &    114.46 &    187.24 &    136.30\\
24.25 ..............  &   2346.24 &   2654.93 &    461.26 &    400.19 &   4839.61 &    770.67 &    674.54\\
24.75 ..............  &   3315.34 &   3871.50 &    563.09 &    502.02 & \multicolumn{1}{c}{...}&\multicolumn{1}{c}{...}&\multicolumn{1}{c}{...}\\
25.25 ..............  &   3366.35 &   4320.23 &    643.00 &    578.23 & \multicolumn{1}{c}{...}&\multicolumn{1}{c}{...}&\multicolumn{1}{c}{...}\\
25.75 ..............  &   3009.31 &   6352.74 &   1143.72 &   1050.85 & \multicolumn{1}{c}{...}&\multicolumn{1}{c}{...}&\multicolumn{1}{c}{...}\\[-0.2cm]
\enddata
\tablecomments{Columns are as in Table \ref{Tab:starcountsb}.}
\end{deluxetable*}

\begin{figure}\begin{center}
\plotone{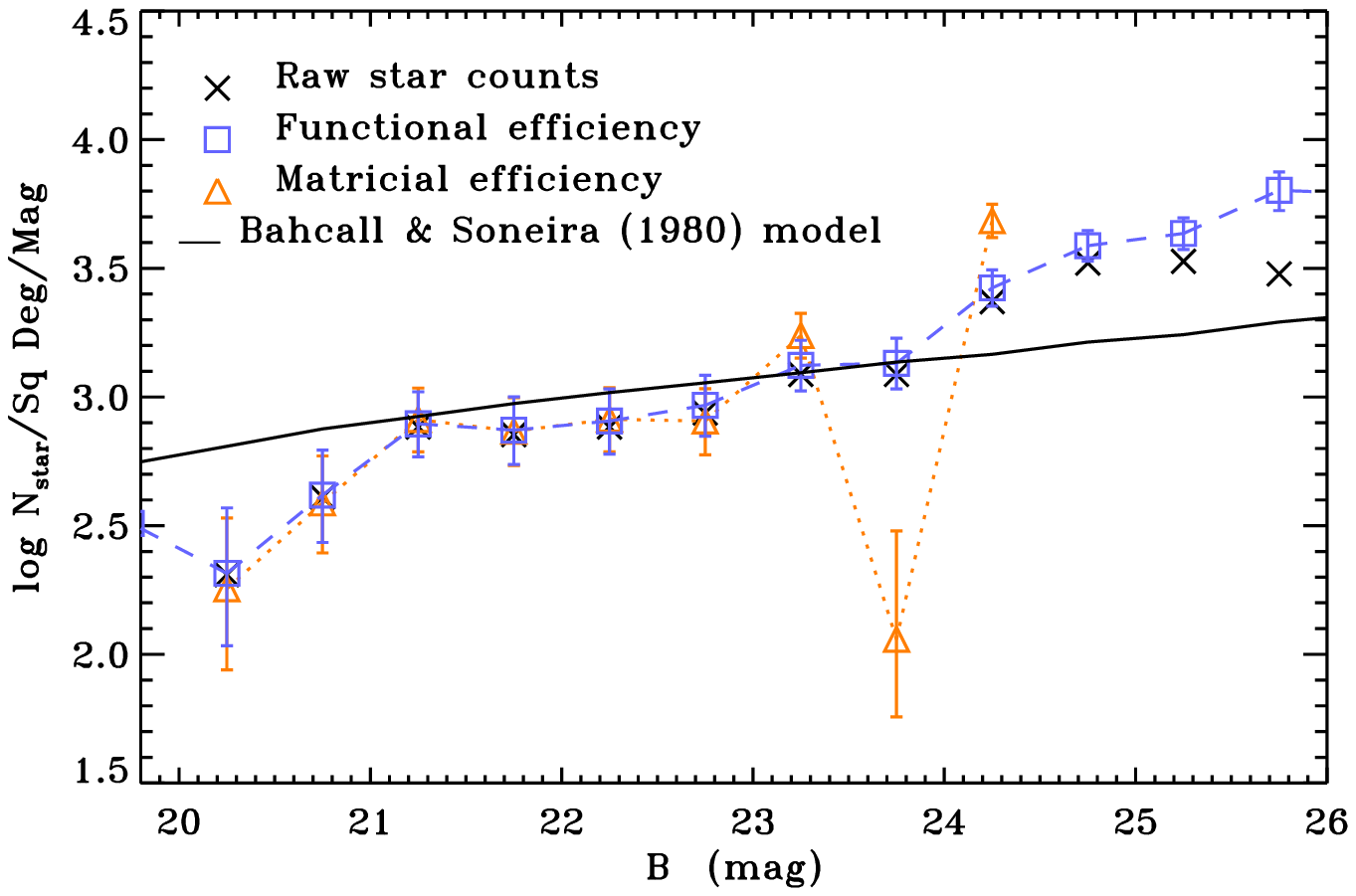}
\plotone{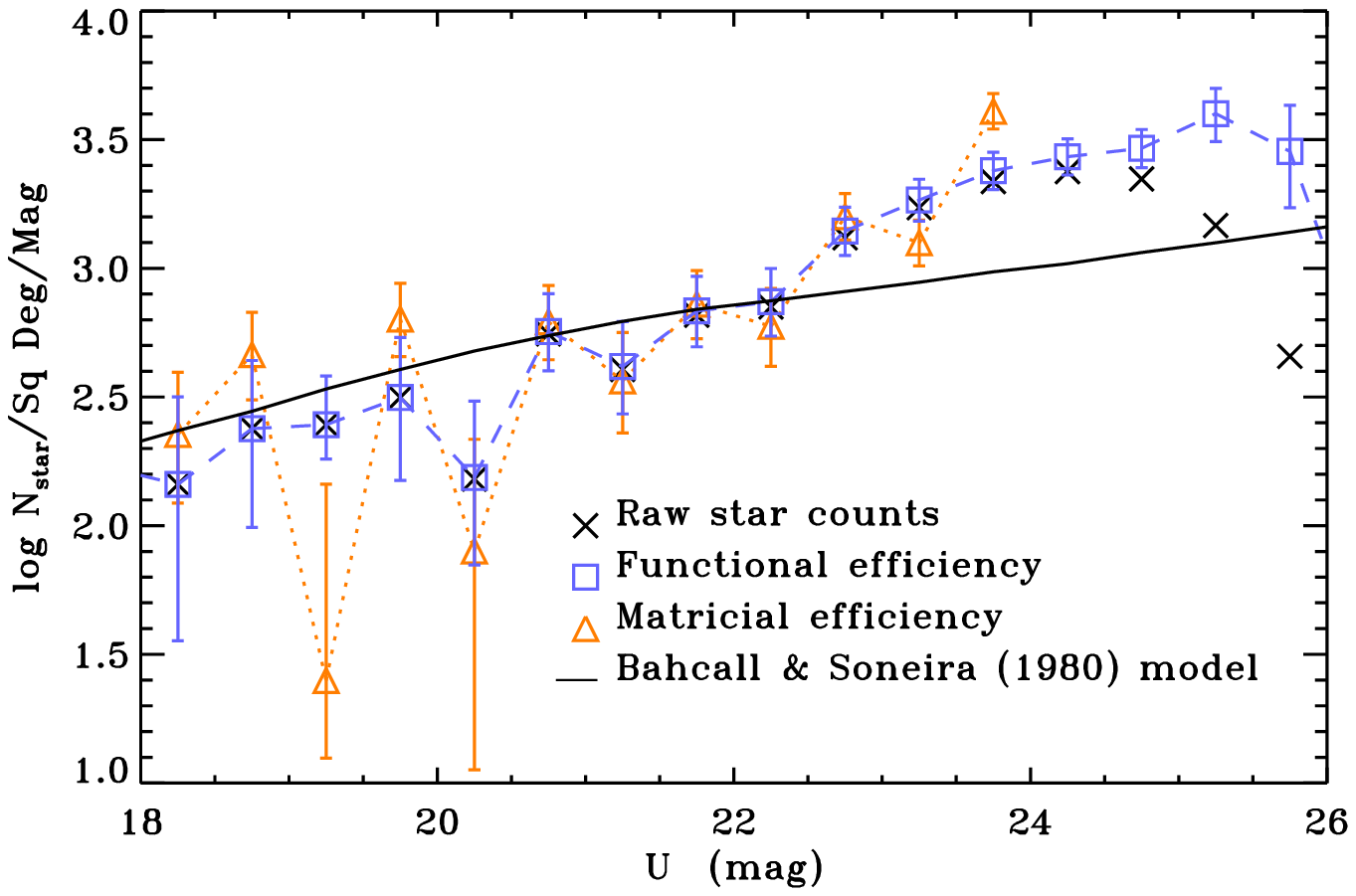}
\end{center}
\caption{
Star counts in GWS field derived using the DEEP catalogue, corrected for 
detection efficiency using "functional" method (squares) and 
"matricial" method (triangles) for both bands. Diagonal crosses are raw star 
counts in each filter. Solid line is the star counts prediction at each band for the GWS ($b=60$\deg, $l=95$\deg) from the \citet{Bahcall80,Bahcall81a} Galaxy model (see text for more details). Error bars correspond to counting statistics, added in quadrature to uncertainties from the efficiency correction.}\label{Fig:starcounts}
\end{figure}

\subsection{Star-galaxy separation}
\label{Sec:stargalaxy}

Number counts for "point-like" sources (defined in \S\ref{Sec:real}) were 
corrected for star counts in order to obtain galaxy number counts.  Star 
identifications were directly obtained over the area of overlap between 
our images and the HST Groth survey by cross-correlating our catalogue with 
the F606W Medium Deep Survey (MDS) catalog of the GWS  
field\footnote{Medium Deep Survey, MDS: \\http://archive.stsci.edu/mds}, using the MDS star identifications \citep{Ratnatunga99}. 
The resulting star counts were 
corrected for detection efficiency using the efficiency functions 
and matrices (see \S\ref{Sec:efficiency}), scaled to the total areas for 
$U$ and $B$, and subtracted from the efficiency-corrected "point-like" 
number counts.  Star number counts in $U$ and $B$ are shown in 
Figure \ref{Fig:starcounts}, corrected for efficiency errors using the 
matricial and functional methods. Both procedures give similar results for 
our entire magnitude range, but the inversion of the efficiency matrices 
became strongly unstable for $B$($U$)$>$24 mag. 
Therefore, we decided to use the functional efficiency method for 
correcting final galaxy number counts. 
We have also compared our star count measurements to the 
predictions from the \citet{Bahcall81a} model of the Galaxy for the coordinates of the GWS\footnote{The source code for computing the star counts for 
different filters and Galactic coordinates is kindly available by the 
author in the Astrophysics Source Code Library Archive, BSGMODEL: The 
Bahcall-Soneira Galaxy Model, 
\\http://ascl.net/bsgmodel.html} 
\citep[see also][]{Bahcall80,Bahcall86,Ratnatunga85}.  
Adopted Galaxy parameters are from 
Table 1 of \citet{Cabanac98}, 
LF parameters are from \citet{Mamon82}, and 
representative color terms are from \citet{Bahcall81b}. 
Predicted counts from the model are overplotted in 
Figure \ref{Fig:starcounts}.  Model and measurements agree in both 
bands at intermediate magnitudes.  Divergences at the faint end are 
to be expected, as the \citet{Bahcall81a} model is only reliable down to 
$B=20$ \citep{Bahcall85,Lasker87,Santiago96,Bath96}. 
At the bright end, the models predict $\sim$10 sources 
per 0.5-magnitude bin at $U$(or $B$)$\sim$20 in the area of the HST GWS data.  
The divergences here are likely due to low number statistics in our 
star count measurements, probably due to the rejection of saturated objects 
in the catalogues. For the bright regime between $U$(or $B$)$\sim$20 mag 
and the saturation limit\footnote{A point source with 
FWHM=1.3\arcsec saturates at $B\sim$20.0 and $U\sim$18.3 mag in 
our images}, we have resorted to {\tt SExtractor\/}'s stellarity parameter 
(CLASS\_STAR $> 0.9$ for both $U$ and $B$), shown by Capaccioli et al\@.
(2001) to reliably identify stars at bright magnitudes in blue filters.

Results from star counts in $U$ and $B$ are listed in Tables 
\ref{Tab:starcountsu} and \ref{Tab:starcountsb}. Columns indicate  
raw star counts with the spurious sources subtracted; and efficiency-corrected star counts using functional efficiencies and matrices, together with the estimated upper and lower limits for both methods. The correction in $\log (N)$ due to stars is lower than 0.05 index at fainter magnitudes. Errors in star counting are computed in the same way as the errors in galaxy counting. Lower and upper errors from star-counting will be cuadratically added to the corresponding upper and lower errors that arise from efficiency corrections and statistical counting to obtain final number counts errors (see Appendix \ref{Append:errors}).

\section{Results and Discussion}
\label{Sec:results}

\begin{deluxetable*}{lll}
\tabletypesize{\scriptsize}
\tablewidth{0pt}
\tableheadfrac{0.01}
\tablecaption{$U$ and $B$ Galaxy Count Slopes from Several 
Surveys\label{Tab:slope}}
\tablehead{\colhead{Reference} & \colhead{d$\log$(N)$/$dm} & \colhead{Magnitude 
Range}} 	  
\startdata 
\cutinhead{$U$-band}\\[-0.2cm]
\citet{Koo86}		&$\sim$0.4    		&$20.0<U(\mathrm{Vega})<22.0$     \\
\citet{Williams96}     	&$\sim$0.05  		& $26.0<F300W(\mathrm{AB})<28.0$	\\
				&$\sim$0.40   		& $23.0<F300W(\mathrm{AB})<26.0$	\\
\citet{Hogg97}         	&$\sim$0.467		& $20.3 <U(\mathrm{AB})<26.3$       \\
\citet{Pozzetti98}     	&$\sim$0.49		& $20.0<U(\mathrm{Vega})<25.0$    \\
\citet{Fontana99}	&$\sim$0.49   		& $20.0<U(\mathrm{Vega})<25.0$    \\
\citet{Volonteri00}    	&0.47$\pm$0.05   	&  $22.25<F300W(\mathrm{AB})<27.25 $  \\
\citet{Metcalfe01}     	&$\sim$0.4    		&  $18.0<U(\mathrm{Vega})<25.0 $  \\
\citet{Radovich04}     &0.54$\pm$0.06  	&  $18.5<U(\mathrm{Vega})<22.5  $     \\
\citet{Capak04}     	&0.526$\pm$0.017  	& $ 20.0<U(\mathrm{AB})<24.5 $      \\
Present work        	&0.48$\pm$0.03	& $ 21.0<U(\mathrm{Vega})<24.0 $  \\

\cutinhead{$B$-band}\\
\citet{Tyson88}        	&$\sim$0.45   		&$22.0<B_j(\mathrm{Vega})<27.0$	       \\
\citet{Jones91}        	&0.442$\pm$0.003	&$19.0<B_j(\mathrm{Vega})<23.5$   \\
\citet{Metcalfe91}	&0.491$\pm$0.009	&$19.0<B_j(\mathrm{Vega})<24.4$        \\
\citet{Metcalfe95}	&0.396$\pm$0.001	&$22.4<B_j(\mathrm{Vega})<26.9$      \\
\citet{Williams96}	&$\sim$0.16      	&$26.0<F450W(\mathrm{AB})<29.0$	       \\
		           	&$\sim$0.39  		&$23.0<F450W(\mathrm{AB})<26.0$ \\
\citet{Bertin97}       	&$\sim$0.464		&$16.0<B_j(\mathrm{Vega})<21.0$     \\
\citet{Pozzetti98}     	&$\sim$0.45		&$20.0<b_j(\mathrm{Vega})<25.0$    \\
\citet{Arnouts99}      	&$\sim$0.31		&$25.0<F450W(\mathrm{AB})<28.0$      \\
\citet{Volonteri00}    	&0.4$\pm$0.1   	& $22.25<F450W(\mathrm{AB})<25.25$   \\
				&0.19$\pm$0.01   	& $25.25<F450W(\mathrm{AB})<27.25$   \\
\citet{Metcalfe01}	&$\sim$0.25		&$15.0<B(\mathrm{Vega})<29.0 $    \\
\citet{Huang01b}       	&0.473$\pm$0.006	&$16.75<B_j(\mathrm{Vega})<24.75 $      \\
\citet{Kummel01}       &0.479$\pm$0.005	&$14.4<B_j(\mathrm{Vega})<23.6$         \\
\citet{Capak04}       	&0.450$\pm$0.008	&$20.0<B_j(\mathrm{Vega})<25.5 $        \\
Present work		&0.497$\pm$0.017  	&$21.0<B(\mathrm{Vega})<24.5$   \\[-0.2cm]
\enddata
\tablecomments{Slopes given in the $B_j$ band have been all taken 
from Table 5 of \citet{Kummel01}.}
\end{deluxetable*}

\begin{figure}\begin{center}
\plotone{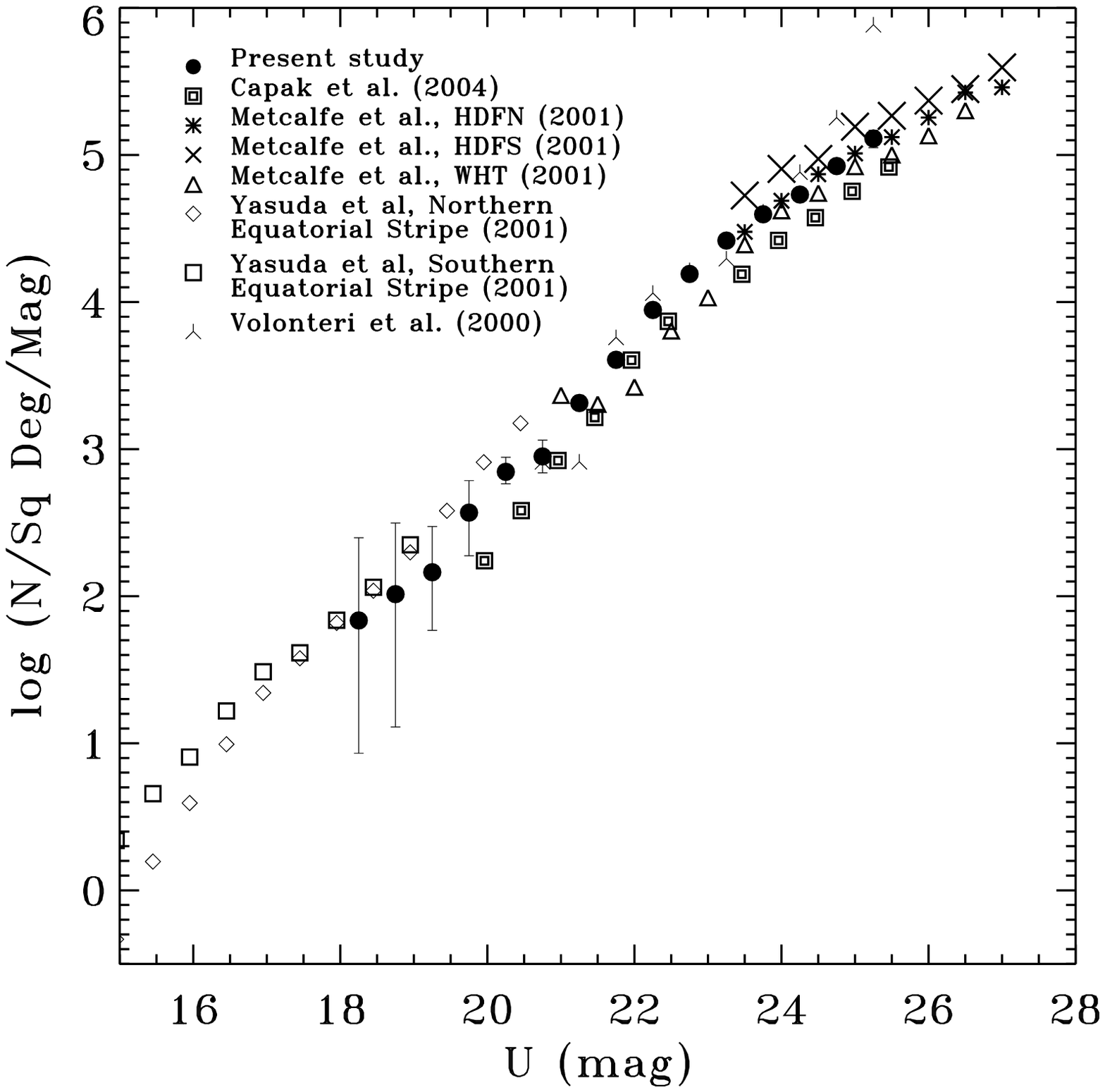}
\plotone{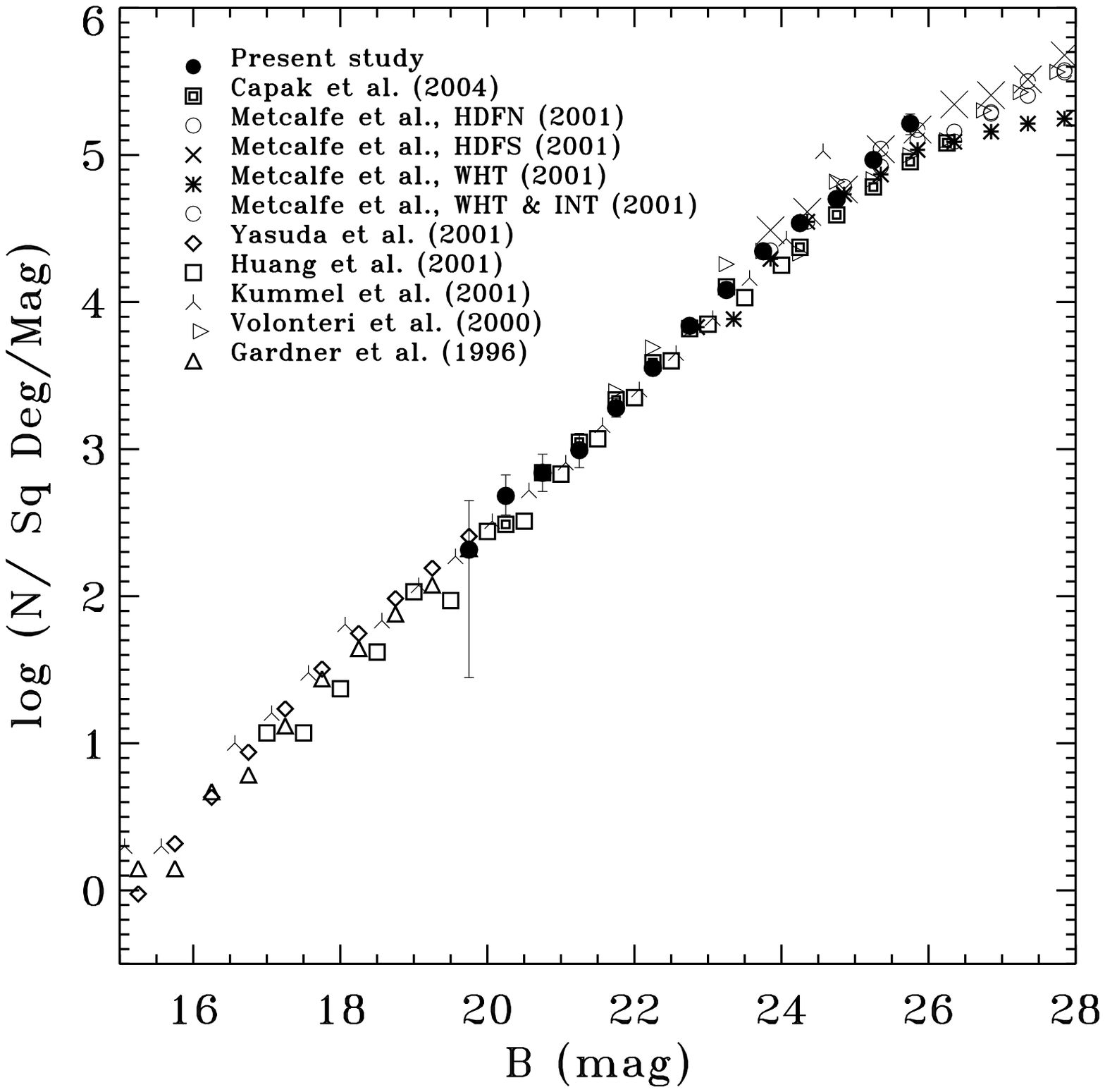}
\caption{
Galaxy number counts from GWS data in $U$ and $B$ over $\sim$900 arcmin$^2$. 
The comparison with previous works includes only tabulated, CCD-based  
counts \citep[see][]{Gardner96,Volonteri00,Metcalfe01,Yasuda01,Huang01b,Kummel01,Capak04}. The faintest point in each band corresponds to the 50\% 
efficiency magnitude, while the brightest end reaches the saturation limit 
of the INT/WFC in each band.
}\label{Fig:counts}
\end{center}\end{figure}

\subsection{Differential number counts}
\label{Sec:ncounts}

The final source catalogues were obtained by running {\tt SExtractor\/} on the $U$ and $B$ images with {\tt DETECT\_THRESH\/}=0.6, excluding spurious candidates according to the criteria outlined in \S\ref{Sec:reliability}. Extinction corrections were applied to all the sources in the catalogues 
(see \S\ref{Sec:extinction}). For the number count study, the edge areas with lower exposure due to the dithering pattern were excluded.  Final useful areas for number counts were 888 arcmin$^2$ for $B$ and 846 arcmin$^2$ for $U$ (see Table \ref{Tab:finalimages}). 
The division into the three size groups defined in \S\ref{Sec:sourceextraction} allowed us to correct for completeness in each size group using the functional 
efficiency method, as described in \S\ref{Sec:efficiency}. For summarizing, final counts are obtained from raw counts through: Galactic extinction correction (\S\ref{Sec:extinction}); correction for spurious 
detections (\S\ref{Sec:reliability}); detection efficiency correction using the functional efficiency method (\S\ref{Sec:efficiency}); and subtraction of the star counts from Tables\,\ref{Tab:starcountsu} and \ref{Tab:starcountsb} 
(\S\ref{Sec:stargalaxy}).   

Our results for $U$ and $B$ galaxy differential number counts are summarized in Tables \ref{Tab:countsu} and \ref{Tab:countsb}, respectively. Magnitudes are in the Vega system. Raw counts in the $U$ and $B$ bands are listed in Col.\,(2)-(4) in these Tables, for each one of the three source size groups defined in \S\ref{Sec:synthetic}. Spurious corrected number counts are shown in Col.\,(5)-(7). Applied efficiency correction factors for each size group appear in Col.\,(8)-(10). Col.\,(11) gives the final, differential galaxy number counts $N$ per magnitude and per deg$^{2}$, and Col.\,(12) and (13) list the upper and lower 1$\sigma$ errors for $N$, computed as explained in Appendix \ref{Append:errors}.  For convenience, we list $\log(N)$ in Col.\,(14). Counts are derived from  0.2350 deg$^2$ and 0.2467 deg$^2$ of sky in $U$ and $B$, respectively. The range of our counts is $18\lesssim U\lesssim 25$ and $19.5\lesssim B\lesssim 25.5$.  The bright limit is set by saturation on our CCD frames.  The faint limit is set by our 50\% detection efficiencies, which are $U_{50\%} = 24.83$ and $B_{50\%} = 25.46$ for point sources (\S\ref{Sec:efficiency}). Very few sources are detected in the large size group (see Col.\,(4) of Tables \ref{Tab:countsu} and \ref{Tab:countsb}), and these are only found in the faintest magnitude bins.  Visual inspection shows that most of these sources are single- or multiple-peak structures embedded in a region of higher background, which {\tt SExtractor\/} takes as single extended sources.  While they may include real sources (note that these detections have survived the spurious source filter), it is very unlikely that sources in the large size bin trace single, faint, extended galaxies.  For consistency, we apply efficiency corrections to them, and add them to our galaxy counts.  Their contribution to the total counts is entirely negligible.

We plot $U$ and $B$ number counts in Figure \ref{Fig:counts}, 
together with literature data. We have only included number count data coming 
from CCD observations which were given in tables, 
and convert $U$ and $B$ data from different 
photometric systems to Vega magnitudes in the Landolt system. 
Our counts are in excellent agreement with the other studies.  Note that literature data near the bright and faint ends of our counts come from independent studies, and that our data bridge the gap between large area, shallow surveys and deep, pencil-beam surveys.  
The dispersion among the different authors, which is greater in the 
$U$-band, is probably due to different completeness and spurious rejection corrections, and to the absence of Galactic extinction correction for the majority of studies. Clustering and cosmic variance can also contribute to the dispersion present in the $U$ and $B$ number counts from the different authors. We have estimated that the contribution due to clustering fluctuation is $\sim$0.1-0.4 the statistical counting errors in the range $20<U\mathrm{(}B\mathrm{)}<24$ mag for both filters \citep[see][ and references therein]{Jones91,Metcalfe95,Volonteri00,Yasuda01}.

\subsection{$U$- and $B$-band Count Slope}
\label{Sec:slope}

The slopes of our number count distributions were obtained by least-squares fits, yielding  
$d\log(N)/dm$= 0.50$\pm$0.02 for $B$=21-24.5, with $\chi^2$=0.018, and 
$d\log(N)/dm$= 0.48$\pm$0.03 for $U$=21-24, with $\chi^2$=0.033.  In Table 
\ref{Tab:slope}, our $U$ and $B$ count slopes are listed together with those from several surveys. Our  
slopes are in good agreement with other studies, with the exception 
of \citet{Williams96} in both bands 
and \citet{Metcalfe01} in $B$. The different photometric bands and magnitude ranges can explain 
the discrepancies. Moreover, differences with \citet{Williams96} could arise from the fact that 
their sample was selected in 
a red band ($F606W$+$F814W$), and hence it is biased towards redder 
objects; while the change of the slope at $B\sim$24 in \citet{Metcalfe01} 
flattens it much more than in other studies, as noticed by \citet{Kummel01}.

\citet{Yasuda01} pointed out that, at the bright magnitudes 
where cosmological and evolutionary corrections 
are relatively small, the shape of the galaxy number counts-magnitude 
relation is well characterised by 
$N(m_{\lambda})\varpropto 10\,^{0.6\,m_{\lambda}}$, the expected relation for 
a homogeneus galaxy distribution in a Euclidean Universe. But since the 
night sky would be infinitely bright in $U$ and $B$ if this trend continued forever, 
at some faint magnitude the $U$-band count slope must break. \citet{Hogg97} 
predicted that $U$-band break should happen at $27<U<28$, where the median 
$U-R$ reaches the value where objects get the UV spectral slope of a 
star-forming galaxy. \citet{Volonteri00} do not find evidence of a turn over 
or flattening down to $F300W(\mathrm{Vega})$=26, contrary to what is claimed by 
\citet{Pozzetti98}. On the other hand, \citet{Williams96} 
report a clear change in slope at $U\sim$25.3 on the HDF-N number counts. 
From our data, we report a change of the slope of the $U$ counts 1.5 
mag brighter: $U\sim 23.25$. 
Considering that, with our combination of depth and area, our survey covers the 
$18.25\lesssim U\lesssim 25.25$ mag range with complete data and good 
statistics better than 
the other existing surveys, this break should be more significant than the one
reported by \citet{Williams96}, whose area is less than 0.5\% ours.

The change of slope at $B$ is reported by 
\citet{Lilly91} at $B\sim 25$. 
In our covered range, our $B$ number counts do not exhibit 
a clear change. At $B\sim 24$, it seems that there is a turnover, 
but we would need deeper images in $B$ in order to corroborate it. 
Nevertheless, several studies found this turnover at $B\sim 24$ 
\citep[see][]{Kummel01,Arnouts99,Williams96,Metcalfe95}, 
so that slope change 
in our $B$ counts is probably real. 
Perhaps more interesting is the fact that from our $B$ data we can confirm 
the slight increase in the slope at $B\sim 23$, reported by 
\citet{Metcalfe95}. This couples with the decrease in the slope at fainter 
magnitudes and leads to an small upward bump in the counts centered at $B\sim 24$, a feature that can be observed in our data. In fact, \citet{Metcalfe95} 
indicate that this "hump" is a characteristic feature of pure luminosity 
evolutionary models (PLE) and is caused by strongly evolving 
early-type galaxies at high redshift.

\section{Modeling the galaxy number counts}
\label{Sec:modeling}

The large area$\times$depth product of our $U$ and $B$ counts, and the simultaneous 
availability of $U$, $B$ and $K$ number counts for the same field, permit a useful comparison with the predictions of galaxy formation models.  We restrict this comparison to traditional number count models which evolve the $z=0$ luminosity function back in time, as opposed to SAM models that evolve galaxies forward in time to z=0.  While number count data in 
blue bands can be reproduced with a fairly wide range of number count model parameters, reproducing number counts in the NIR has proven more challenging, due to the peculiar change in the number count slope at $\Ks = 17.5$ (Gardner et al.\ 1993; CH03). 
CH03 showed that a fairly recent formation epoch for massive galaxies ($z\approx 1.5$) yielded a number count slope change similar to that observed; allowing red, massive galaxies at higher redshifts would yield higher predicted faint counts than seen in the data.   

In this paper we extend the number count modeling presented in CH03 by comparing the models to both the NIR counts and the $U$ and $B$ counts.
The combination of the blue number count distributions, which are almost featureless over our magnitude range, with the \Ks\ number count distribution, which shows a knee at intermediate magnitudes, provides useful constraints on the formation history of the various galaxy types.  We provide the first number count model that accounts for both blue ($U$, $B$) and NIR 
(\Ks) number counts.  Working on the same area of the sky ensures 
that the distinctions between blue and NIR number count profiles are 
not a reflection of cosmic variance.  

To build galaxy number count predictions, we have used the {\tt ncmod\/} code from \citet{Gardner98}, made available by the author at his web site\footnote{http://survey.gsfc.nasa.gov/$\sim$gardner/ncmod/}.  The reader is referred to the above publication for details of the model.
Briefly, the code evolves the local LF back in time, for a number of galaxy types, using SEDs from the Galaxy Isochrone Synthesis Spectral Evolution Library (GISSEL96) model \citep{Bruzual93,Leitherer96}.  The star formation history for each galaxy type is parametrized by the redshift of galaxy formation \zf\ and the timescale $\tau$ of the (exponential) decay of the SFR.  The code allows for the inclusion of extinction by dust internal to the galaxies; dust is modeled as an absorbing layer, symmetric around the midplane of the galaxy, whose thickness is a fraction of the total thickness of the stellar disk \citep{Bruzual88,Wang91}. A power-law $\propto \lambda^{-2}$ extinction law is adopted, and galaxies are assumed to have an extinction $\tau_{4500} = 0.6\,(L_{z=0}/L^\star)^{0.5}$ at 4500~\AA.  (Gardner sets the coefficient in the previous equation to 0.2.  As discussed below, 0.6 is more justified by observations, and we modified the code accordingly.)
Two recipes are provided to account for the effects of merging, namely, a $z$-evolution of the LF parameters, i.e., $\Phi^* \propto (1+z)^\beta$, which conserves the luminosity density by setting $L^*\propto (1+z)^{-\beta}$, and the formulation proposed by \citet{Broadhurst92}, 
with $\Phi^* \propto \exp \{-Q/\beta[(1 + z)^{-\beta} - 1]\}$.  Here, $Q$ is approximately the number of objects at $z=1$ that will merge to form a typical galaxy today, and $\beta$ is a function of the look-back time.

The main inputs of the model are: LFs, SEDs and formation redshift \zf\ for each of the galaxy classes, extinction and merging switches, and cosmological parameters. 
We have used the local, morphologically-dependent luminosity functions (MDLF) from \citet{Nakamura03}, which are derived from about 1500 bright galaxies of the Sloan Digital Sky Survey (SDSS) northern equatorial strips. 
A literature search from 1988 to 2003 showed that \citet{Nakamura03} provide the morphologically-dependent LFs with best statistics.  
After correcting for Galactic extinction, the limiting magnitudes are $r^*\leq15.7$ or $B\leq17.3$ Vega mag.  With these depths, the MDLFs include local blue compact dwarf galaxies, which have typical magnitudes in the range $B\simeq12-17$ mag.  We adopt the galaxy classes from \citet{Nakamura03}, who classify galaxies into four groups: E-S0, S0/a-Sb, Sbc-Sd, and Im.  In Table \ref{Tab:luminosityfunction} we give the Schechter parameters  of these MDLF in the SDSS $r^*$ filter, for our adopted cosmology.  

We adopt fairly standard population parameters to describe each galaxy class (see Table~\ref{Tab:sed}).  A Salpeter IMF is used for all classes.  In view of the mass-metallicity relation \citep{Tremonti04}, Solar metallicities are adopted for the E-S0 and S0/a-Sb groups, and lower metallicities for later types.   Star formation is instantaneous (SSP models) for E-S0, exponentially-decaying for spirals, and constant for the Im class.  

We include number evolution, as ample evidence shows that merger fractions increase with look-back time \citep{Lefevre00,Conselice03,Cassata05}.  
We parametrize merger-driven number evolution as $\Phi^* \propto (1+z)^\beta$, with $\beta = 2.0$ providing good results, and we explore also $\Phi^* \propto  \exp \{-Q/\beta[(1 + z)^{-\beta} - 1]\}$, with $Q\approx 1$ giving reasonable results\footnote{Note that the choice of exponents is not based on exponents $\eta$ found when parametrizing  merger fractions as $f_m \sim (1+z)^\eta$: the latter does not lead to a $z$-dependency of $\Phi^*$ of the form $\propto (1+z)^\beta$.}


An accurate description of extinction is critical for comparing model predictions with data for wavebands from $U$ through \Ks.  We find Gardner's assumption that $\tau_B=0.2$ too conservative in view of the evidence from studies of local disk galaxies.  \citet{Peletier92} infer a face-on $\tau_B\approx1.0$ for spirals; \citet{Boselli94} find $\tau_B>1$, while \citet{Xilouris99} infer a central $\tau_B=0.8$.  We have adopted $\tau_B=0.6$, an intermediate value between Gardner's and the above references.  Extinction is assumed for all galaxy classes.  


\begin{figure}[!]\begin{center}
\plotone{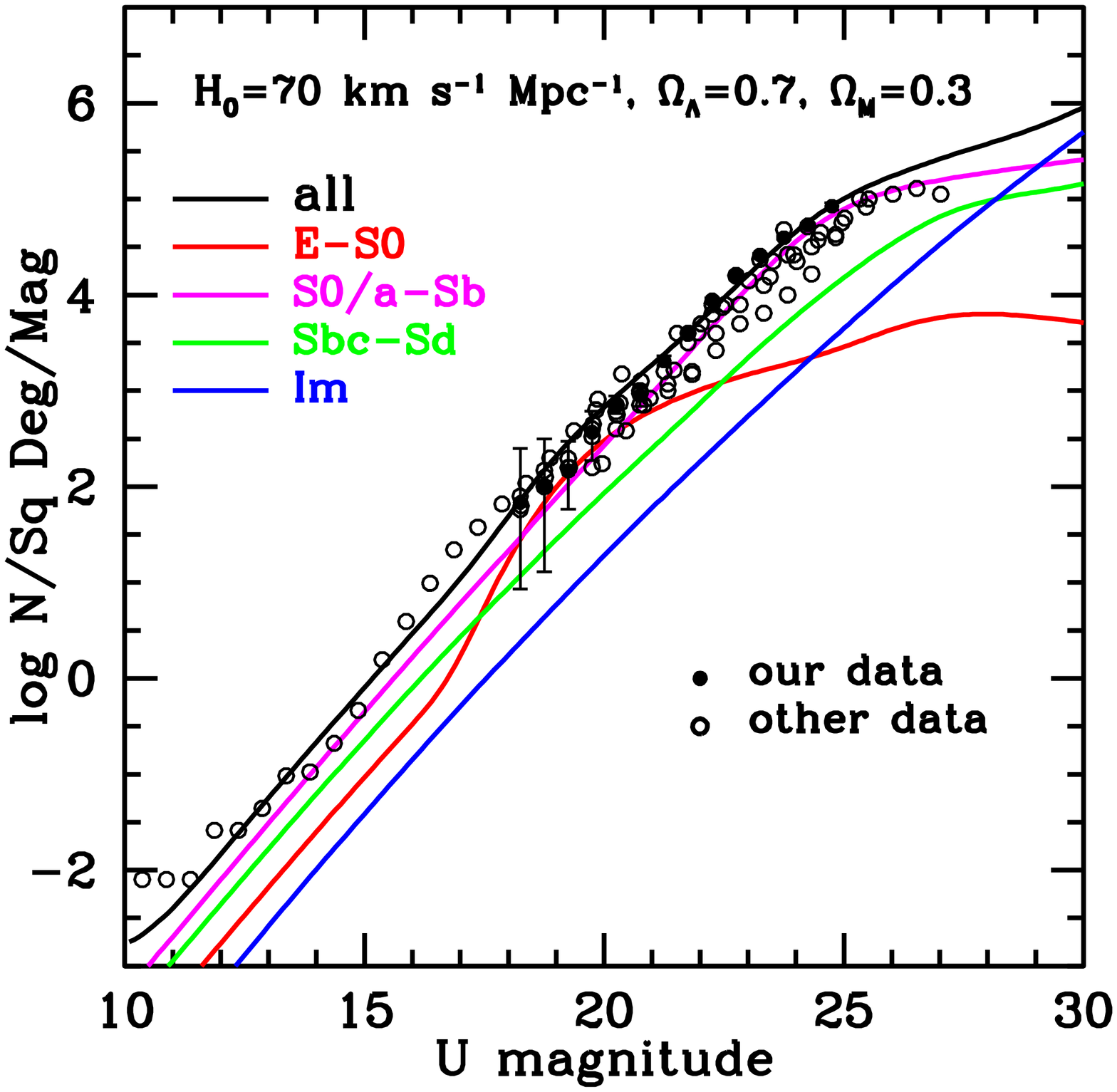}
\plotone{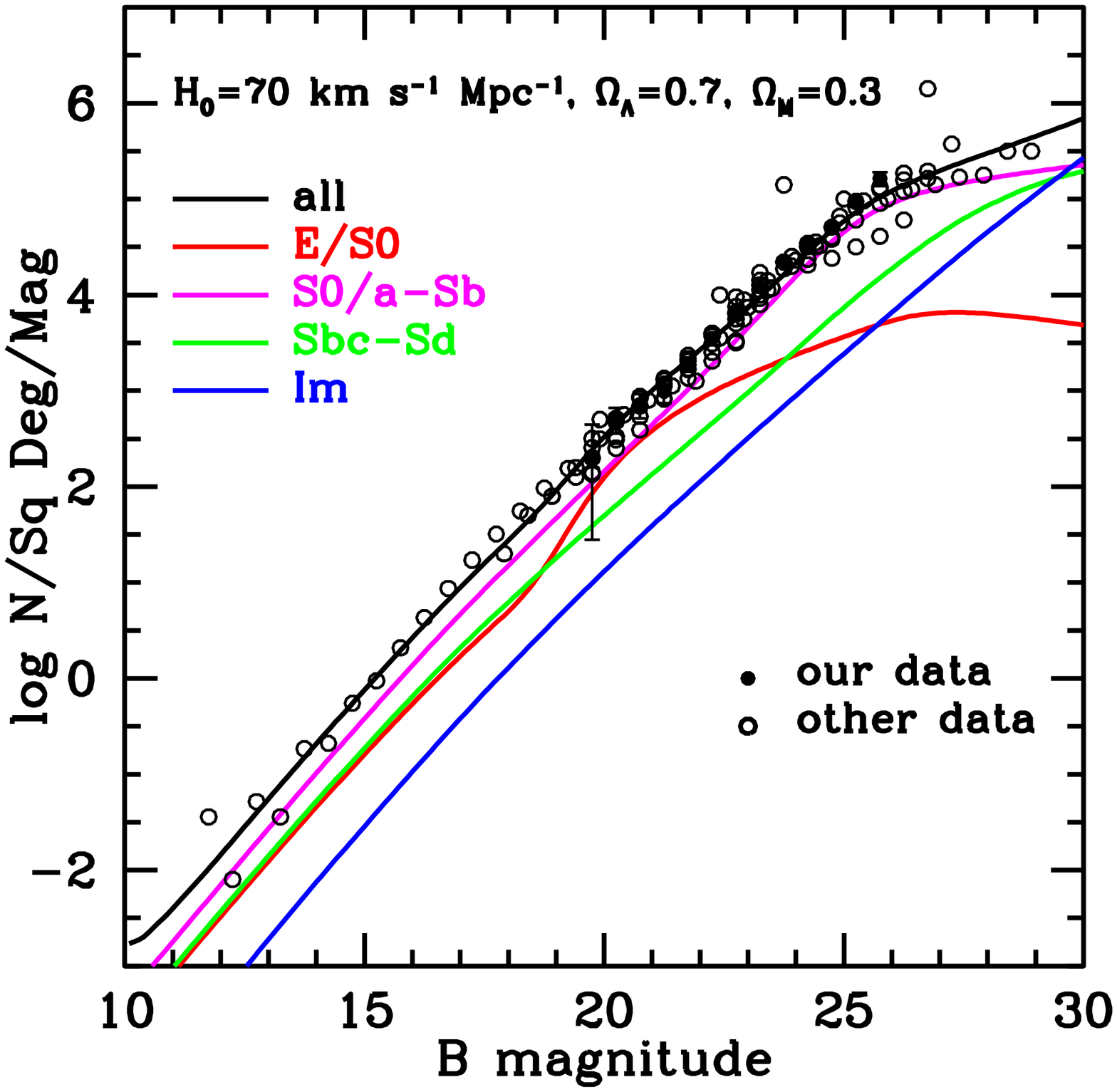}
\plotone{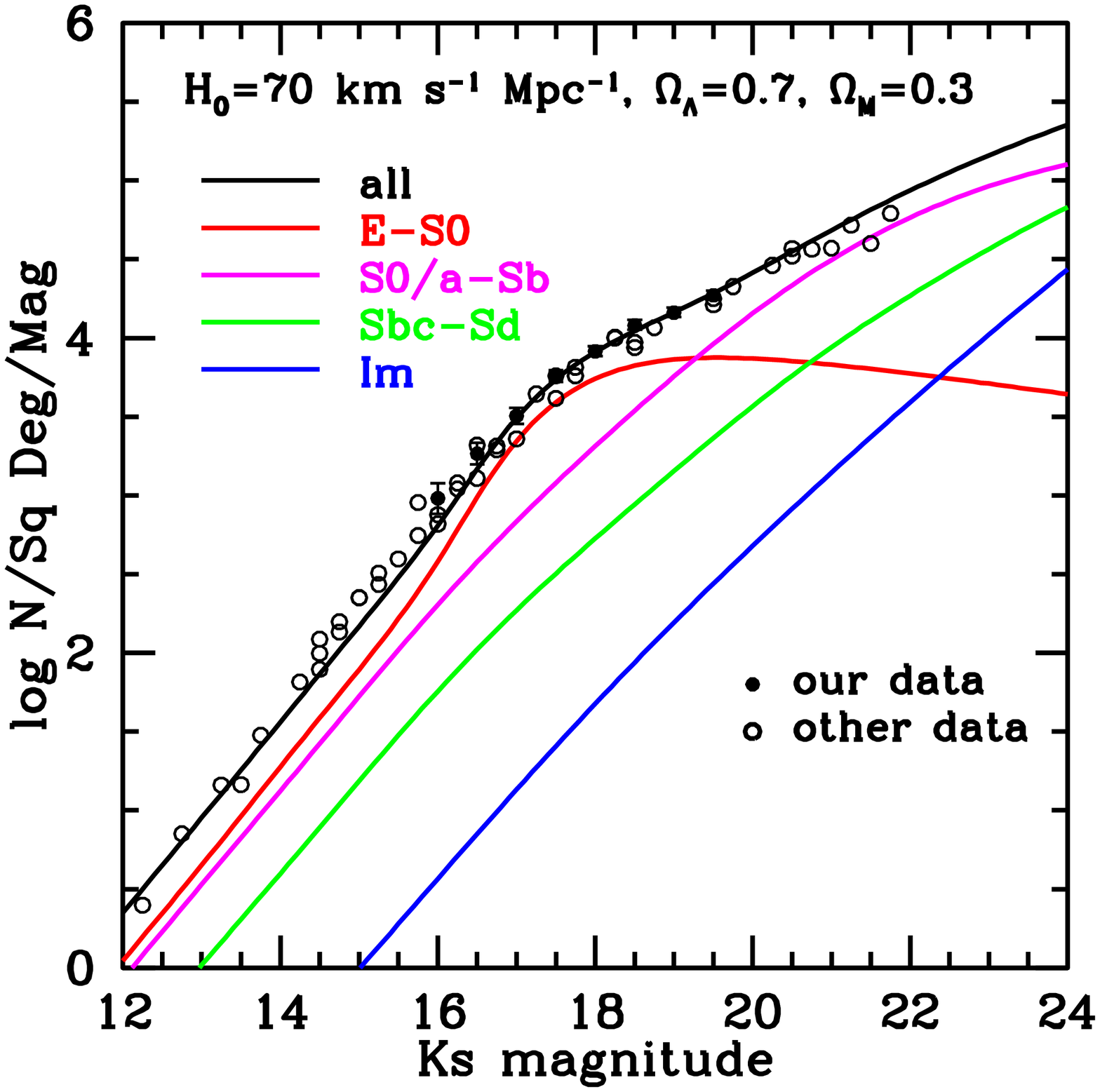}
\caption{From top to bottom, evolution models overplotted on the $U$, $B$, and \Ks\ galaxy number counts, with  $\zf = 1.5$ for ellipticals and early spirals and $\zf = 4$ for other types. Number evolution of the luminosity function is modeled using  $\phi^*\propto (1 + z)^\beta$, with $\beta=2.0$. \emph{Filled circles}: Our GWS data. $U$ and $B$ counts are from Tables~\ref{Tab:countsu} and \ref{Tab:countsb}, respectively, and \Ks\ counts are from CH03.  \emph{Open circles}: Counts from other authors. See references in the text. \emph{Solid lines}: Total galaxy count prediction. \emph{Dotted lines}: Prediction for ellipticals. \emph{Short-dashed lines}: Prediction for early spirals.
\emph{Long-dashed lines}: prediction for late spirals.
\emph{Long-short dashed lines}: Prediction for Im galaxies.}
\label{Fig:modelsUBKzf1.5}
\end{center}\end{figure}

\begin{figure}[!]\begin{center}
\plotone{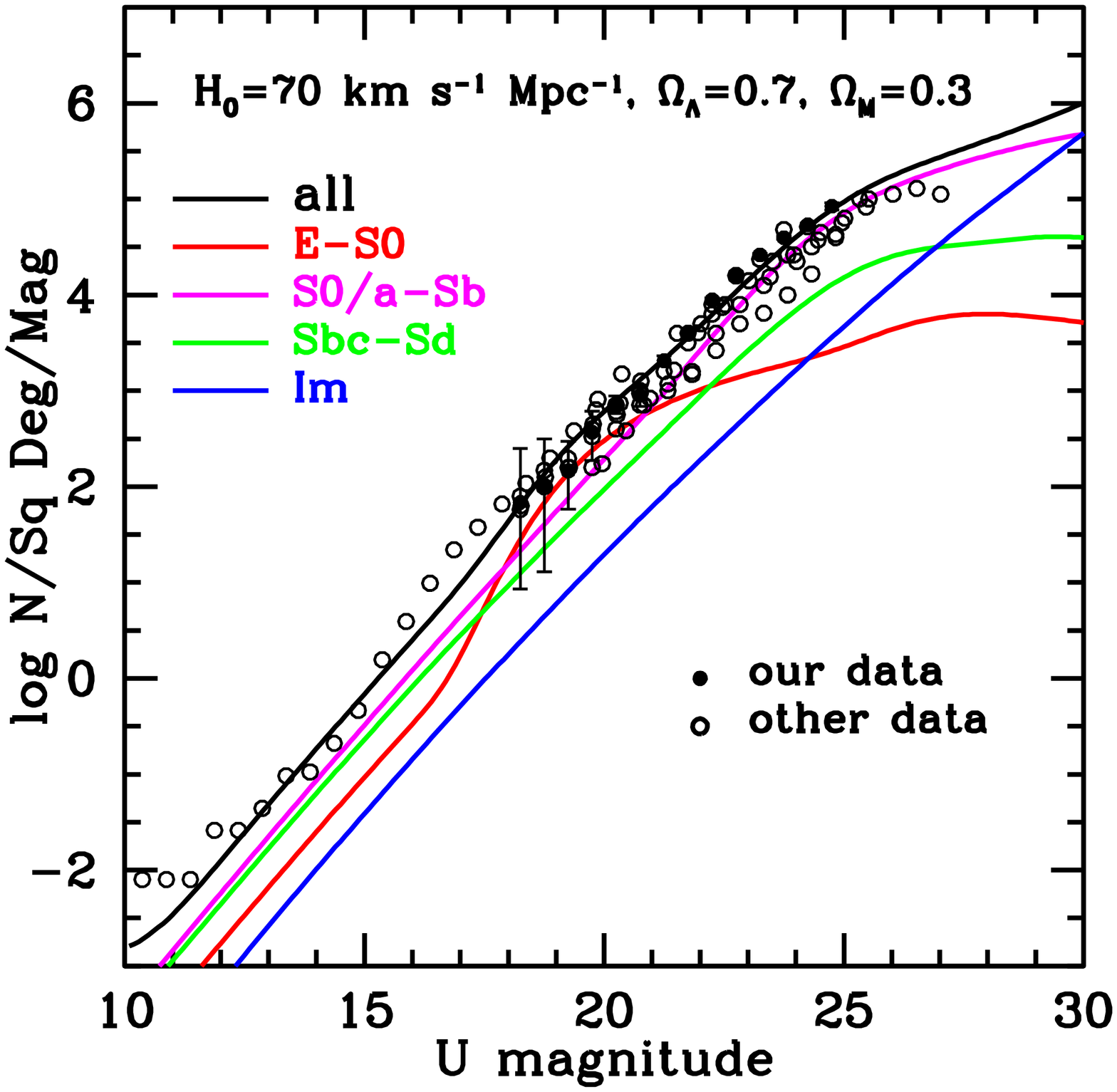}
\plotone{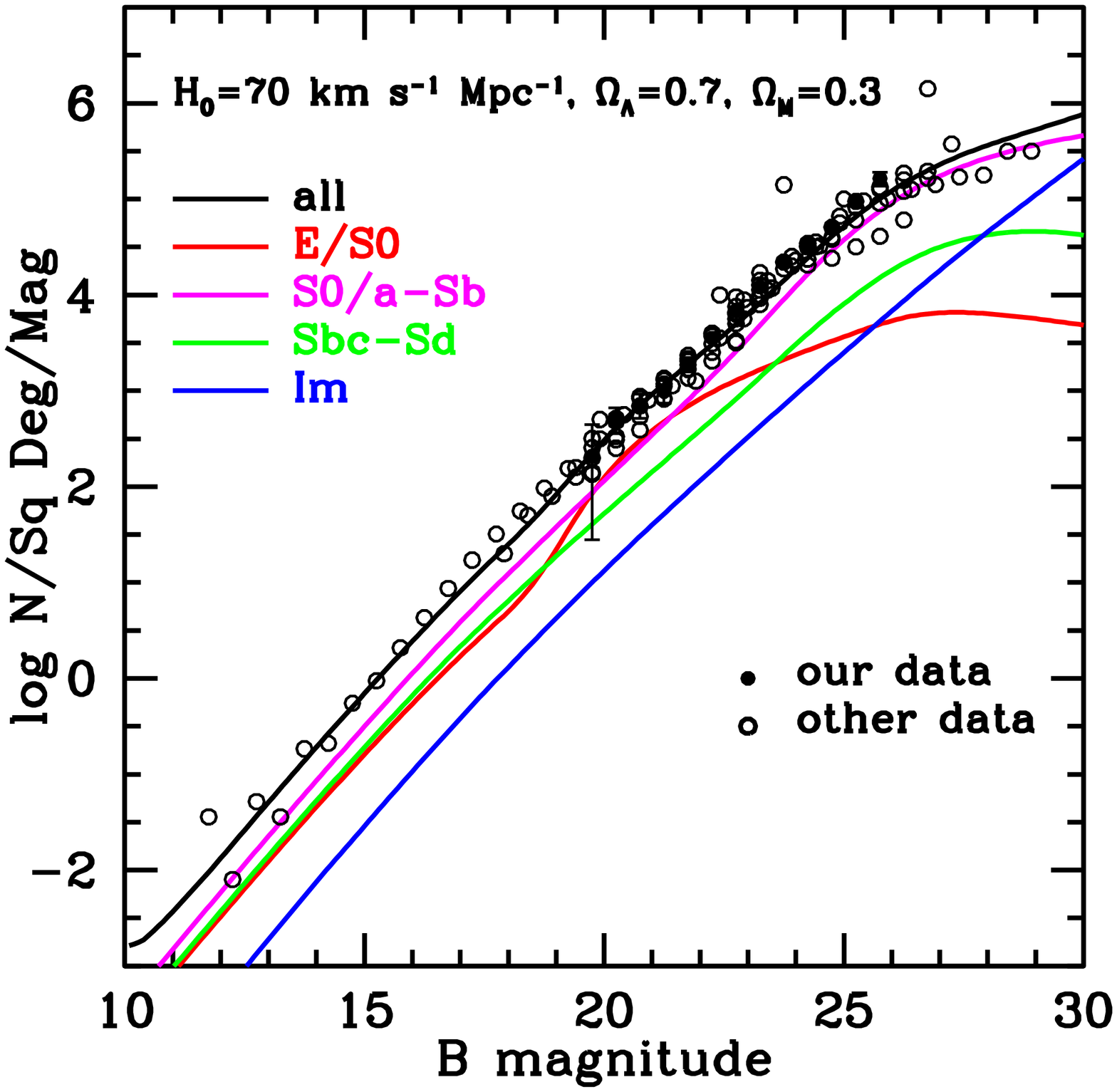}
\plotone{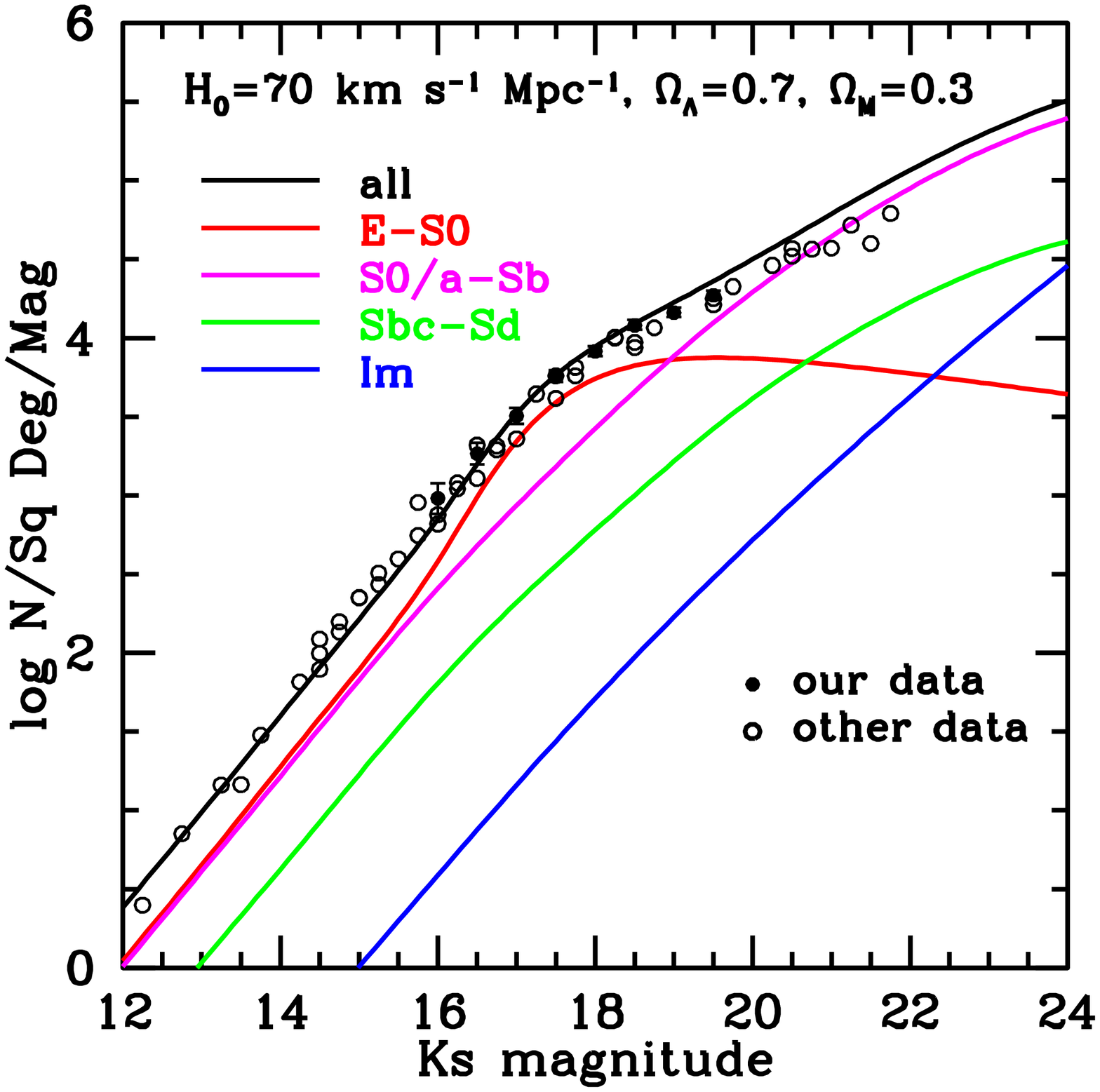}
\caption{From top to bottom, evolution models overplotted on the $U$, $B$, and \Ks\ galaxy number counts, with  $\zf = 1.5$ for ellipticals,  and $\zf = 4$ for other types including early-type spirals. Number evolution is modeled using  the \citet{Broadhurst92} prescription, with $Q_\mathrm{b}=1.0$.  \emph{Filled circles}: Our GWS data. $U$ and $B$ counts are from Tables~\ref{Tab:countsu} and \ref{Tab:countsb}, respectively, and \Ks\ counts are from CH03.  \emph{Open circles}: Counts from other authors. See references in the text. \emph{Solid lines}: Total galaxy count prediction. \emph{Dotted lines}: Prediction for ellipticals. \emph{Short-dashed lines}: Prediction for early spirals.
\emph{Long-dashed lines}: prediction for late spirals.
\emph{Long-short dashed lines}: Prediction for Im galaxies.}
\label{Fig:modelsUBKzf4}
\end{center}\end{figure}

\begin{deluxetable}{lccc}
\tablewidth{0pt}
\tabletypesize{\scriptsize}
\tableheadfrac{0.01}
\tablecaption{Schechter Parameters of the Adopted Luminosity Functions
\label{Tab:luminosityfunction}}
\tablehead{\colhead{LF Parameters} & \colhead{$M^*(r^*)\, $\tablenotemark{a,b}} 
& \colhead{$\alpha$ } 
& \colhead{$\Phi^*(0.001\,{\rm Mpc^{-3}})$\,\tablenotemark{b} } \\ 
\colhead{(1)}&\colhead{(2)}&\colhead{(3)}&\colhead{(4)}}  
\startdata
E-S0   			&-21.53$\pm$ 0.17  &-0.83$\pm$0.26 &1.61 $\pm$ 0.09\\
S0$/$a-Sb  			&-21.08 $\pm$ 0.19 &-1.15$\pm$0.26 & 3.26 $\pm$ 0.15\\
Sbc-Sd 			&-21.08$\pm$ 0.20  &-0.71$\pm$0.26 & 1.48 $\pm$ 0.05\\
Im\tablenotemark{c}	&-20.78                       &-1.90	             & 0.37\\[-0.2cm]
\enddata
\tablenotetext{a}{\/$M^*(r^*)$ in AB magnitudes \citep{Stoughton02}.}
\tablenotetext{b}{\/Converted to $h\equiv H_0/100 = 0.7$.}
\tablenotetext{c}{\/Irregular, star-forming galaxies.}
\end{deluxetable}

\begin{deluxetable}{llccr}
\tablewidth{0pt}
\tabletypesize{\scriptsize}
\tableheadfrac{0.01}
\tablecaption{Adopted Properties for each Galaxy Type\label{Tab:sed}}
\tablehead{
&  \multicolumn{2}{c}{SFR} & &\\\cline{2-3} \\[-0.1cm]
\colhead{Galaxy Type}	& \colhead{Type} & 
\colhead{$\tau$\tablenotemark{a}}& \colhead{Z/Z$_{\odot}$} &
\colhead{IMF} \\ 
\colhead{(1)}&\colhead{(2)}&\colhead{(3)}&\colhead{(4)}&\colhead{(5)}} 
\startdata   
 E-S0 &    SSP  &     0    &  1   &        Salpeter\\                 
   S0$/$a-Sb &  exponential &    4   & 1 &         Salpeter\\              
   Sbc-Sd &  exponential  &   7   &  2/5&       Salpeter \\        
   Im          &  constant  &   ...     &  1/5   &      Salpeter\\[-0.2cm]
\enddata
\tablenotetext{a}{\,The star-formation e-folding time in Gyr.}
\end{deluxetable}

\subsection{Number Count Model Predictions}
\label{Sec:modelresults} 

When evolving galaxy populations back in time, a characteristic power-law 
behavior is found for the counts.  This is given by the intrinsic 
brightening of galaxies with look-back time together with the 
$z$-evolution of the volume element, and it extends faintward until we 
reach magnitudes for which galaxies have yet to build a large fraction of 
their final stellar masses.  Breaking such power-law behavior, as is 
necessary at intermediate magnitudes for the \Ks\ counts, can only be 
accomplished by setting the formation redshift \zf\ of a dominant 
population to a moderately low value; other model parameters, such as the 
evolution of the merger fraction or the extinction, are unable to yield 
the observed knee in the \Ks\ counts.  A model that reproduces the $U$, 
$B$, and \Ks\ counts can be obtained by setting $\zf =1.5$ for ellipticals 
and early spirals, and $\zf = 4$ for other galaxy classes, and assuming 
a merger-driven number evolution as $(1 + z)^{2.0
.}$.  
Figure \ref{Fig:modelsUBKzf1.5} shows the  number count predictions of 
this model in $U$, $B$ and \Ks. Our total number counts as well as the 
contributions of each galaxy class are shown. Number count data from other 
authors have been plotted: in the $U$-band \citep[][]{Volonteri00,Metcalfe01,Yasuda01}; 
in the $B$-band \citep[][]{Gardner96,Volonteri00,Metcalfe01,Yasuda01,Huang01b,
Kummel01}; and in the \Ks-band  \citep{Gardner93,Djorgovski95,McLeod95,Huang97,
Minezaki98,Totani01}.  

A second model which avoids the uncomfortably low \zf\ for spirals is shown in Figure~\ref{Fig:modelsUBKzf4}.  This model, with $\zf = 4$ for early-type spirals, also provides a good description of the observed counts in $U$, $B$, and \Ks.  The merger prescription by \citet{Broadhurst92} was used, with $Q=1$, in order to lower the slope in the faint \Ks\ counts, which seems steeper than the trend defined by the data;  shallower faint-end slopes are obtained for lower values of $Q$ (\S\,\ref{Sec:modeling}).   For ellipticals, we need to keep $\zf = 1.5$, otherwise we fail to reproduce the $\Ks = 17.5$ knee 
(\S~\ref{Sec:formationredshifts}).  While differing in the details, both models presented in Figures \ref{Fig:modelsUBKzf1.5} and \ref{Fig:modelsUBKzf4} provide similarly accurate overall descriptions of the data.  

These models capture the essential ingredients of the modeling of \Ks\ counts we presented in CH03, by setting \zf\ to a low value, and indicate that the solution proposed by CH03 to the knee in the \Ks\ counts does reproduce the counts in blue passbands as well.  But the new models differ from CH03 
in several details, including the LF, here derived from SDSS data; the addition of extinction, which was unnecessary for the NIR counts of CH03; the inclusion of merging evolution; and the exclusion of any galaxy population not present in the local LF.  



\subsubsection{Dominant populations}
\label{Sec:dominantpopulations}

Over the magnitude range sampled by our data, the $UB$ counts are dominated by early- and late-type spirals, especially so in $U$.  At bright magnitudes, the lower contributions of E-S0 and Im galaxies reflect their lower $\Phi^*$ at $z=0$ (see Table\,\ref{Tab:luminosityfunction}).  Im galaxies contribute $\sim$1/20 of the total counts throughout our magnitude range, while the contributions of E-S0s decreases with magnitude owing to the strong K-corrections of their red stellar populations: neither has a measurable contribution to the counts.  In \Ks, the counts are dominated by E-S0 at bright magnitudes, and by early-type spirals at $\Ks > 19.5$.  We note that, contrary to many previous works \citep{Babulrees92,Phillips95,Driver95,Babul96,Ellis97,Driver98,Phillips00}, we do not need to introduce ad-hoc populations of star-forming galaxies at intermedite redshifts to explain the observed number counts; our model counts only comprise galaxies in the local LFs evolved back in time.  

\subsubsection{Formation redshifts}
\label{Sec:formationredshifts}

As discussed in CH03, only by adopting a late $\zf \approx 1.5$ for ellipticals do the models reproduce the $\Ks = 17.5$ knee.  In the context of the number count models we are using, this feature requires the disappearance, or significant downsizing, of a dominant red population with look-back time at intermediate redshifts.  Only the ellipticals can play this role given the $z=0$ LFs.  We searched for alternative schemes which did not involve a late \zf\ for ellipticals, by varying either the star formation time scales, the merger recipes, or the extinction.  We found no combination of parameters that simultaneously yields the knee in \Ks\ \emph{and} the lack of a knee in blue counts.  In particular, postulating a specific dusty phase in galaxy evolution at intermediate redshifts barely affects the \Ks\ counts, and instead introduces features in the blue number counts which are not seen in the data.   

The merger rate in hierarchical $\Lambda$CDM models, and the merger-driven star formation assumed to derive from it in SAM models, peaks at $z \sim 2$ \citep[e.g.,][]{Cole00}.  Ellipticals are assumed to result from such mergers.  We conjecture that the knee at $\Ks = 17.5$ in the NIR number counts may reflect the onset of this red population of merger origin.  Such merger origin cannot be properly accounted for by \ncmod, due to its simplified galaxy formation recipes, in which mergers change galaxy numbers and luminosities, but not galaxy types.   The appearance of the red population needs to be introduced by setting the appropriate \zf.  Setting the population to form instantaneously (SSP model) helps \ncmod\ mimic the onset of the merger-induced red population by minimizing the phase during which the galaxies contain young stars.

The late ($z \sim 1.5$) formation of ellipticals predicted by \ncmod\ can be reconciled with the existence of a significant population of evolved galaxies at $2<z<4$ \citep{Franx03,Labbe03,Daddi03,Glazebrook04,deMello04}. The mass density of red, dead galaxies at z=1.8 was  $\sim$16\% of the local
density \citep{McCarthy04}; \zf\ = 1.5 in our number count models reflects
such downsizing of the evolved population. Also, the rapid increase in the clustering of high-$z$ galaxies toward redder colors \citep{Daddi03} suggests that red sources at $2 < z < 4$ trace a cluster population; our Groth number counts predominantly trace a field population, in which elliptical formation may occur later than in clusters.  A similarly late formation for the majority of ellipticals/S0 ($\zf < 2$) is derived from the analysis of color gradients in early-type galaxies in the HDFN using multi-zone single collapse models \citep{Menanteau01}.   


In contrast with the ellipticals, \zf\ for all the star-forming galaxy classes is poorly constrained in the models.  For early spirals, values in the range $1.5 < \zf < 10$ yield accepable results; for late spirals, the acceptable range is $0.5 < \zf < 10$. 

\subsubsection{Dust}
\label{Sec:dust}

Adoption of extinction for all galaxy classes is critical to the success of the model.  In particular, we need to assume extinction for ellipticals.  Otherwise, the predicted blue counts show a strong feature, at $U \sim 18$ and $B \sim 19$.  Our extinctions ($\tau_B = 0.6$ for $L^\star$ galaxies, see \S\,\ref{Sec:modeling}) are moderate, if we take into account that precursors of ellipticals may include dusty EROs at $1<z<2$ \citep[e.g.][]{Smail02}.  Red, massive galaxies at higher redshifts show much higher extinctions; e.g., $A_V \sim 2.7$ at $2< z < 3.5$ in the HDF-S \citep{ForsterSchreiber04}.  

\begin{figure*}\begin{center}
\epsscale{0.8}
\plotone{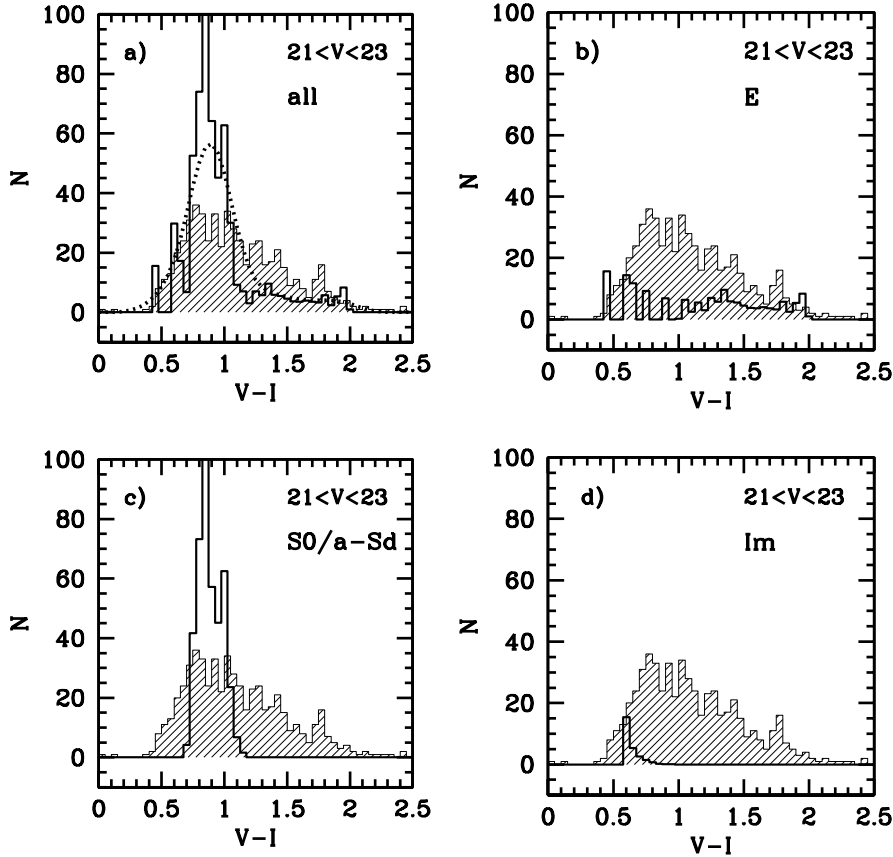}
\caption{The $V-I$ color distributions for the data (\emph{hashed histogram}), overlaid on four representations of the color distributions for the model with $\zf =1.5$ for spirals (\emph{Thick-lined empty histogram}).  The model is normalized to the number of galaxies in each V magnitude bin. Only galaxies with $21< V < 23$ are shown.   
\emph{Panel (a)}: color distribution for the entire $21<V<23$ model population.  The effect of  photometric errors has been modeled by convolving the model histogram with a Gaussian kernel of $\sigma_{VI} = 0.13$ (\emph{dotted line}).  
\emph{Panel (b)}: Color distribution for E/S0. 
\emph{Panel (c)}: Color distribution for S0/a-Sd. 
\emph{Panel (d)}: Color distribution for Im.}
\label{Fig:colorsbytype}
\end{center}
\end{figure*}

\begin{figure*}\begin{center}
\epsscale{0.8}
\plotone{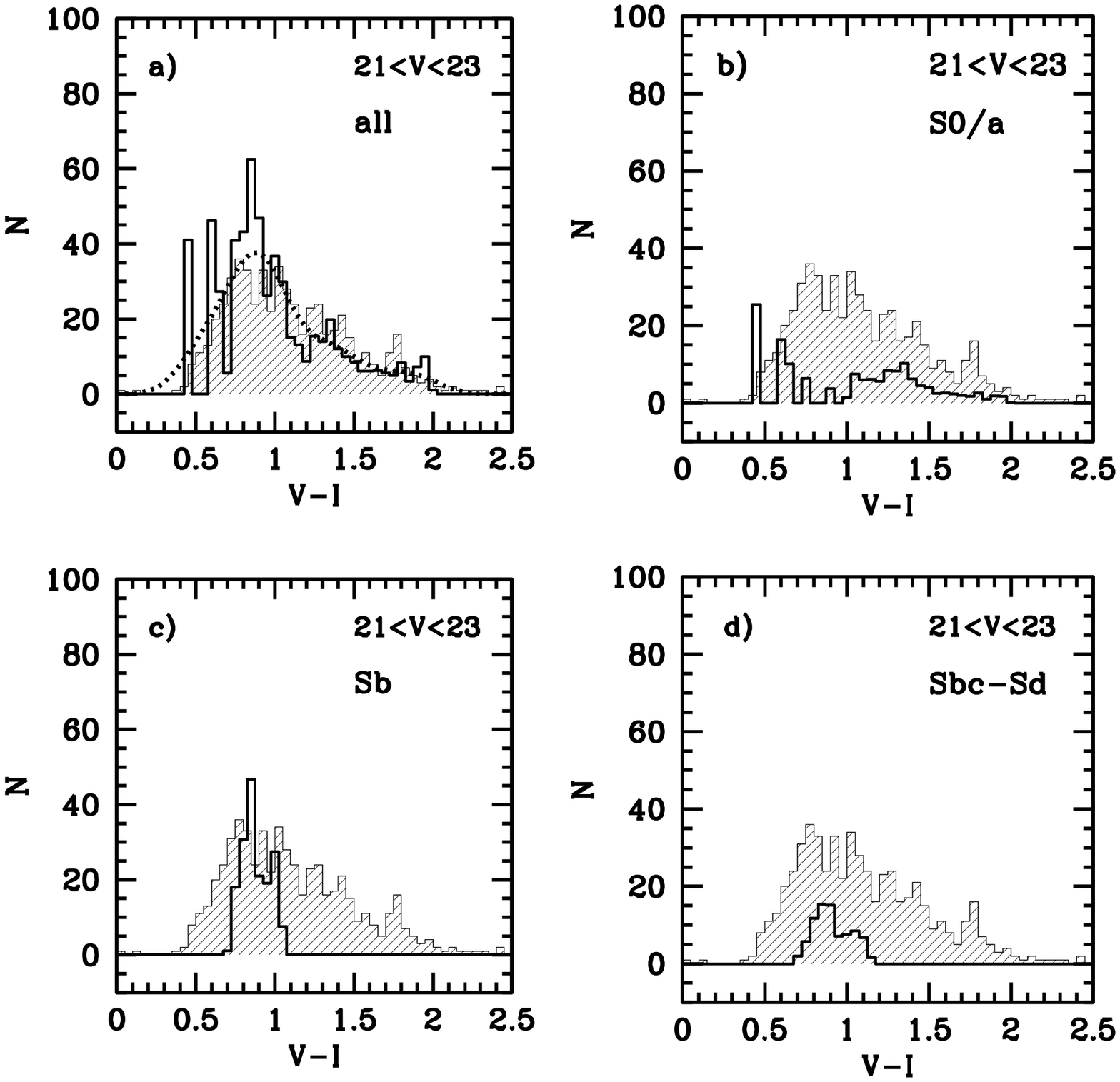}
\caption{The $V-I$ color distributions for the data (\emph{hashed histogram}), overlaid on four representations of the color distributions for the toy model in which the early-type spiral class has been split in two halves, corresponding to ad-hoc S0/a-Sa and Sb populations, respectively.   The model is normalized to the number of galaxies in each V magnitude bin. Only galaxies with $21< V < 23$ are shown. 
\emph{Thin-lined filled histogram}: Color distributions of our data. \emph{Thick-lined empty histogram}: Color distributions predicted by the model. 
\emph{Panel (a)}: color distribution for the entire $21<V<23$ model population.  The effect of  photometric errors has been modeled by convolving the model histogram with a Gaussian kernel of $\sigma_{VI} = 0.13$ (\emph{dotted line}).  
\emph{Panel (b)}: Color distributions for the ad-hoc S0/a-Sa population. 
\emph{Panel (c)}: Color distributions for the ad-hoc Sb population. 
\emph{Panel (d)}: Color distributions for Sbc-Sd. 
For this model, the color distribution for E-S0 galaxies is identical to that shown in 
Fig.~\ref{Fig:colorsbytype}b.  }
\label{Fig:colorsSsplit}
\end{center}
\end{figure*}

\subsubsection{Color distributions}
\label{Sec:colors}

As pointed out by \citet{Gardner98}, color distributions provide checks on the evolutionary processes of galaxies that cannot be obtained from number counts.  We focus on the $V-I$ color,  as
our observed catalog is most complete in those bands.
The analysis starts with our model with $\zf = 1.5$ for spirals, and focusses on galaxies with $21<V<23$; similar conclusions are derived from other magnitude ranges.   $V-I$ color distributions for the model and the data are compared in Figure~\ref{Fig:colorsbytype}a.  Clearly, the model distribution is narrower than seen in the data, and it shows a deficit of red galaxies.  Photometric errors cannot account for the discrepancy, as shown by the dotted distribution in Figure~\ref{Fig:colorsbytype}a.  \citet{Gardner98} noted that the color predictions from the models are often too narrow owing to the discrete nature of the modeled populations. 
We find that this is the case for spirals (see Fig.~\ref{Fig:colorsbytype}c, where we have plotted early- and late-type spirals together) and for irregulars (Fig.~\ref{Fig:colorsbytype}d).  However, the model distribution for E-S0 is broad (Fig.~\ref{Fig:colorsbytype}b).  Such width arises from the strong K-corrections of old populations, which appears to be responsible for the width in the observed $V-I$ histogram.  

Figure~\ref{Fig:colorsbytype} suggests that the discrepancy between observed and model colors arises from the failure to include red, evolved populations in the precursors of present-day early-to-intermediate type spirals.  
Possibly, our early-spiral class (S0/a - Sb) is too broad to account
for the range of colors present in galaxies of types S0/a to Sb. The models appear to miss the strong contribution of bulges to the integrated light of early-type spirals.  
We are constrained by the morphological classes of the luminosity function defined by Nakamura et al., hence we cannot properly split the early S class.
To estimate the potential effects of splitting the early S class, we generate a toy-model in which we simply divide the early spirals
in two equal groups.  For the first ("S0/a-Sa" group) we assign the formation history
corresponding to ellipticals (SSP Solar metallicity, Scalo IMF, and
formation redshift \zf\ = 1.5).  For the second ("Sb group"), we assign a formation redshift  $\zf = 4.0$, an exponentially-decaying SFH with $\tau = 7$ Gyr, and Solar metallicity.  The color distribution for such model (Figure~\ref{Fig:colorsSsplit}a), provides a close match to the observed colors, specially when smoothed to account for photometric errors (dotted line in Figure~\ref{Fig:colorsSsplit}a).  This exercise suggests that the deficit of red galaxies in the model shown in 
Figure~\ref{Fig:colorsbytype}a traces an overly simplified description of the
populations in early-type spirals.  The single-age nature of the population  might affect as well; a broad color distribution for the S group would be obtained as well if an important fraction of the precursors of present-day spirals became dominated by old populations earlier than assumed in our models.  
Finally, we note that splitting the early-S population does not affect significantly to the fits to the number counts.

\subsubsection{Cosmological parameters}
\label{Sec:cosmologicalparameters}

When the cosmology is changed to an Einstein-de Sitter model, our \ncmod\ model reproduces the $U$ and $B$ counts, but fails to reproduce the \Ks\ counts by a large margin.  Hence, the combined number counts in $UBK$ appear to favor a $\Lambda$-cosmology over an Einstein-de Sitter one, a result also found by  \citet{Totani00}.

\section{Summary}
\label{Sec:summary}

We have presented $U$ and $B$ number counts from a field of 
the GOYA Survey that covers 
$\sim$900 arcmin$^2$ over the Groth-Westphal Strip. Counts are derived from  0.2350 deg$^2$ and 0.2467 deg$^2$ of sky in $U$ and $B$, respectively. Achieved limiting magnitudes (50\% detection efficiency for point sources) are $U = 24.8$ mag and $B = 25.5$ mag, in
the Vega system. The counts have been corrected for 
detection efficiency as a function of source size, and for spurious 
detections, using the method of $S/N$ threshold in two complementary half-time images detailed in \citet{Eliche05}. 
Star-galaxy separation has been performed using \emph{HST} images in the GWS and stellarity indexes from {\tt SExtractor\/}. 

Counts are given over $18.0<U<25.0$ and $19.5<B<25.5$.  
These wide ranges (7 mag in $U$ and  6 mag in $B$) result from the combination of wide area and depth of our survey. In both bands, our number counts are in good agreement with other studies that cover fainter and brighter magnitudes.  The slopes of the number count distributions are very 
similar to those reported by other authors: 
$d\log (N)/dm= 0.50 \pm 0.02$ for $B$=21.0-24.5, and 
$d\log (N)/dm= 0.48 \pm 0.03$ for $U$=21.0-24.0.

When combined with \Ks\ number counts, the data provide strong constraints on galaxy formation models, due to the presence of a knee at $\Ks = 17.5$ in the NIR counts, and the absence of such feature in blue passbands.  
Adopting a $\Lambda$-dominated cosmological model ($\Omega_M = 0.3$, 
$\Omega_\Lambda = 0.7$, $H_0 = 70$ km s$^{-1}$ Mpc$^{-1}$), 
a simple number count model including luminosity evolution from GISSEL SEDs and number evolution as $(1+z)^{2.7}$ accurately reproduces the observed counts in $U$, $B$, and \Ks\ in a consistent way.  
Extensive modeling suggests that, only by assuming a moderately low formation redshift ($\zf \approx 1.5$) for the dominant NIR population (ellipticals) does the model reproduce the $\Ks = 17.5$ knee;  while, reproducing the lack of a knee in $U$ and $B$ counts in turn requires the adoption of  a moderate optical depth for all galaxy types, including ellipticals ($\tau_B = 0.6$ for $L^\star$ galaxies).  
Neither of the two assumptions is at odds with current ideas on galaxy formation and evolution in hierarchical Universes.  
Future comparison with SAM galaxy formation models will tell whether the $\zf = 1.5$ required by the model reflects, as we suspect, a major epoch of early-type \emph{field} galaxy formation through mergers of disk galaxies, rather than the epoch for the formation of the stellar content of ellipticals.


\acknowledgements
We thank the anonymous referee for suggestions and comments that helped to improve the manuscript. 
This research has made use of the Digitized Sky Survey, 
the Medium Deep Survey online database, the DEEP project archive, 
and the Guide Star Catalogue-II.
The Digitized Sky Survey was produced at the Space Telescope Science 
Institute under U.S. Government grant NAG W-2166. 
The Medium Deep Survey catalog is based on 
observations with the NASA/ESA Hubble Space Telescope, obtained at the 
Space Telescope Science Institute, which is operated by the Association 
of Universities for Research in Astronomy, Inc., under NASA contract 
NAS5-26555. The Medium-Deep Survey analysis was funded by the HST WFPC2 
Team and STScI grants GO2684, GO6951, GO7536, and GO8384 to Prof.\ Richard 
Griffiths and Dr.\ Kavan Ratnatunga at Carnegie Mellon University.
The DEEP database is maintained with support of the National 
Science Foundation grants AST 95-29028 and AST 00-71198.
The Guide Star Catalogue-II is a joint project of the Space Telescope Science 
Institute and the Osservatorio Astronomico di Torino. 

\begin{appendix}
\section{Efficiency error estimation by functional and matricial methods}
\label{Append:errors}

Our error treatment is analogous to that described in the Appendix of CH03. 
Number count errors arise from detection efficiency errors 
and statistical counting, once the spurious have been removed from the catalogues. The latter errors can be approximated by \citep{Gehrels86}:
\begin{eqnarray}   \label{eq:statistical}
\sigma _{\mathrm{u},s}^{\mathrm{N,stat}}(\mout)&=&\sqrt{N_{\mathrm{det},s} (\mout)+\frac{3}{4}}\\ 
\sigma _{\mathrm{l},s}^{\mathrm{N,stat}}(\mout)&=& N_{\mathrm{det},s} (\mout)- N_{\mathrm{det},s} (\mout)\left[ 1-\frac{1}{9N_{\mathrm{det},s}(\mout)}-\frac{1}
{3\sqrt{ N_{\mathrm{det},s} (\mout)}}\right] ^{3},    \nonumber  
\end{eqnarray}   
\noindent with $s$ being the index that refers to the size group (see 
\S\ref{Sec:synthetic}), $ N_{\mathrm{det},s} (\mout)$ the raw number counts in  
\mout\ for the size group $s$ after spurious subtraction,
$\sigma _{\mathrm{u},s}^{\mathrm{N,stat}}$ the upper error, and 
$\sigma _{\mathrm{l},s}^{\mathrm{N,stat}}$ the lower error. 
Depending on the method used for completeness correction, 
errors have been estimated as follows:

\begin{enumerate}   
\item \emph{Using the method of functional efficiency, \Emout}. 
Considering the counting errors from equations (\ref{eq:statistical}) and errors arising from the efficiency corrections, 1$\sigma$ error 
propagation of equation (\ref{eq:efficiency}) for the size group $s$ 
leads to \citep[see][]{Bevington69}

\begin{equation}   
\sigma _{s}^{\mathrm{N}}(\mout)=\frac{1}{0.5\cdot A}\sqrt{\left[ 
\frac{\sigma _s^{\mathrm{N,stat}}(\mout)}{E(\mout)}\right] ^{2}+\left[ \sigma ^\mathrm{E}(\mout)\cdot\frac{N_{\mathrm{det},s}(\mout)}{E^{2}(\mout)}\right] ^{2}},   
\end{equation}   
\noindent where $\sigma _{s}^{\mathrm{N,stat}}(\mout)$ is the statistical error associated to counting (upper or lower), $E(\mout)$ is the efficiency at that magnitude bin for the size group $s$, $A$ is the area of counting, and 
$\sigma ^\mathrm{E}(\mout)$ is the efficiency error. The factor 0.5 accounts for the magnitude interval used for counting (0.5 mag in our case). 

Using upper and lower statistical errors from equations 
(\ref{eq:statistical}), we can obtain the corresponding total upper and lower errors 
$\sigma _{s,\mathrm{u}}^{\mathrm{N}}(\mout)$ and  $\sigma_{s,l}^{\mathrm{N}}(\mout)$. Then the propagated errors 
from the addition of the number counts of the three size groups are

\begin{equation}   \label{eq:error}
\sigma _{\mathrm{l,total}}(\mout)=\sqrt{\sum_{s=1}^{3}\left[ \sigma _{\mathrm{l},s}^{\mathrm{N}}(\mout)\right] 
^{2}}\; ,\; \sigma _{\mathrm{u,total}}(\mout)=\sqrt{ \sum_{s=1}^{3}\left[ \sigma _{\mathrm{u},s}^{\mathrm{N}}(\mout) \right] ^{2}}.   
\end{equation}   
  
Finally, this error is quadratically added to that associated to 
star counting (see the definition of this last in \S\ref{Sec:stargalaxy}):
 
\begin{eqnarray}   \label{eq:errorfinal}
\sigma _{\mathrm{l,final}}(\mout) &=&\sqrt{\left[ \sigma _{\mathrm{l,total}}(\mout)\right]  
^{2}+\left[ \sigma _{\mathrm{l,stars}}(\mout)\right] ^{2}} \\  
\sigma _{\mathrm{u,final}}(\mout) &=&\sqrt{\left[ \sigma _{\mathrm{u,total}}(\mout)\right]  
^{2}+\left[ \sigma _{\mathrm{u,stars}}(\mout)\right] ^{2}}.  \nonumber  
\end{eqnarray}  

\item \emph{For the method of the efficiency matrices, \Pminmout}. We have 
considered counting and efficiency errors separately:

\begin{itemize}   
\item  We have estimated \emph{statistical counting} errors by adding lower and upper statistical errors from equation (\ref{eq:statistical}) to the raw counts in equation (\ref{eq:inversion}) in each size group:
 
\begin{eqnarray}  
N_{\mathrm{l},s}(\min)^{\prime} = \frac{1}{0.5\cdot A}\cdot \left\{ \sum_{\forall \mout } 
\left[N_{\mathrm{det},s}(\mout)+\sigma_{\mathrm{u},s}^{\mathrm{N,stat}}(\mout)\right] \cdot {P}^{-1}_{s} (\mout,\min) \right\}  \\   
N_{\mathrm{u},s}(\min)^{\prime } = \frac{1}{0.5\cdot A}\cdot \left\{   
\sum_{\forall \mout }\left[ N_{\mathrm{det},s}(\mout)+\sigma _{\mathrm{l},s}^{\mathrm{N,stat}}(\mout) \right] \cdot   
{P}^{-1}_{s} (\mout,\min) \right\},     
\nonumber   
\end{eqnarray}   
\noindent where $N_{\mathrm{l},s}(\min)^{\prime}$ and $N_{\mathrm{u},s}(\min)^{\prime }$ are 
the lower and upper values to the corrected number counts in equation (\ref{eq:inversion}) respectively; $N_{\mathrm{det},s}(\mout)$ is the raw number counts in \mout\ after spurious subtraction; and 
${P}^{-1}_{s} (\mout,\min)$ is the element $ (\mout,\min) $ of the inverse 
of the efficiency matrix that corresponds to the size group $s$, 
\Psminmout. Notice that the addition of the upper limit to the raw counts would 
produce the lower limit in the efficiency-corrected counts, because of the 
inversion of the matrix, and vice-versa. Therefore, upper and lower 
errors are defined as the difference of the 
resulting values with respect to those obtained without adding errors:
  
  \begin{equation}\label{eq:error1} 
  \begin{array}{clc}
  \sigma^{\mathrm{N}} _{\mathrm{l},s}(\min)^{\prime }&=&
 N_{\mathrm{orig},s}(\min)^{\prime}-N_{\mathrm{l},s}(\min)^{\prime}  \\
   \sigma^{\mathrm{N}} _{\mathrm{u},s}(\min)^{\prime }&=& N_{\mathrm{u},s}(\min)^{\prime }- N_{\mathrm{orig},s}(\min)^{\prime},
\end{array}  
    \end{equation}
   
   \noindent where $N_{\mathrm{orig},s}(\min)^{\prime}$ is obtained from equation  (\ref{eq:inversion}).
    
\item  The errors from \emph{efficiency correction} are estimated 
in a similar way: upper and lower 
limits to the efficiency-corrected counts are computed by adding or 
subtracting the RMS of the efficiency matrices for the size 
group $s$ in equation (\ref{eq:inversion}). Then, corresponding upper and lower errors due to efficiency determination errors are the difference of these  
resulting values with respect to those obtained in equation (\ref{eq:inversion}), as follows:
    
\begin{equation} \label{eq:error2} 
\begin{array}{clc}
\sigma _{\mathrm{l},s}^{\mathrm{Ef}}(\min)^{\prime } &=&
N_{\mathrm{orig},s}(\min)^{\prime}-\frac{1}{0.5\cdot A}\cdot \left\{   

\sum_{\forall \mout}N_{\mathrm{det},s}(\mout)
\cdot \left[ (\widehat{P_{s}}+\widehat{\sigma _{\mathrm{P},s}})^{-1}  (\mout,\min) \right] \right\}  \\   
\sigma _{\mathrm{u},s}^{\mathrm{Ef}}(\min)^{\prime } &=&\frac{1}{0.5\cdot A}\cdot \left\{   
\sum_{\forall \mout}N_{\mathrm{det},s}(\mout)\cdot \left[ (\widehat{P_{s}}-\widehat{\sigma _{\mathrm{P},s}})^{-1}  (\mout,\min) \right] \right\}  -N_{\mathrm{orig},s}(\min)^{\prime },
  \end{array}  
    \end{equation}
    \noindent where $\widehat{\sigma _{\mathrm{P},s}}$ is the RMS matrix of the efficiency matrix for the size group $s$ (see Figure \ref{Fig:matrices}). 
\end{itemize}   
 
Thus, the errors from equations (\ref{eq:error1}) and 
(\ref{eq:error2}) are added quadratically in each case to obtain 
the upper and lower values:
 
\begin{equation}   
\begin{array}{clc}
\sigma _{\mathrm{u},s}(\min)&=&\sqrt{\left[ \sigma _{\mathrm{u},s}^{\mathrm{Ef}}(\min)^{\prime }\right]   
^{2}+\left[ \sigma _{\mathrm{u},s}^{\mathrm{N}}(\min)^{\prime }\right] ^{2}} \\

\sigma _{\mathrm{l},s}(\min)&=&\sqrt{\left[ \sigma _{\mathrm{l},s}^{\mathrm{Ef}}(\min)^{\prime }\right]   
^{2}+\left[ \sigma _{\mathrm{l},s}^{\mathrm{N}}(\min)^{\prime }\right] ^{2}}.   
 \end{array}  
    \end{equation}
The errors associated to star counts are estimated in an 
analogous way. The quadratic addition of values from this last expression 
for the three size groups and the 
errors associated to star count subtraction would estimate 
the final total error, as in equations (\ref{eq:error}) and 
(\ref{eq:errorfinal}).

\end{enumerate}

\section{INT/WFC offsets}
\label{Append:offsets}

In \S\ref{Sec:astrometry}, we report the difficulties we found when we 
tried to fit an independent astrometric solution for each CCD of the INT/WFC, using the TNX projection from IRAF astrometric tasks. 
The instrument exhibits an important "pincushion" distortion introduced by the 
telescope optics (see an instrument description 
in \S\ref{Sec:observations}). 
We concluded that a single fit to 
the entire field of view was needed in order to produce a continuous 
solution in the CCD edges. 

Our procedure for creating single images 
of each exposure is based on the method described 
by \citet{Taylor00}. He did an astrometric solution of the 
INT/WFC, correct to an accuracy of 1 or 2 pixels ($\sim 0.5\arcsec$). 
The corrected coordinates 
($x\prime$, $y\prime$) were obtained from each set of pixel coordinates 
($x_i$, $y_i$) by first translating so that the origin is on the optical 
axis, then rotating to the correct angle, and finally correcting for the radial 
distortion effect, as follows:

\begin{equation}  \label{eq:astrometry}
\begin{array}{lcl}
x^{\mathrm{shift}}_i&=&x_i-X_i\\
y^{\mathrm{shift}}_i&=&y_i-Y_i\\
x^{\mathrm{rot}}_i&=&x^{\mathrm{shift}}_i\,\cos\theta_i-y^{\mathrm{shift}}_i\,\sin\theta_i\\
y^{\mathrm{rot}}_i&=&x^{\mathrm{shift}}_i\,\sin\theta_i+y^{\mathrm{shift}}_i\,\cos\theta_i\\
x\prime&=&x^{\mathrm{rot}}_i\,\{1+D\,[\, (x^{\mathrm{rot}} _i)^2 +(y^{\mathrm{rot}}_i)^2 \,]\}\\
y\prime&=&y^{\mathrm{rot}}_i\,\{1+D\,[\, (x^{\mathrm{rot}} _i)^2 +(y^{\mathrm{rot}}_i)^2 \,]\},\\
\end{array}
\end{equation}
\noindent where ($X_i$, $Y_i$) are the coordinates of the optical centre of the 
instrument in the pixel coordinate system of CCD\#i, $\theta_i$ is the 
angle at which CCD\#i sits on the focal plane, and $D$ is the pincushion 
distortion coefficient. The values of the coefficients in these 
equations reported by \citet{Taylor00} are listed in Col.\,(2)-(4) of 
Table \ref{Tab:coefficients}. The optical axis is the final origin of 
the reference frame.

Single images of each exposure were created by positioning each CCD in an 
empty frame wide enough to include the four chips, and using the 
coefficients reported by \citet{Taylor00}. The optical axis was 
situated in coordinates (2223.0,72.0) in the global frame, and coordinates ($X_i$, $Y_i$) were 
transformed to this new origin, taking into account the angles 
between the CCDs. Taylor's transformed coefficients are tabulated in 
Col.\,(5)-(6) of Table \ref{Tab:coefficients}. After that, we fitted a 
third-order astrometric solution with the TNX projection, in order to 
describe remaining, non-corrected linear distortions and 
non-linear optical aberrations. Discontinuities in the 
astrometric solutions indicated that errors of $\sim$1\arcsec are 
present in the chip separations reported by \citet{Taylor00}.

In order to improve the astrometric solution, we 
estimated corrections to Taylor's positions by offseting the relative CCD positions until we found a 
combination which minimized the astrometric discontinuities at the chip-chip interface. 
The corrected values and offsets are listed at Table \ref{Tab:coefficients}. 

Discontinuities in the astrometric solutions were removed, and 
the RMS astrometric error after a new third-order fit to the astrometry has been reduced to 0.3\arcsec over the $\sim$36\arcmin$\times$36\arcmin\ field of 
the camera.

\begin{deluxetable}{crrrcrrcrrcrr}   
\tabletypesize{\scriptsize}
\tableheadfrac{0.01}
\tablecaption{Relative Positions and Rotations of INT/WFC CCDs
\label{Tab:coefficients}}
\tablehead{\multicolumn{1}{c}{CCD ID}& \multicolumn{3}{c}{Original coeff.} &
\multicolumn{1}{c}{}& \multicolumn{2}{c}{Transformed coeff.}&
\multicolumn{1}{c}{}& \multicolumn{2}{c}{Corrected coeff.}&
\multicolumn{1}{c}{}& \multicolumn{2}{c}{Offsets}\\
\cline{2-4}\cline{6-7}\cline{9-10}\cline{12-13}\\[-0.2cm]
\multicolumn{1}{c}{} &\multicolumn{1}{c}{$\theta_i$} & 
\multicolumn{1}{c}{$X_i$}& \multicolumn{1}{c}{$Y_i$} & 
\multicolumn{1}{c}{} &
\multicolumn{1}{c}{$X_i$}& \multicolumn{1}{c}{$Y_i$} & 
\multicolumn{1}{c}{} &
\multicolumn{1}{c}{$X_i$}& \multicolumn{1}{c}{$Y_i$}&  
\multicolumn{1}{c}{}& \multicolumn{1}{c}{$\Delta X_i$}& 
\multicolumn{1}{c}{$\Delta Y_i$}\\
\colhead{(1)}&\colhead{(2)}&\colhead{(3)}&\colhead{(4)}&
\colhead{}&\colhead{(5)}&\colhead{(6)}&\colhead{}&
\colhead{(7)}&\colhead{(8)}&\colhead{}&\colhead{(9)}&\colhead{(10)}}
\startdata

\#1&  0.01868 &-336.74 & 3039.14&&4338.73 &  61.97 && 4339.73 & 57.47  &&1.0&-4.5\\

\#2&-90.62115 &3180.68 & 1729.67&&2305.90 &6298.23 && 2308.90 & 6296.73&&3.0&-1.5\\
\#3&  0.11436 &3876.73 & 2996.30&& 130.26 &  96.97 &&  134.76 &99.47   &&4.5&2.5\\

\#4&  0.00000 &1778.00 & 3029.00&&2223.00 &  72.00 && 2223.00&  72.00  &&0.0&0.0\\[-0.2cm]
\enddata
\tablecomments{Col.\,(2)-(4) are original values for the rotation angles in 
degrees and positions  of the optical axis in the frame of each CCD, reported by \citet{Taylor00}. Col.\,(5)-(6) are these positions in a global reference frame where the optical axis is situated at coordinates (2223.0,72.0). 
Col.\,(7)-(8) are our corrected coefficients in the same global frame. 
Col.\,(9)-(10) show the computed offsets. All positions are given in pixels.}
\end{deluxetable}

  \end{appendix}

\clearpage
\begin{landscape}
\begin{deluxetable}
{lcccccc
  cccccc
  cccccc
  cccccc}   
\tabletypesize{\tiny}
\tableheadfrac{0.01}
\tablewidth{0pt}
\tablecaption{Estimated Fractions of Real Sources Rejected as "False Detections", Truly Spurious Sources Rejected as "False Detections", and Truly Spurious Non-Rejected from the $U$ Catalogue, for Different DETECT\_THRESH-$(S/N)_\mathrm{lim}$ Combinations\label{Tab:fractions}}
\tablehead{\multicolumn{1}{c}{{\tt DETECT\_THRESH\/}} &\multicolumn{1}{c}{}& 
\multicolumn{5}{c}{0.4} &\multicolumn{1}{c}{}&
\multicolumn{5}{c}{0.5} &\multicolumn{1}{c}{}& \multicolumn{5}{c}{0.6} &
\multicolumn{1}{c}{}& \multicolumn{5}{c}{0.7}\\[-0.2cm]\\
\cline{3-7}\cline{9-13}\cline{15-19}\cline{21-25}\\[-0.1cm]
\multicolumn{1}{c}{$(S/N)_\mathrm{lim}$} &\multicolumn{1}{c}{}& 
\multicolumn{1}{c}{1.8} & 
\multicolumn{1}{c}{2.2} & \multicolumn{1}{c}{2.5} & \multicolumn{1}{c}{7} & \multicolumn{1}{c}{10} &\multicolumn{1}{c}{}& 
\multicolumn{1}{c}{1.8} & 
\multicolumn{1}{c}{2.2} & \multicolumn{1}{c}{2.5} &\multicolumn{1}{c}{7} & \multicolumn{1}{c}{10} &\multicolumn{1}{c}{}& 
\multicolumn{1}{c}{1.8} & 
\multicolumn{1}{c}{2.2} & \multicolumn{1}{c}{2.5} &\multicolumn{1}{c}{7} & \multicolumn{1}{c}{10} &\multicolumn{1}{c}{}& 
\multicolumn{1}{c}{1.8} & 
\multicolumn{1}{c}{2.2} & \multicolumn{1}{c}{2.5}& \multicolumn{1}{c}{7} & \multicolumn{1}{c}{10} }
\startdata
\multicolumn{1}{l}{\% Rej.\,Real }  &
\multicolumn{1}{c}{}& $\sim$5 & $\sim$11 &  $\sim$17 &$\sim$48  & $\sim$54  & 
\multicolumn{1}{c}{}& $\sim$3 & $\sim$6 & $\sim$9   &$\sim$45  & $\sim$53  & 
\multicolumn{1}{c}{}& $\sim$1 & $\sim$3 & $\sim$5    &$\sim$43  & $\sim$52  & 
\multicolumn{1}{c}{}& $\sim$0.5 &$\sim$1 & $\sim$3     &$\sim$45  & $\sim$55  \\    
 \multicolumn{23}{c}{}\\   
\begin{tabular}{l}   
\multicolumn{1}{l}{\% Rej.\,Truly Spurious}  
\end{tabular}   
 &\multicolumn{1}{c}{}& $\sim$26 & $\sim$29 & $\sim$29 & $\sim$29  & $\sim$29  & 
\multicolumn{1}{c}{} &$\sim$21 & $\sim$25 &$\sim$28    & $\sim$29  & $\sim$29  & 
\multicolumn{1}{c}{}& $\sim$17 & $\sim$22 &  $\sim$25 &  $\sim$29  & $\sim$29  & 
\multicolumn{1}{c}{}&$\sim$13 & $\sim$18 & $\sim$21   &  $\sim$25  & $\sim$25 \\   
 \multicolumn{23}{c}{}\\  
\begin{tabular}{l}   
\multicolumn{1}{l}{\% Non-Rej.\,Truly Spurious}
\end{tabular}   &
\multicolumn{1}{c}{}&$\sim$3 & $\sim$0 & $\sim$0 & $\sim$0  & $\sim$0  & 
\multicolumn{1}{c}{}&$\sim$8 & $\sim$4 & $\sim$1 & $\sim$0  & $\sim$0  & 
\multicolumn{1}{c}{}&$\sim$12 &$\sim$7 &$\sim$4  & $\sim$0  & $\sim$0  & 
\multicolumn{1}{c}{}&$\sim$12 &$\sim$7 & $\sim$4 & $\sim$0  & $\sim$0 \\[-0.2cm]
\enddata
\tablecomments{All percentages are relative to the total number of 
detections in the $U$ catalogue. The fraction of non-rejected truly spurious sources (last row) has been computed subtracting the number of rejected truly spurious detections (penultimate row) from the estimated maximum number of truly spurious sources that are in the catalogue, which is $\sim 29$\% of the total number of detections for {\tt DETECT\_THRESH\/}=0.4, 0.5, and 0.6, and $\sim 25$\% for 
{\tt DETECT\_THRESH\/}=0.7 (see the text).}
\end{deluxetable}

\begin{deluxetable}{lrrrrrrrrrrrrr}
\tabletypesize{\scriptsize}
\tableheadfrac{0.01}
\tablewidth{0pt}
\tablecaption{$U$ Differential Number Counts in the GWS Field\label{Tab:countsu}}
\tablehead{\colhead{$U$}  & \colhead{$N^\mathrm{Raw}_1$} & \colhead{$N^\mathrm{Raw}_2$} & \colhead{$N^\mathrm{Raw}_3$} & \colhead{$N^\mathrm{Spur}_1$} &\colhead{$N^\mathrm{Spur}_2$} & \colhead{$N^\mathrm{Spur}_3$} & \colhead{$E_1(U)$} & \colhead{$E_2(U)$} & \colhead{$E_3(U)$} & \colhead{$N$} &  \colhead{$\sigma _{\mathrm{u}}$} & 
\colhead{$\sigma _{\mathrm{l}}$}& \colhead{log($N$)}\\
\colhead{(mag)} &  \colhead{($N$)} & \colhead{($N$)}& \colhead{($N$)}&
\colhead{($N$)}& \colhead{($N$)}& \colhead{($N$)}&
 & & & \colhead{($N$ mag$^{-1}$ deg$^{-2}$)}& \colhead{($N$ mag$^{-1}$ deg$^{-2}$)} & \colhead{($N$ mag$^{-1}$ deg$^{-2}$)} & \\
\colhead{(1)} &\colhead{(2)} &\colhead{(3)} &\colhead{(4)} &\colhead{(5)} &
\colhead{(6)} &\colhead{(7)} &\colhead{(8)} &\colhead{(9)} &\colhead{(10)} &\colhead{(11)}&\colhead{(12)}&\colhead{(13)}&\colhead{(14)}
}     
\startdata 
 18.25 ..............   &    25    &   0   &   0  &    25 &	0 &	0 &    0.999 &    0.997 &    0.995 &	   68.38 &	180.91 &       59.84 &  1.835 \\
 18.75 ..............   &    40    &   0   &   0  &    40 &	0 &	0 &    0.997 &    0.995 &    0.989 &	  103.23 &	211.11 &       90.32 &  2.014 \\
 19.25 ..............   &    46    &   0   &   0  &    46 &	0 &	0 &    0.998 &    0.993 &    0.987 &	  145.47 &	151.87 &       86.98 &  2.163 \\
 19.75 ..............   &    80    &   0   &   0  &    80 &	0 &	0 &    0.994 &    0.986 &    0.978 &	  369.82 &	240.85 &      181.83 &  2.568 \\
 20.25 ..............   &    99    &   0   &   0  &    99 &	0 &	0 &    0.986 &    0.982 &    0.971 &	  700.43 &	179.72 &      120.06 &  2.845 \\
 20.75 ..............   &    167   &   1   &   0  &   167 &	1 &	0 &    0.980 &    0.973 &    0.961 &	  891.53 &	260.83 &      203.41 &  2.950 \\
 21.25 ..............   &    278   &   5   &   0  &   278 &	5 &	0 &    0.974 &    0.964 &    0.944 &	 2057.71 &	262.29 &      208.84 &  3.313 \\
 21.75 ..............   &    505   &   30  &   0  &   505 &    30 &	0 &    0.963 &    0.951 &    0.921 &	 4048.07 &	341.39 &      291.94 &  3.607 \\
 22.25 ..............   &    983   &   90  &   0  &   983 &    89 &	0 &    0.955 &    0.933 &    0.897 &	 8830.89 &	440.50 &      399.18 &  3.946 \\
 22.75 ..............   &    1589  &   276 &   0  &  1589 &   276 &	0 &    0.942 &    0.918 &    0.872 &	15517.46 &	627.92 &      590.18 &  4.191 \\
 23.25 ..............   &    2313  &   724 &   1  &  2313 &   723 &	1 &    0.934 &    0.890 &    0.817 &	26166.13 &	897.11 &      866.13 &  4.418 \\
 23.75 ..............   &    3082  &   1309&   1  &  3075 &  1303 &	1 &    0.908 &    0.849 &    0.690 &	39491.81 &     1385.45 &     1361.52 &  4.597 \\
 24.25 ..............   &    3418  &   1972&   3  &  3409 &  1937 &	3 &    0.876 &    0.707 &    0.339 &	53784.13 &     2176.80 &     2158.07 &  4.731 \\
 24.75 ..............   &    3347  &   2380&   4  &  3346 &  2088 &	3 &    0.760 &    0.360 &    0.166 &	84023.84 &     6954.28 &     6943.38 &  4.924 \\
 25.25 ..............   &    2857  &   2338&   24 &  2855 &  1200 &	9 &    0.368 &    0.153 &    0.107 &   129374.22 &    17466.33 &    17449.77 &  5.112 \\[-0.2cm]

\enddata
\tablecomments{Col.\,(2), (3), and (4) show source numbers (raw counts) for size groups $r_{\mathrm{e}}\le 1.5$\arcsec, 1.5\arcsec$\le r_{\mathrm{e}}\le 3$\arcsec, and $r_{\mathrm{e}}\ge 3$\arcsec, respectively, in intervals of 0.5 mag centered on the magnitudes shown in Col.\,(1). Col.\,(5), (6), and (7) show raw counts corrected for spurious detections. Col.\,(8), (9), and (10) indicate the applied functional efficiency correction for the spurious-corrected number counts in  each one of the previous size groups. Col.\,(11) presents differential galaxy  number counts per unit magnitude and area (spurious-corrected,  efficiency-corrected, and with stars subtracted). Col.\,(12) and (13) show the  1-$\sigma$ confidence upper and lower limits on corrected differential counts of Col.\,(11). Col.\,(14) lists the logarithm of Col.\,(11).}
\end{deluxetable}
\clearpage
\end{landscape}

 \begin{landscape}
\begin{deluxetable}{lrrrrrrrrrrrrr}
\tabletypesize{\scriptsize}
\tableheadfrac{0.01}
\tablecaption{$B$ Differential Number Counts in the GWS Field 
\label{Tab:countsb}}
\tablewidth{0pt}

\tablehead{\colhead{$B$}  & \colhead{Raw1} & \colhead{Raw2} & 
\colhead{Raw3} & \colhead{$N^1_\mathrm{spur}$} &\colhead{$N^2_\mathrm{spur}$} &\colhead{$N^3_\mathrm{spur}$} & \colhead{$E_1(U)$} & \colhead{$E_2(U)$} & \colhead{$E_3(U)$} &\colhead{N} &  \colhead{$\sigma _{\mathrm{u}}$} & 
\colhead{$\sigma _{\mathrm{l}}$}& \colhead{log(N)}\\
\colhead{(mag)} &  \colhead{($N$)} & \colhead{($N$)}& \colhead{($N$)}&
\colhead{($N$)}& \colhead{($N$)}& \colhead{($N$)}&
 & & & \colhead{($N$ mag$^{-1}$ deg$^{-2}$)}& \colhead{($N$ mag$^{-1}$ deg$^{-2}$)} & \colhead{($N$ mag$^{-1}$ deg$^{-2}$)} & \\
\colhead{(1)} &\colhead{(2)} &\colhead{(3)} &\colhead{(4)} &\colhead{(5)} &
\colhead{(6)} &\colhead{(7)} &\colhead{(8)} &\colhead{(9)} &\colhead{(10)} &\colhead{(11)}&\colhead{(12)}&\colhead{(13)}&\colhead{(14)}
}           
\startdata 
 19.75 .............. &     64    &    0   &	0  &	64&	0 &	0 &    1.000 &    1.000 &    1.000 &	  206.90 &	238.93 &      178.96 &  2.316 \\
 20.25 .............. &     82    &    0   &	0  &	82&	0 &	0 &    0.988 &    0.979 &    0.970 &	  480.55 &	186.23 &      124.56 &  2.682 \\
 20.75 .............. &     130   &    1   &	0  &   130&	1 &	0 &    0.981 &    0.969 &    0.962 &	  689.34 &	233.03 &      173.83 &  2.838 \\
 21.25 .............. &     206   &    2   &	0  &   206&	2 &	0 &    0.973 &    0.951 &    0.941 &	  983.33 &	294.54 &      237.09 &  2.993 \\
 21.75 .............. &     302   &    5   &	0  &   302&	5 &	0 &    0.961 &    0.936 &    0.927 &	 1902.86 &	308.81 &      253.34 &  3.279 \\
 22.25 .............. &     484   &    16  &	0  &   484&    16 &	0 &    0.948 &    0.909 &    0.899 &	 3563.79 &	353.15 &      301.57 &  3.552 \\
 22.75 .............. &     808   &    70  &	0  &   808&    70 &	0 &    0.934 &    0.891 &    0.878 &	 6883.04 &	439.10 &      393.41 &  3.838 \\
 23.25 .............. &     1267  &    213 &	0  &  1267&   212 &	0 &    0.922 &    0.865 &    0.847 &	12073.11 &	583.52 &      542.64 &  4.082 \\
 23.75 .............. &     1941  &    581 &	3  &  1941&   581 &	3 &    0.906 &    0.841 &    0.808 &	22130.66 &	864.11 &      833.89 &  4.345 \\
 24.25 .............. &     2741  &    1114&	4  &  2741&  1114 &	3 &    0.884 &    0.812 &    0.744 &	34401.66 &     1258.68 &     1230.61 &  4.537 \\
 24.75 .............. &     3377  &    1747&	5  &  3376&  1743 &	4 &    0.856 &    0.677 &    0.483 &	50142.56 &     2040.21 &     2017.83 &  4.700 \\
 25.25 .............. &     3891  &    2175&	4  &  3884&  2140 &	4 &    0.779 &    0.319 &    0.211 &	92527.11 &     7562.70 &     7551.66 &  4.966 \\
 25.75 .............. &     3520  &    2382&	11 &  3505&  2134 &	9 &    0.474 &    0.164 &    0.136 &   163445.76 &    26245.47 &    26235.68 &  5.213 \\[-0.2cm]
\enddata
\tablecomments{Columns are as in Table \ref{Tab:countsu}.}
\end{deluxetable}
\clearpage

\end{landscape}

\end{document}